\documentclass[prd,twocolumn,a4paper,nofootinbib,superscriptaddress,eqsecnum]{revtex4}

\usepackage{geometry} 
\usepackage{graphicx,psfrag} 
\usepackage{mathrsfs}
\usepackage{amsmath,amsfonts,amssymb} 
\usepackage{multirow}
\usepackage{bm} 
\usepackage{comment} 
\usepackage{hyperref}
\usepackage{placeins} 
\usepackage[normalem]{ulem} 
\usepackage{booktabs} 
\usepackage{xspace}

\newcommand{\be}{\begin{equation}} 
\newcommand{\ee}{\end{equation}}
\newcommand{\bea}{\begin{eqnarray}} 
\newcommand{\eea}{\end{eqnarray}}
\newcommand{\bel}{\begin{align}} 
\newcommand{\eel}{\end{align}}
\newcommand{\bse}{\begin{subequations}}
\newcommand{\ese}{\end{subequations}} 
\newcommand{\vv}[1]{\bm{#1}}

\DeclareMathOperator{\Real}{Re} 
\DeclareMathOperator{\Imag}{Im}
\usepackage{color}
\definecolor{cyan}{rgb}{0,0.9,0.9}
\definecolor{orange}{rgb}{0.9,0.5,0} 
\definecolor{magenta}{rgb}{1,0,1}
\definecolor{purple}{rgb}{0.8,0.4,0.8}
\definecolor{gray}{rgb}{0.87,0.87,0.87}
\definecolor{mgreen}{rgb}{0.1,0.8,0.1}
\definecolor{myblue}{rgb}{0.2,0.6,0.8}

\newcommand{\SGRID}{{SGRID}\xspace}

\newcommand{\mean}[1]{\langle#1\rangle}

\begin{document}

\title{ Binary Neutron Stars with Generic Spin, Eccentricity, Mass
  ratio, and Compactness - Quasi-equilibrium Sequences and First
  Evolutions }

\author{Tim \surname{Dietrich}}
\affiliation{Theoretical Physics Institute, University of Jena, 07743
  Jena, Germany}

\author{Niclas \surname{Moldenhauer}}
\affiliation{Theoretical Physics Institute, University of Jena, 07743
  Jena, Germany}

\author{Nathan~K.\ \surname{Johnson-McDaniel}}
\affiliation{International Centre for Theoretical Sciences, Tata
  Institute of Fundamental Research, Bengaluru 560012, India}

\author{Sebastiano \surname{Bernuzzi}}
\affiliation{Theoretical Astrophysics, California Institute of
  Technology, 1200 E California Blvd, Pasadena, California 91125, USA}
\affiliation{DiFeST, University of Parma, and INFN Parma, I-43124
  Parma, Italy}  
  
\author{Charalampos~M.\ \surname{Markakis}}
\affiliation{Mathematical Sciences, University of Southampton,
  Southampton SO17 1BJ, United Kingdom}

\author{Bernd \surname{Br\"ugmann}}
\affiliation{Theoretical Physics Institute, University of Jena, 07743
  Jena, Germany}

\author{Wolfgang \surname{Tichy}}
\affiliation{Department of Physics, Florida Atlantic University, Boca
  Raton, FL 33431 USA}

\date{\today}

\begin{abstract} 

Information about the last stages of a binary neutron star inspiral
and the final merger can be extracted from quasi-equilibrium configurations 
and dynamical evolutions. In this article, we
construct quasi-equilibrium configurations for different spins,
eccentricities, mass ratios, compactnesses, and equations of state.
 For this purpose we employ the \SGRID code, which allows us to
construct such data in previously inaccessible regions of
the parameter space. In particular, we consider spinning neutron
stars in isolation and in binary systems; we
incorporate new methods to produce highly eccentric and eccentricity
reduced data; we present the possibility of computing data for significantly
unequal-mass binaries with mass ratios $q \simeq 2$; and we create
equal-mass binaries with individual compactness up to
$\mathcal{C}\simeq0.23$.
As a proof of principle, we explore the dynamical evolution of three new
configurations.
First, we simulate a $q=2.06$ mass ratio which is the highest 
mass ratio for a binary neutron star evolved in numerical relativity to date. 
We find that mass transfer from the companion star sets in a few revolutions before
merger and a rest mass of $\sim10^{-2}M_\odot$ is transferred between
the two stars. This amount of mass accretion corresponds to $\sim 10^{51}$~ergs
of accretion energy. This configuration also ejects a large amount of material 
during merger ($\sim 7.6 \times 10^{-2} M_\odot$), imparting a substantial kick to the
remnant neutron star.
Second, we simulate the first merger of a precessing binary neutron star.
We present the dominant modes of the gravitational waves for the
precessing simulation, where a clear imprint of the precession is
visible in the $(2,1)$ mode. 
Finally, we quantify the effect of an eccentricity reduction procedure
on the gravitational waveform. The procedure improves the waveform
quality and should be employed in future precision studies. However,
one also needs to reduce other errors 
in the waveforms, notably truncation errors, in order for the improvement due to
eccentricity reduction to be effective.
\end{abstract}

\maketitle

\tableofcontents

\section{Introduction}
\label{sec:Introduction}

The majority of coalescing binary neutron star systems are often
expected to have negligible eccentricity, low spins, and be very
close to equal-mass (with masses around $1.35M_\odot$).  These
expectations are based on the masses and spins of the population of
observed binary neutron stars where at least one star is seen as a
radio pulsar (see, e.g.,~\cite{Lat12,LatXX}), combined with the
efficient shedding of eccentricity due to gravitational waves (GWs)
during the long inspiral \cite{Pet64,KowBulBel11}. However, this
observed population is quite small, currently consisting of
around $12$ systems.  For many of these systems, the evidence that the
companion of the pulsar is in fact another neutron star is indirect,
at best---the companion could still be a fairly massive white dwarf (see, e.g.,
Sec.~8 in~\cite{LeeKasSta15} for some discussion of this
issue). Moreover, of these $12$ systems, only $7$ have well-determined
masses, and only $6$ systems (all with well-determined masses) will
merge within a Hubble time, and thus contribute directly to merger
rate calculations; see, e.g., Table~2 of~\cite{SwiRosMcL15} and
Table~1 in~\cite{Lat12,LatXX}. It is thus unclear to what extent these
small spins, medium masses, and small mass ratios are just a selection
effect. 

On the one hand, population synthesis models for binaries formed ``\emph{in situ}'' (e.g.,~\cite{DomBelFry12}) predict a much
wider range of masses and mass ratios than those currently observed; see the
discussion in Appendix~\ref{sec:Population_Synthesis_M}. 
Also the spins at merger could potentially be considerably
higher than those observed in binary pulsars to date, which is discussed in
Appendix~\ref{sec:BNS_spins}. 
On the other hand, dynamical capture in dense stellar
regions, such as globular clusters, offers the possibility of forming
quite exotic objects, such as double millisecond pulsars~\cite{BenDow14,SigHer92}. 
In fact, there is good evidence that dynamical capture and exchange interactions involving neutron stars
are a frequent occurrence in globular clusters~\cite{VerFre14}. Binaries formed by dynamical capture might 
also have nonnegligible eccentricity at merger~\cite{LeeRuiVen09}.

Numerical simulations in full general relativistic hydrodynamics are
the only way to make accurate theoretical predictions for the
properties of these systems in the time period around merger. 
A prerequisite for all simulations are accurate initial data that
solve the Einstein constraint equations along with the Euler equation
on the initial hypersurface and also describe the physical system one
wants to study at some instant of time. Generally, one also wants to
have this time be not too far from merger, to avoid an excessively
expensive computation to reach the merger. 
Additionally, quasi-equilibrium sequences of initial data at different
separations can 
be used to study certain pre-merger properties of these systems 
without a full dynamical evolution.

Given the potential diversity of the population of coalescing neutron
stars in the universe, it is important to be able to generate accurate
initial data for as much of the potential parameter space as possible.
In particular, even relatively small spin and eccentricity can
significantly bias measurements of the neutron star tidal
deformabilities, affecting their ability to constrain the nuclear
equation of state~\cite{Fav14,AgaMeiDel15}.  Also note that higher
mass-ratio neutron star systems, even if rare, are quite interesting
from a gravitational wave data analysis standpoint, since
the individual masses of a $q = 2.5$ system could be measured 
much more precisely than for equal-mass systems~\cite{RodFarRay14}.

There are a number of well-developed codes for computing binary neutron star
initial data in certain portions of the parameter space, most notably the open source spectral code
LORENE~\cite{LORENE}. Other codes include the Princeton group's
multigrid solver~\cite{EasRamPre12}, BAM's multigrid
solver~\cite{MolMarJoh14}, the COCAL code~\cite{TsoUryRez15}, the SpEC
code's spectral solver~\cite{FouKidPfe08,2015APS..APRY13004P,TacYY}, 
and our spectral code \SGRID~\cite{Tic09a,Tic11,Tic12}. All
these codes are incapable of reaching certain portions of the possible
binary neutron star parameter space.  In particular, they cannot
generate consistent initial data with (noncorotating) spin with
specified eccentricities.  They also generally have difficulty
reaching large compactnesses and high mass ratios. Additionally, when
one evolves quasicircular initial data computed using these codes, one
obtains an eccentricity of $\sim 10^{-2}$, which is orders of
magnitude larger than we would expect in most binary neutron star
systems (see, e.g.,~\cite{KowBulBel11}).
There are standard methods for iteratively reducing eccentricity in
binary black hole initial data
(e.g.,~\cite{BuoKidMro10,TicMar10,MroPfe12,PueHusHan12}). However, it is
not possible to apply these methods to binary neutron star initial data
generated with the standard helical Killing vector technique. Such
methods can only be applied to consistent initial data if one
appropriately generalizes the helical Killing vector, as was only done
recently in~\cite{KyuShiTan14,MolMarJoh14}. Finally, there is no known
way of including magnetic fields consistently in binary neutron star
initial data---they are added by hand in all studies that include them.

Of course, it is possible to generate less accurate initial data in
large regions of parameter space if one does not solve the Euler
equation and possibly not even the Einstein constraint equations
(e.g., using superpositions of boosted isolated star solutions, possibly
with constraint solving, as was done in,
e.g.,~\cite{GolBerThi11,EasRamPre12,KasGalAli13,TsaMar13,KasGal14,EasPasPre15}). However,
if one uses inconsistent initial data to initialize a simulation, it
is always unclear how accurate the resulting simulation will be. 

Recently, there has been progress in generating consistent initial
data in all these portions of parameter space (except for magnetic
fields).  One of us presented a method for constructing consistent
binary neutron star initial data with arbitrary spin using the
constant rotational velocity (CRV) approach in~\cite{Tic11} and
implemented the method in \SGRID in~\cite{Tic12}. This method has now
also been implemented by other groups~\cite{TsoUryRez15,TacYY}. 
There has also been some work on obtaining somewhat high mass ratios (up to $q =
1.5$~\cite{ShiTan06a,DieBerUje15}) and high compactnesses (up to
$\mathcal{C} \simeq 0.26$~\cite{TanShi10}),
but neither of these are close to the maximum mass ratios (at least
$\sim 2$; possibly up to $\sim 3$ for a large maximum neutron star
mass) and compactnesses (up to $\sim 0.3$) that are (at least in
principle) possible.  We also presented a
method for generating consistent binary neutron star initial data with
arbitrary eccentricity, including the possibility of reducing the
eccentricity present in standard quasicircular data,
in~\cite{MolMarJoh14}.
Concurrently,~\cite{KyuShiTan14} 
applied a similar method for eccentricity reduction of binary neutron
star initial data.

Here, we use \SGRID to construct binary neutron star initial data
pushing in all these directions.  We have implemented the eccentric
neutron star binary initial data construction method
from~\cite{MolMarJoh14} in \SGRID, allowing us to solve for the
velocity potential, which was not feasible in our initial multigrid implementation. We have also
implemented piecewise polytropes in \SGRID, allowing us to construct
initial data with more realistic equations of state (EOSs). We thus show the improvement in the initial
density oscillations of the simple polytropic highly eccentric data
from~\cite{MolMarJoh14} with the new \SGRID data, and also construct
eccentricity-reduced initial data for a simple polytropic and a
realistic EOS. We also compute aligned spin initial data with somewhat
larger spins than in~\cite{BerDieTic13} (as well as for more realistic
EOSs). Finally, we illustrate the ability of \SGRID to compute binary initial
data with compactnesses up to $\mathcal{C}\sim 0.23$ as well as to
compute high mass-ratio initial data ($q = 2.06$) with a realistic
EOS.

Of course, one would like to study the phenomena around merger in these newly accessible portions of the 
binary neutron star parameter space: Around merger, the strongest gravitational wave, electromagnetic, and
neutrino emission happens; see~\cite{FabRas12} for a review of binary
neutron star simulations. Such a study requires dynamical evolutions, which will start from 
initial data provided by a member of a quasi-equilibrium sequence.
In recent years there has been significant
work on improving binary neutron star simulations on many fronts,
notably by including more realistic equations of state from
piecewise polytropes~\cite{ShiTanUry05} to finite temperature
EOSs~\cite{SekKiuKyu11b,SekKiuKyu11},
magnetohydrodynamics~\cite{AndHirLeh08a,LiuShaEti08,GiaRezBai09,KiuKyuSek14,DioAliRez15},
and neutrino cooling~\cite{SekKiuKyu11b,GalKasRez13,SekKiuKyu15}. There are now
some simulations that include all three of these improvements at
once~\cite{PalLieNei15}.
For our simulations, we use the BAM code in 
its newest version~\cite{BruGonHan06,ThiBerBru11,DieBerUje15}.

In this work, we present three different dynamical simulations. 
We consider a binary configuration with a mass ratio of $q=2.06$, 
which is the largest mass ratio binary neutron star system ever evolved in full general relativity. 
This large mass ratio is particularly interesting, since the system undergoes mass transfer prior 
to merger and during the merger process a large amount of material 
gets unbound and is ejected from the system. 
The second example we consider is an unequal mass configuration, 
where the two neutron stars have spins misaligned with the orbital angular momentum; 
this is the first simulation of the merger of a precessing binary neutron star system. Here we find the
same close relation between the precessing system (viewed in the nonprecessing frame) and the
aligned-spin analogue found for binary black holes in previous studies (e.g.,~\cite{SchHanHus10}).
As a third test, we perform a simulation of an equal mass setup, 
with and without the eccentricity reduction procedure, where 
one can see a clear improvement in the waveform quality due to the 
eccentricity reduction procedure. 

The article is structured as follows: In Sec.~\ref{sec:Method}, we
recall the most important equations for our initial data construction and the general framework employed
in \SGRID. In Sec.~\ref{sec:NumericalMethods} we describe the
implementation and the numerical methods focusing on the recent
upgrades to the code.  We summarize the main results of our work in
Sec.~\ref{sec:Results}, where we compute binary neutron star (BNS)
systems in quasi-equilibrium sequences varying the spin, eccentricity,
mass ratio, and compactness and show convergence of the numerical method. 
In Sec.~\ref{sec:Dynamical_Evolutions},
we evolve a $q=2.06$ nonspinning system, an unequal mass precessing configuration, 
and an equal-mass system with eccentricity reduction. 
We conclude in Sec.~\ref{sec:Conclusion}. 
The appendices summarize binary neutron star
population synthesis predictions for more extreme systems, along with some
issues in predicting the spins expected in binary neutron stars at
merger; an alternative derivation of the CRV approach; 
and results from our study of single CRV-stars.

Throughout this work we use geometric units, setting $c=G=M_\odot=1$,
though we will sometimes include $M_\odot$ explicitly or quote values
in cgs units for better understanding.  Spatial indices are denoted by
Latin letters running from 1 to 3 and Greek letters are used for
spacetime indices running from 0 to 3.  We always raise and lower
indices with the physical metric ($3$-metric for spatial indices and
$4$-metric for spacetime indices).  We shall also use index-free
notation when convenient, denoting vectors (spatial or spacetime)
using boldface.

\section{Method}
\label{sec:Method}

The following section summarizes the fundamental framework of 
the initial data construction in \SGRID. 
Since no new development was made in the evolution method in this work, 
we refer the reader to~\cite{BruGonHan06,ThiBerBru11,HilBerThi12,DieBerUje15}
for details of the methods used in the BAM code.

\subsection{General Framework}
\label{sec:general_framework}

In this article we investigate BNS systems in quasi-equilibrium.  We
construct such configurations with the help of a~$3+1$ decomposition
of Einstein's field equations \cite{Yor79}. In this section we recast important
equations and derive the specific system of partial differential
equations we solve.

We start writing the spacetime metric in the form
\begin{align}
  {\rm d}s^2 & = g_{\mu \nu} {\rm d}x^\mu {\rm d}x^\nu \nonumber
  \\ & = -\alpha^2 {\rm d}t^2 + \gamma_{ij} ({\rm d}x^i +
  \beta^i{\rm d}t)({\rm d}x^j + \beta^j {\rm d}t),
\end{align}
where $\alpha$ is the lapse function and $\beta^i$ the shift. The
spatial metric induced on 3-dimensional hypersurfaces of constant
$t$ is denoted by $\gamma_{ij}$.  By performing the~$3+1$
decomposition, the field equations split into two sets, namely the
Hamiltonian and momentum constraints, \bse
\begin{eqnarray}
  R- K_{ij} K^{ij} +K^2 &=& 16 \pi \rho, \\ D_j \left( K^{ij}
  -\gamma^{ij} K \right) & = & 8 \pi j^i,
\end{eqnarray}
\ese and the evolution equations \bse
\begin{align}
  \partial_t \gamma_{ij} & = - 2 \alpha K _{ij} + \mathcal{L}_\beta
  \gamma_{ij}, \\ \partial_t K_{ij} & = \alpha (R_{ij} - 2 K_{il}
  K^l{}_j + K K_{ij}) - D_i D_j \alpha \nonumber \\ & \quad
  +\mathcal{L}_\beta K_{ij} - 8 \pi \alpha S_{ij} + 4 \pi \alpha
  \gamma_{ij} (S-\rho).
\end{align}
\ese Here the Ricci tensor $R_{ij}$ and Ricci scalar $R$ are
computed from the spatial metric $\gamma_{ij}$ with compatible
covariant derivative operator $D_i$, and the extrinsic curvature
$K_{ij}$ is given by
\begin{equation}
  K_{ij} = - \frac{1}{2 \alpha} \left( \partial_t \gamma_{ij}
  -\mathcal{L}_\beta \gamma_{ij}\right),
\end{equation}
where $\mathcal{L}_\beta$ denotes the Lie derivative along the
vector field $\beta^i$. We use $K := \gamma^{ij}K_{ij}$.  The source
terms (energy density, flux, and stress tensor) contained in the
right hand side of the equations are \bse
\begin{eqnarray}
  \rho & := & \ \ T_{\mu \nu} n^\nu n^\mu, \label{eq:defrho} \\ j^i
  & := & - T_{\mu \nu} n^\mu \gamma^{\nu i}, \label{eq:defj}
  \\ S^{ij} & := & \ \ T_{\mu \nu} \gamma^{\mu i } \gamma^{\nu
    j}, \label{eq:defS}
\end{eqnarray}
\ese where $n^\mu$ is the normal vector to the hypersurface; we also
write $S := \gamma^{ij}S_{ij}$.  In this work we assume that the
matter can be described as a perfect fluid with stress-energy tensor
\begin{equation}
  T^{\mu \nu} = \left[ \rho_0 (1+\epsilon) + p \right] u^\mu u ^\nu
  + p g^{\mu \nu}, \label{eq:Tmunu}
\end{equation}
where $\rho_0$, $\epsilon$, $p$, and $u^\mu$ denote the mass
density, the internal energy, the pressure, and the four-velocity of
the fluid, respectively.  Inserting \eqref{eq:Tmunu} in the
definition of $\rho$, $j^i$, and $S^{ij}$
[Eqs.~\eqref{eq:defrho}--\eqref{eq:defS}], one can obtain explicit
expressions for the matter quantities entering the right-hand sides
of the constraint and evolution equations in terms of the perfect
fluid variables.

To ensure consistency, we have to solve the constraint equations
along with the matter equation
\begin{equation}
  \nabla_\nu T^{\mu \nu} = 0 \label{eq:energyconservation}
\end{equation}
and the continuity equation
\begin{equation}
  \nabla_\nu (\rho_0 u^\nu) = 0 , \label{eq:continuity}
\end{equation}
which comes from the conservation of the baryon number.
Equation~\eqref{eq:energyconservation} can be written as the
relativistic Euler equations
\begin{equation}
  \left[\rho_0 (1+\epsilon) + p \right] u ^\nu \nabla_\nu u^\mu = -
  (g^{\mu \nu}+u^\mu u^\nu) \nabla_\nu p. \label{eq:euler}
\end{equation}
In many cases it is useful to introduce the specific enthalpy
\begin{equation}
  h = 1 + \epsilon + p/\rho_0 .
\end{equation}
Then the Euler equations can be written as
\begin{equation}
  u ^\nu \nabla_\nu (h u_{\mu}) + \nabla_\mu h = 0 .
  \label{eq:Euler_h}
\end{equation}

To further simplify the equations, we split the three-metric into a
conformal factor $\psi$ and the corresponding conformal metric
$\bar{\gamma}_{ij}$, writing
\begin{equation}
  \gamma_{ij} = \psi^4 \bar{\gamma}_{ij}. \label{eq:confdecomp1}
\end{equation}
Similarly, we express the extrinsic curvature in terms of a
trace-free piece $A_{ij}$, writing
\begin{equation}
  K_{ij} = A_{ij} +\frac{1}{3}\gamma_{ij} K.  \label{eq:confdecomp2}
\end{equation}
Inserting~\eqref{eq:confdecomp1} and~\eqref{eq:confdecomp2} in the
constraint equations gives their final general form before making
any additional assumptions.

\subsubsection{Assumptions for metric variables}

In order to obtain a stationary configuration that is appropriate
for initial data, we include additional assumptions which bring the
entire system in an elliptic form.  The first assumption is the
existence of an approximate symmetry vector\footnote{Notice that in
  this work we employ a different notation than presented
  in~\cite{Tic11,Tic12} where the Killing vector was denoted with
  $\xi$.  We change our notation to emphasize that we assume a
  symmetry vector, not necessarily a helical Killing vector.}
\begin{equation}
  \vv{k} = \vv{\partial_t} + g_x \vv{\partial_x} + g_y
  \vv{\partial_y}, \label{eq:killing_vector_fundament}
\end{equation}
where the functions $g_x$, $g_y$ are chosen according to the problem
we want to tackle, allowing us to construct quasicircular,
eccentric, and eccentricity-reduced configurations, as discussed in
Sec.~\ref{Sec:symmetry_vector}.  Together with the existence of $\vv{k}$
and the assumption
\begin{equation}
  \mathcal{L}_{\vv{k}} g_{\mu \nu} = 0, \label{eq:geomassump1}
\end{equation}
we assume spatial conformal flatness\footnote{See, e.g., Sec.~III~A
  in~\cite{MolMarJoh14} for a discussion of the limitations of the
  conformal flatness assumption, but note that these caveats are
  quite mild for the systems we are considering. In particular,
  while the conformal flatness assumption is a significant
  obstruction to constructing high-spin binary black hole initial
  data (with dimensionless spin $j \gtrsim 0.93$) as discussed in, e.g., \cite{LovOwePfe08}, neutron stars cannot spin
  rapidly enough for this to be a problem. Specifically, neutron stars
  have maximum dimensionless spins of at most $0.7$, except for strange
  quark stars, which can have dimensionless spins greater
  than~$1$; see, e.g., Figs.~3 and~6 in~\cite{LoLin11}. Since we are
  not considering strange quark stars in this work, we do not expect
  that the assumption of conformal flatness places any restrictions
  on the parameter space we can cover.} and maximal slicing \bse
\begin{eqnarray}
  \bar \gamma_{ij} & = & f_{ij}, \\ K := \gamma_{ij} K^{ij} & = & 0,
\end{eqnarray}
\ese where $f_{ij}$ denotes the flat metric which simplifies to
$\delta_{ij}$ in Cartesian coordinates.  We preserve these
conditions in time (at least infinitesimally), that is, $\partial_t
\bar{\gamma}_{ij} = \mathcal{L}_{\vv{k}} \bar{\gamma}_{\mu \nu} =0$ and
$\partial_t K = \mathcal{L}_{\vv{k}} K = 0$.  Note that we have not used
the assumption of an approximate symmetry vector in obtaining these
last equalities, which is the usual approach, with conformal
flatness and maximal slicing imposed afterwards, as discussed in
Sec.~III~A of~\cite{MolMarJoh14}.  We also obtain
\begin{equation}
  A^{ij} = \frac{1}{2\psi^4 \alpha} (\textbf{L} \beta)^{ij},
\end{equation}
with $(\textbf{L} \beta)^{ij}= \bar{D}^i \beta^j + \bar{D}^j \beta^i
- \frac{2}{3} \delta^{ij} \bar{D}_k \beta^k$, where $\bar{D}_i$
denotes the flat-space covariant derivative.

Finally, these assumptions lead to the following partial
differential equations \bse
\label{eq:final_eqs}
\begin{align}
  \bar{D}^2 \psi & = - \frac{\psi^5}{32 \alpha^2} (\textbf{L}
  B)^{ij} (\textbf{L} B)_{ij} - 2 \pi \psi^5 \rho \label{eq:metric1}
  \\ \bar{D}_j (\textbf{L} B)^{ij} &= (\textbf{L} B)^{ij} \bar{D}_j
  \ln\left(\frac{\alpha}{\psi^6}\right) + 16 \pi \alpha \psi^4
  j^i \label{eq:metric2} \\ \bar{D}^2(\alpha \psi) & = \alpha \psi
  \left( \frac{7 \psi^4}{32 \alpha^2} (\textbf{L} B)^{ij}
  (\textbf{L} B)_{ij} + 2 \pi \psi^4 (\rho + 2S)
  \right), \label{eq:psialpha}
\end{align}
\ese with $\bar{D}_i =\partial_i$ in Cartesian coordinates and $B^i
= \beta^i+k^i + \Omega \epsilon_{ijl} (x^j-x^j_\text{CM}) {a}^l$,
where $x^j_\text{CM}$ is the center of mass, $\Omega$ the orbital
frequency, $\epsilon_{ijl}$ is the Levi-Civita symbol, and ${a}^l$
is a unit vector pointing along the direction of the orbital angular
momentum.

\subsubsection{Assumptions for matter variables}

Similarly to the metric variables, we also need to make assumptions
for the matter fields.  These assumptions are discussed in more
detail in~\cite{Tic11,MolMarJoh14} and briefly described below.  We
start by splitting the four-velocity into a piece along $k^\mu$, and
one orthogonal to it, which we call $V^\mu$. Specifically, we write
\begin{equation}
  u^\mu = u^0 (k^\mu + V^\mu), \label{eq:matterasm0}
\end{equation}
with $u^0=-u^\nu n_\nu/\alpha$.  Next we define
\begin{equation}
  p_\mu = h u_\mu .
  \label{eq:pdef}
\end{equation}
While we can assume that $\mathcal{L}_{\vv{k}} p_\mu = 0$ for irrotational
binaries, this equation is, in general, not satisfied for spinning
neutron stars---see appendix~A of~\cite{Tic11}.  Thus, we introduce
the canonical momentum 1-form of a fluid element
\begin{equation}
  {p}_i = \gamma_i{}^\mu p_\mu
\end{equation}
and split $p_i$ into an irrotational part  which can be
written as the gradient of a potential, 
 $D_i \phi$, and a rotational part $w_i$:
\begin{equation}
  p_i = D_i \phi + w_i \label{eq:matterasm0.5}
\end{equation}
or equivalently in four-dimensions
\begin{equation}
  p_\mu = \nabla_\mu \phi + w_\mu .
  \label{eq:p_of_phi_w}
\end{equation}
Although $\mathcal{L}_{\vv{k}} p_\mu \neq 0 $, we assume \bse
\begin{eqnarray}\textbf{\textbf{}}
  \mathcal{L}_{\vv{k}} (\rho u^0) &=&
  0, \label{eq:matterasm1}\\ \gamma_i{}^\mu \mathcal{L}_{\vv{k}}
  (\nabla_\mu \phi) & = & 0, \label{eq:matterasm2} \\ \gamma_i{}^\mu
  \mathcal{L}_{\bar{\vv{k}}} w_\mu & = & 0, \label{eq:matterasm3}
\end{eqnarray}
\ese with
\begin{equation}
  \bar{k}^\mu := \frac{\nabla^\mu \phi}{h u^0} =: k^\mu - \Delta
  k^\mu ,
  \label{eq:kbardef}
\end{equation}
which is parallel to the worldline of the star's center.  At this
point useful relations can be derived immediately, \bse
\begin{align}
  \gamma_i{}^\nu \mathcal{L}_{\vv{k}} p_\nu
  \overset{\text{\eqref{eq:matterasm2}}}{=} &\gamma_i{}^\nu
  \mathcal{L}_{\vv{k}} w_\nu = \gamma_i{}^\nu \mathcal{L}_{\bar{\vv{k}}+\vv{\Delta
    k}} w_\nu \nonumber
  \\ \overset{\text{\eqref{eq:matterasm3}}}{=}&\gamma_i{}^\nu
  \mathcal{L}_{\vv{\Delta k}} w_\nu = {}^{(3)} \mathcal{L}_{\vv{\Delta}
    \tilde{\vv{k}}} w_\nu, \label{eq:matrel1} \\ V^i + \Delta \tilde{k}^i
  \overset{\text{\eqref{eq:matterasm0}}}{=}&
  \frac{u^i}{u^0}-k^i+\Delta k^i = \frac{u^i}{u^0}-\bar{k}^i
  \overset{\text{\eqref{eq:matterasm0.5}}}{=} \frac{w^i}{h
    u^0} \label{eq:matrel2}
\end{align}
\ese with the three-dimensional Lie derivative ${}^{(3)}
\mathcal{L}$ and $\Delta k^\mu = (0,\Delta \tilde{k}^i)$.
Additionally, from the fact that $h u^0$ and $\gamma_{ij}$ are
approximately constant along $\frac{w^i}{h u^0}$ we obtain
\begin{equation}
  ^{(3)} \mathcal{L}_{\vv{V}+\vv{\Delta} \tilde{\vv{k}}} w_i = \frac{w_i}{h u^0}
  {}^{(3)}\mathcal{L}_{\frac{\vv{w}}{h u^0}} h u^0 + w^j \ ^{(3)}
  \mathcal{L}_{\frac{\vv{w}}{h u^0}} \gamma_{ij} \approx
  0. \label{eq:matrel3}
\end{equation}

Plugging~\eqref{eq:matterasm0} into the continuity
equation~\eqref{eq:continuity} and using~\eqref{eq:geomassump1},
\eqref{eq:matterasm1} we get
\begin{equation}
  D_i \left( \rho_0 \alpha u^0 V^i \right) = 0.
\end{equation}
Similarly, the Euler equation~\eqref{eq:euler} together with
\eqref{eq:matrel1}, \eqref{eq:matrel2}, and \eqref{eq:matrel3} can
be simplified to
\begin{equation}
  D_i \left( \frac{h}{u^0} + V^j D_j \phi \right) = 0,
\end{equation}
which can be integrated to obtain
\begin{equation}
  \frac{h}{u^0} + V^j D_j \phi = - C =
  \text{const.} \label{eq:frist_integral}
\end{equation}
Note that a simple derivation of this first integral, which
makes use of the Cartan identity, can be found in Appendix
\ref{sec:appendix_firstintegral}.

The constant $C$ is chosen during the numerical iteration process in
such a way that the baryonic mass of each star stays constant; see
Sec.~\ref{sec:NumericalMethods}.

In general the velocity is given by
\begin{equation}
  V^i = \frac{D^i \phi + w^i}{h u^0} - (\beta^i + k^i),
\end{equation}
which brings the continuity equation in the form
\begin{equation}\label{eq:velocity_potential}
  D_i \left[ \frac{\rho_0 \alpha}{h} (D^i \phi + w^i) - \rho_0
    \alpha u^0 (\beta^i+k^i) \right] = 0.
\end{equation}
This equation can be seen as a nonlinear elliptic equation in $\phi$
and especially needs known boundaries at the star's surface to be
solved.  To handle this issue we introduce surface-fitted
coordinates in the subsequent section.\\ Integrating and using
$u^\mu u_\mu = -1$ leads to
\begin{equation}
  h = \sqrt{L^2 - (D_i \phi +w_i) (D^i \phi +
    w^i)}, \label{eq:h_star}
\end{equation}
with \bse
\begin{align}
  L^2 & = \frac{b + \sqrt{b^2 - 4 \alpha^4 \left[ (D_i\phi + w _i )
        w^i\right]^2}}{2 \alpha^2},\\ b & = \left[ (k^i +\beta^i)
    D_i \phi -C \right]^2 + 2 \alpha^2 (D_i \phi+w_i) w^i.
\end{align} 
\ese

For the data constructed with the CRV-approach we choose throughout
the entire paper
\begin{equation}
  \label{w_choice}
        {w}^i = \epsilon^{ijk} \omega^j (x^k - x_{C*}^k),
\end{equation}
where $x_{C*}^i$ gives the coordinate position of the center of the
star and $\omega^i$ is an arbitrary angular velocity vector.

\subsection{Specifying the symmetry vector}
\label{Sec:symmetry_vector}

Helical Killing vectors are well known and commonly used constructs
in numerical relativity to construct binaries on circular orbits that are stationary in
a corotating frame. The general expression for these vectors is
given by
\begin{equation}
  \label{eq:helicalKV}
  k_{\rm{qc}}^\alpha = t^\alpha +\Omega_{\rm{qc}} \varphi^\alpha =
  t^\alpha + \Omega_{\rm{qc}} (x y^\alpha - y x^\alpha),
\end{equation}
where we used the vectors $\bm{t} = \partial_t$, $\bm{x} =
\partial_x$, $\bm{y} = \partial_y$ and $\bm{\varphi} =
\partial_\varphi$ that generate translations in the $t$, $x$, and
$y$ directions, respectively, and rotations in the $\varphi$
direction. In \cite{MolMarJoh14}, we showed how to generalize this
vector to incorporate eccentricity as well as radial
velocity. However, because our initial numerical implementation of
the method used a Cartesian grid, without the surface-fitted coordinates needed to solve for the velocity potential, we settled on a constant
fluid-velocity approximation 
instead of solving Eq.~\eqref{eq:velocity_potential}. We are now
able to solve the full set of equations for the first time. In the
following, we will briefly summarize how we generalize the
standard approximate helical Killing vector to an approximate
helical symmetry vector that incorporates radial velocity and
eccentricity.

To find a vector $k^\alpha$ that approximately Lie-derives the flow
we make the following two assumptions: (i) Such a $k^\alpha$ exists. (ii) $k^\alpha$
is along the motion of the star center.

In order to describe eccentric orbits we make the additional
assumption that (iii) each star center moves along a segment of an
elliptic orbit at apoapsis.  Since we only need a small segment of
an orbit near apoapsis, we will approximate this segment by the
circle inscribed into the elliptical orbit there.  Then the radii of
the inscribed circles are
\begin{equation}
  \label{eq:r_c_of_d}
  r_{{\rm c}_{1,2}} = (1-e) d_{1,2} ,
\end{equation}
where $d_1$ and $d_2$ are the distances of the particles from the
center of mass at apoapsis and $e$ is the eccentricity parameter for
the elliptic orbit~\cite{MolMarJoh14}.  These two inscribed circles
are not centered on the center of mass, but on the points \be
\label{eq:x_circlecenters}
x_{{\rm c}_{1,2}} = x_{1,2} \mp r_{{\rm c}_{1,2}} = x_\text{CM} +
e (x_{1,2} - x_\text{CM}) , \ee where we have used $d_{1,2} =
|x_{1,2} - x_\text{CM}|$ and assumed that apoapis occurs on the
$x$-axis. (The upper and lower signs correspond to the subscripts
$1$ and $2$, respectively.)  Assumption (ii) then tells us that
the approximate Killing vector for elliptic orbits must have the
form \be
\label{eq:ellipKV}
k^\alpha_{\text{ecc} 1,2} = t^ \alpha + \Omega \, [(x-x_{{\rm
      c}_{1,2}}) y^\alpha - y \,x^\alpha] \ee near each
star~\cite{MolMarJoh14}.

The next step is to allow a slow inspiral of the orbit due to energy
loss because of GW emission. This means the orbital velocity will
have a small radial component in the direction of the center of
mass.  Assumption (ii) then tells us that assumption (iii) above
needs to be modified to include a radial piece. We assume that the
approximate Killing vector now is 
\be
\label{eq:ellinspiralKV}
k^\alpha_{1,2} = k^\alpha_{\text{ecc} 1,2} + \frac{v_r}{r_{12}}
r^{\alpha} = t^\alpha + \Omega \, [(x-x_{{\rm c}_{1,2}}) y^\alpha
  - y \,x^\alpha] + \frac{v_r}{r_{12}} r^{\alpha} , 
\ee 
which we also refer to as a \textit{helliptical approximate} symmetry vector. Here
$r^{\alpha} = (0,x,y,z)$ points in the radial direction,
$r_{12}=|x_1-x_2|$ is the distance between the star centers, and
$v_r$ is a radial velocity parameter. The radial velocity $v_r$
could be chosen corresponding to the radial velocity of an
inspiralling binary from post-Newtonian calculations, or it can be
obtained from an iterative procedure aimed at reducing the orbital
eccentricity such as the one described in
Sec.~\ref{sec:eccentricity_reduction}. In this case we also have
to adjust the eccentricity parameter $e$ that appears in $x_{{\rm
    c}_{1,2}}$.  The reason is that changing $e$ amounts to 
changing the tangential orbital velocity, which is needed when we
want non-eccentric inspiral orbits.

To have a more physical quantity that will be useful for comparisons, we consider the mean motion,
as in \cite{MolMarJoh14}:
\begin{equation}
 \bar\Omega = 2\pi/T = \Omega (1+e)\sqrt{1-e^2},
\end{equation}
where $T$ is the orbital period.

\section{Code description}
\label{sec:NumericalMethods}

To construct initial data with \SGRID we use the numerical framework
described in~\cite{Tic09a,Tic12}.  In this section we recall important
aspects to give an almost complete picture.  In particular, we
describe the grid configuration, the iteration procedure, and recent
changes which allow us to use more realistic EOS and also compute
configurations with a relatively large mass ratio.
A short discussion about the BAM code will 
be given in Sec.~\ref{sec:Dynamical_Evolutions}.

\subsection{Grid configuration}

\begin{figure*}[t]
  \includegraphics[width=0.98\textwidth]{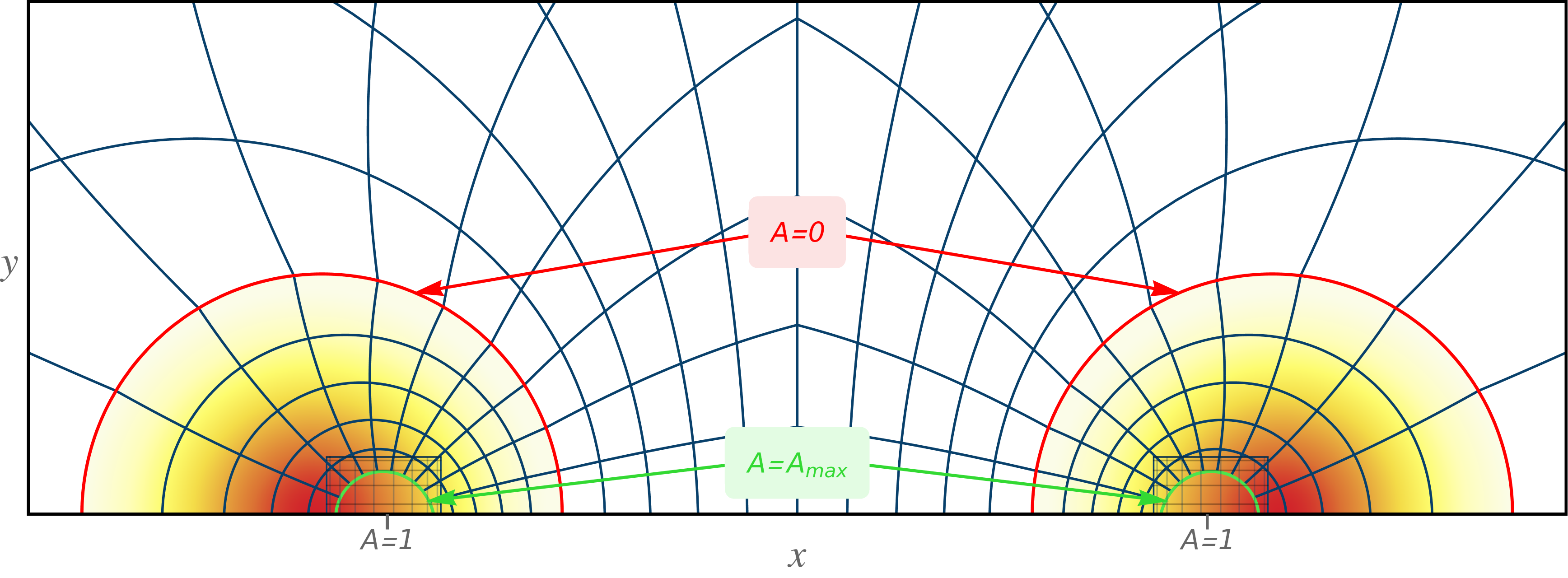}
  \caption{The grid structure in the $xy$-plane for an equal-mass
    configuration.  One can see lines of constant $A$ and $B$ (dark blue
    lines) for $b =16, \sigma_{+} = - \sigma_{-} = 1.304$ and $A_{\rm{max}}=0.5$ with an
    overlay of the density profile of the cross section. Moreover one
    can see the Cartesian boxes with Chebychev grids inside the
    stars.}
  \label{fig:grid}
\end{figure*}

We place the neutron stars along the $x$-axis.  Figure~\ref{fig:grid}
illustrates the part of the computational domain with $y>0$, $z=0$.
The entire grid is built out of six individual domains.  The grid
configuration is not fixed and changes during the computation in
response to changes in the positions of the stars' surfaces.

We follow the approach of~\cite{Ans06,Tic09a} and express Cartesian
coordinates as \bse
\begin{eqnarray}
 x & = & \frac{b}{2} \left( \frac{1}{(X^2 + R^2)^2}+1 \right) (X^2 -
 R^2), \\ y & = & b \left( \frac{1}{(X^2 + R^2)^2}-1 \right) X R
 \cos(\varphi), \\ z & = & b \left( \frac{1}{(X^2 + R^2)^2}+1 \right)
 X R \sin (\varphi),
\end{eqnarray}
\ese with $X\in[0,1]$, $R \in [0,\sqrt{1-X^2}]$, $\varphi \in
     [0,2\pi)$. Furthermore, we transform to coordinates
     $A,B,\varphi$ for the different domains.  The domains covering
     the exterior of the stars and including spatial infinity
     $(A,B)=(1,0)$ employ \bse
\begin{align}
 X &= (1-A) \left[ \Real(C_{\pm}(B,\varphi)) - B
   \Real(C_{\pm}(1,\varphi)) \right] \nonumber \\ &\quad + B
 \cos\left(\frac{A \pi}{4} + (1-A) \arg(C_\pm (1,\varphi))\right),
 \\ R &= (1-A) \left[ \Imag(C_{\pm}(B,\varphi)) - B
   \Imag(C_{\pm}(1,\varphi)) \right] \nonumber \\ &\quad + B
 \sin\left(\frac{A \pi}{4} + (1-A) \arg(C_\pm (1,\varphi))\right).
\end{align}
\ese Since spatial infinity is included in our domain, we can impose
exact Dirichlet boundary conditions: \bse
\begin{eqnarray}
 \lim_{r\rightarrow \infty} \psi &=& 1,\\ \lim_{r\rightarrow \infty}
 B^i & = & 0,\\ \lim_{r\rightarrow \infty} \alpha \psi & =& 1,
\end{eqnarray}
\ese
where $r$ denotes the coordinate distance from the origin.

The inner domain boundary $A=0$ is the star's surface, with
\begin{equation}
 C_\pm (B,\varphi) = \sqrt{ \tanh \left( \frac{\sigma_\pm (B,\varphi)
     + i \pi B}{4} \right)},
\end{equation}
where $\sigma_\pm$ is a function that determines the
shape of the star's surface, and $\pm$ denotes the sign of the
$x$-coordinate, i.e., the left or the right star.  At each star's
surface Eq.~\eqref{eq:velocity_potential} is subject to the boundary
conditions
\begin{equation}
 \left[ (D^i + w^i \phi) - h u^0 (\beta^i + k^i ) \right] D_i \rho =
 0.
\end{equation}
The coordinate transformations inside the stars are
\begin{subequations}
\begin{align}
 X &= (1-A) \left[ \Real(C_{\pm}(B,\varphi)) - B
   \Real(C_{\pm}(1,\varphi)) \right] \nonumber \\ &\quad + B
 \cos(D_\pm) + \delta_\pm (1-B) A, \\ R &= (1-A) \left[
   \Imag(C_{\pm}(B,\varphi)) - B \Imag(C_{\pm}(1,\varphi)) \right]
 \nonumber \\ &\quad + B \sin\left(D_\pm\right)+ \delta_\mp (1-B) A,
\end{align}
\end{subequations}
with
\begin{equation}
 D_\pm := (1-A) \arg(C_\pm(1,\varphi)) + \delta_\mp \frac{\pi}{2}A,
\end{equation}
where $\delta_\pm = 1$ for the star with $x > 0$ and is zero for the
other, and vice versa for $\delta_\mp$.  Unfortunately, the
transformation to $(A,B,\varphi)$ coordinates is singular for $A=1$
(i.e., at the star's center).  To cure this problem, we cover the
center by a Cartesian box with grid points at
\begin{equation}
 {x^i}_k=\frac{x^i_\text{min}-x^i_\text{max}}{2} {\rm cos}\left(
 \frac{k \pi}{n^i-1}\right) + \frac{x^i_\text{min}+x^i_\text{max}}{2},
\end{equation}
where $x^i=(x,y,z)$, with $0 \leq k < n_\text{Cart}$. The Cartesian
boxes cover a region for $A>A_\text{max}$.  The choice of
$A_\text{max}$ allows us to specify the clustering of the grid points.
For large $A_\text{max}$ the Cartesian box is smaller, while for small
$A_\text{max}$ the box is larger.  Thus, introducing a small
$A_\text{max}$ increases the resolution in the outer region of the
stars.  This will be important when piecewise polytropes are employed,
where it is crucial to resolve the crust with a sufficient number of
grid points.  The collocation points in the other regions of the grid
are
\begin{subequations}
\begin{eqnarray}
 A_i &= & \frac{A_\text{max}}{2}\left[ 1- \cos\left( \frac{\pi i }{
     n_A -1} \right) \right], \\ B_j & = & \frac{1}{2}\left[ 1-
   \cos\left( \frac{\pi j }{ n_B -1}\right) \right], \\ \varphi_k & =
 & \frac{2 \pi k }{ n_\varphi},
\end{eqnarray}
\end{subequations}
with $0 \leq i <n_A$, $0 \leq j < n_B$, $0 \leq \varphi < n_\varphi$.
In the $A,B$-directions we use Chebyshev polynomials, while for the
$\varphi$-direction a Fourier expansion is used.  For a typical
configuration, we employ between 20 and 28 points in $A,B$ and 8
points in the $\varphi$-direction.  The Cartesian box is covered with
$n_x=n_y=n_z=n_\text{Cart}=16,\ldots,24$ points, where typically we
choose $n_\text{Cart} = n_A-4$.

Finally, some regularity conditions along the $x$-axis have to be
imposed: In the domains where $(A,B,\varphi)$-coordinates are employed,
we set
\begin{subequations} 
\begin{eqnarray}
 \partial_\varphi F & = & 0,\\ \partial_s F + \partial_s
 \partial_\varphi \partial_\varphi F & = & 0,
\end{eqnarray}
\end{subequations}
with $F\in\{\psi, B^i, \alpha, \phi\}$ and
$s:=\sqrt{y^2+z^2}$.

\subsection{Iteration procedure}

\begin{figure}[t]
 \includegraphics[width=0.38\textwidth]{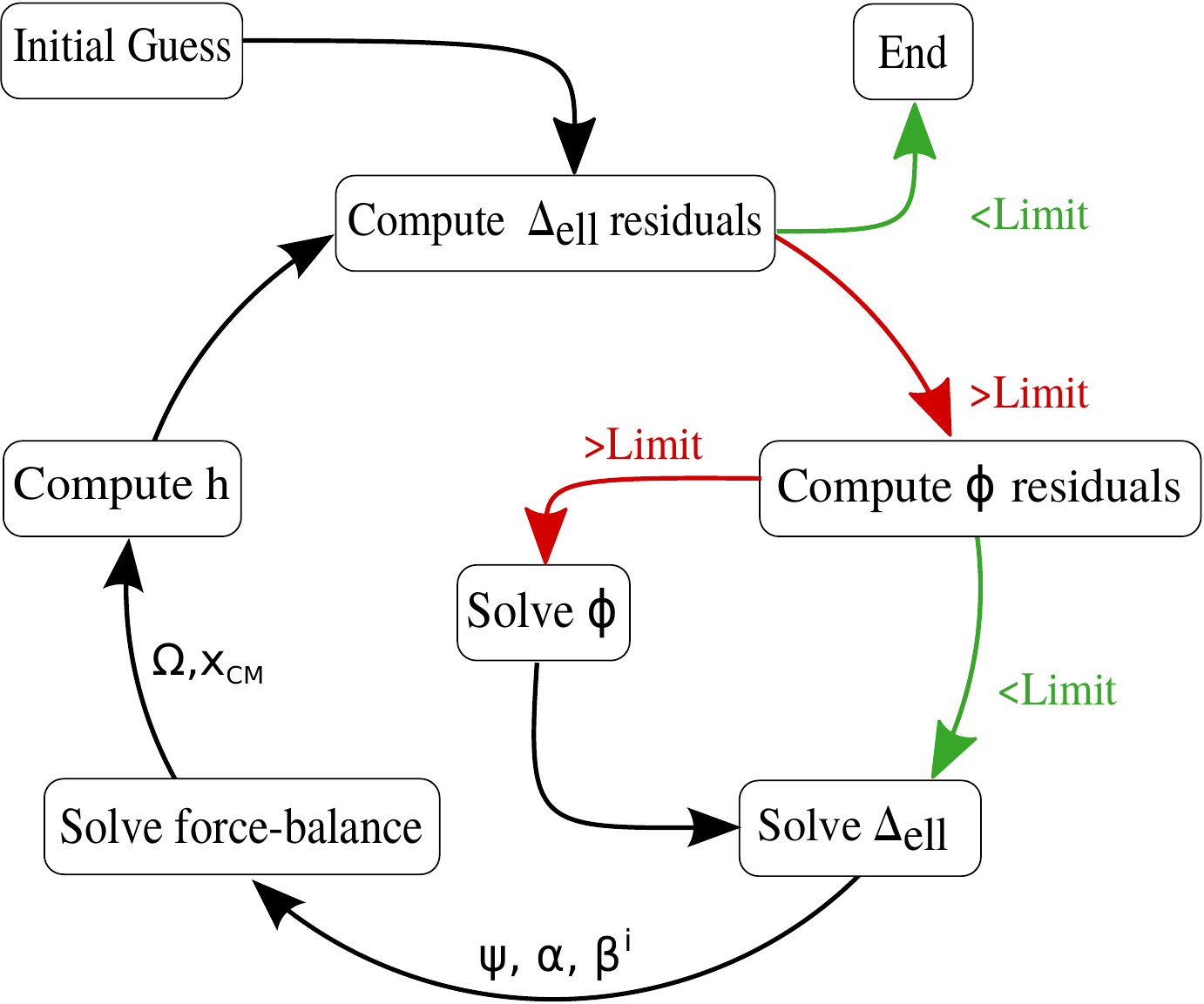}
 \caption{Iteration scheme as outlined in the text.
}
  \label{fig:IterationScheme}
\end{figure}
To solve the coupled system of partial differential equations we
perform a specific iteration procedure first introduced and described
in detail in~\cite{Tic12}.  The scheme is sketched in
Fig.~\ref{fig:IterationScheme} and we describe it in detail in the
following: \\

(i) We start with an initial guess.  This guess can be obtained by
solving the Tolman-Oppenheimer-Volkoff (TOV)~\cite{Tol39,OppVol39a}
equation (and superposing, if we are considering a binary) or given by a previously constructed configuration.  The
velocity potential $\phi$ in each star is set to $\phi=\Omega
(x_{C*}-x_\text{CM}) y$, where $x_{C*}$ is the $x$-coordinate of the
star's center (this initial guess corresponds to a spatially constant velocity field and is exact for rigidly rotating nonrelativistic stars).\\

(ii) In the second step we evaluate the residuals of all elliptic
equations (denoted by $\Delta_\text{ell}$ in
Fig.~\ref{fig:IterationScheme}) and stop if these residuals are below
the prescribed tolerance.\\

(iii) If the residual of Eq.~\eqref{eq:velocity_potential} is bigger
than the combined residuals of Eqs.~\eqref{eq:final_eqs}, we solve
\eqref{eq:velocity_potential} for $\phi$ and use a softening procedure
$\phi=\zeta \phi_{\rm solved} + (1-\zeta) \phi_{\rm old}$, where for
this iteration procedure $\zeta=0.2$ is applied.\\

(iv) We solve the elliptic equations for $\psi,B^i,\alpha$
\eqref{eq:final_eqs} with a softening of
$\zeta=0.4$. \\

(v) The positions of the stars' centers, $x_{C*,\pm}$, are determined
by the maximum of $h$ along the $x$-axis.  We determine $\Omega$ and
$x_\text{CM}$ with the help of the force balance
equation~\eqref{eq:forcebalance}; see~\ref{sec:omega_and_xCM} for more
details. \\
     
(vi) We then compute $h$ and choose $C_\pm$ such that the baryonic
mass of each star remains constant. Afterwards, we update $\sigma_\pm$
to reflect the changes in the shape of the stars' surfaces and adjust
the domain boundaries accordingly. In most cases we filter out high
frequencies in $\sigma_\pm$ for overall stability and apply
$\partial_B \sigma_{\pm} (B,\varphi)|_{B=0,1}=0$ to keep the stars on
the $x$-axis.\\
     
(vii) We go back to step (ii). 
     
\subsection{Code Improvements} 
\label{sec:Code_Novelties}

\subsubsection{Including piecewise polytropes}

In order to easily incorporate more realistic equations of state we follow
the approach in~\cite{ReaLacOwe09} and approximate them by piecewise 
polytropic equations of state.

For a simple polytropic equation of state (EOS), with $p=\kappa
\rho_0^\Gamma$ (where $\Gamma=1+1/n$), the matter-variables inside the star
are $C^\infty$ and only the star's surface needs special attention. However,
this is no longer the case when dealing with piecewise polytropes.

\begin{figure}[t]
  \includegraphics[width=0.48\textwidth]{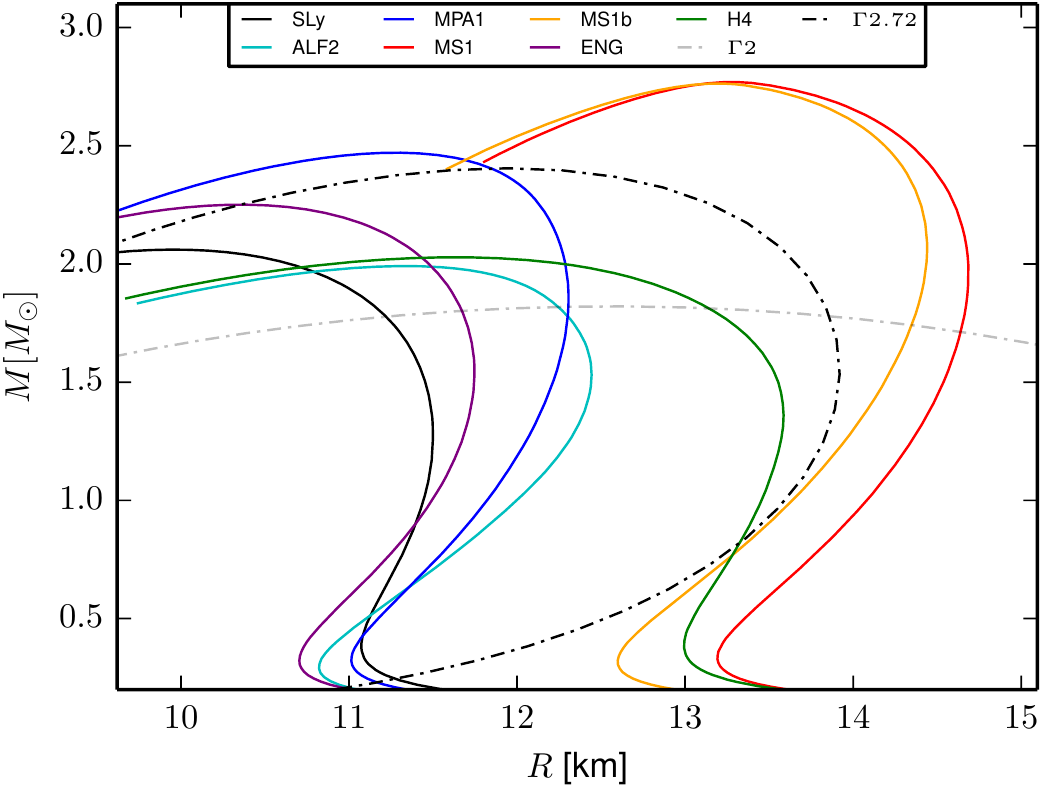}
  \caption{The mass-radius relation for the EOSs employed in this
    work.  See Table~\ref{tab:listEOS} for further details.}
  \label{fig:EOS_overview}
\end{figure}

For a polytrope with polytropic index $n_I$ and constant $\kappa_I$,
the pressure $p$, specific enthalpy $h$, and energy density $\rho_E$
are related to the rest mass or baryonic mass density $\rho_0$ via
\begin{subequations}
  \begin{eqnarray}
     \label{eq:poly_of_rho0}
     p & = & \kappa_I \rho_0^{1+1/n_I} , \\
     h & = & (n_I+1) \kappa_I \rho_0^{1/n_I} + K_I , \\
\rho_E & = & (n_I \kappa_I \rho_0^{1/n_I} + K_I) \rho_0 ,
  \end{eqnarray}
\end{subequations}
where $K_I$ is a constant that determines the specific enthalpy at the star
surface.
For piecewise polytropic EOSs we divide the range of possible
$\rho_0$ into intervals $[0,\rho_{0,1}]$, $[\rho_{0,1},\rho_{0,2}]$, etc. We
label the intervals by $I=0,1,...$. Within each interval $I$ we use the
polytropic relations of Eq.~(\ref{eq:poly_of_rho0}), but with a different
$n_I$, $\kappa_I$ and $K_I$. In interval $0$ we must choose $K_0=1$ to
ensure that the specific enthalpy is unity at the star surface. We can
freely choose all the $n_I$ and $\kappa_0$ to closely approximate some
desired EOS. However, in order to assure continuity of $p$,
$h$ and $\rho_E$ across interval boundaries the remaining $\kappa_I$ and
$K_I$ must be related by
\begin{subequations}
\begin{eqnarray}
\kappa_I &=& \kappa_{I-1} \rho_{0,I}^{1/n_{I-1} - 1/n_I} , \\
K_0 &=& 1 , \\
K_I &=& K_{I-1} + n_{I-1} \kappa_{I-1} \rho_{0,I}^{1/n_{I-1}}
                - n_I \kappa_I \rho_{0,I}^{1/n_I} .
\end{eqnarray}
\end{subequations}
In this case we only know that all the matter
variables are at least $C^0$ (i.e., continuous), but not necessarily
differentiable.\footnote{We have not used the spectral fits to
  realistic equations of state from~\cite{Lin10}, even though these
  would give $C^\infty$ matter variables inside the star, since we
  need to have the same implementation of the EOS as in the BAM code
  that we use for evolutions: Slight differences between the two
  fits to a given EOS would lead to unphysical effects upon starting
  the evolution.}  
We use the parameters presented
in~\cite{ReaLacOwe09} and employ four different pieces consisting of
a crust and three inner regions.  All piecewise polytropic EOSs we use 
give a maximum mass of $ M_\text{max} \geq 1.99M_\odot$ (so they
are $1\sigma$ compatible with the precise high-mass neutron star measurements
in~\cite{DemPenRan10,AntFreWex13}), and have an adiabatic sound
speed $c_s \le 1$ for densities up to the maximum density of a
stable TOV star. These EOSs also span a range of microphysical
content, including hyperons (H4) and the hadron-quark mixed phase
(ALF2), as well as the standard $npe\mu$ composition, and were
obtained using a variety of calculational methods (see Sec.~II
in~\cite{ReaLacOwe09} for further details). Important parameters for
the EOSs we employ are given in Tab.~\ref{tab:listEOS} and the
mass-radius relations for TOV stars are shown in
Fig.~\ref{fig:EOS_overview}.  
Additionally, we use two simple
polytropic EOS, one ($\Gamma2$) with $\kappa = 123.6489$ and
$\Gamma=2$ and one ($\Gamma2.72$) with $\kappa= 23841.43$ and
$\Gamma=2.7203$. The latter EOS is constructed as an ``average fit" to
certain of the realistic EOSs we consider,
where we have fitted $p(\rho_0) = \sum_{i \in S} p_i(\rho_0)/6$ for the
six EOS $S = \text{(SLy, ALF2, MPA1, MS1, MS1b, ENG)}$ with a simple
polytropic EOS in the density interval $\rho_0 \in
[0,3.24\times10^{-3}]$.  The resulting EOS tends to be more
``realistic" than the $\Gamma2$ polytrope widely used in the
literature.  In particular, it allows a maximum mass $>2M_\odot$ and
has a maximum adiabatic sound speed $<1$ for the maximum mass
TOV star. The overall qualitative behavior of its mass-radius
curve is also more similar to those of realistic EOSs than that of the $\Gamma 2$ EOS (see~Fig.~\ref{fig:EOS_overview}).

In the past, \SGRID used $\bar{q}:=p/\rho_0$ as the fundamental
variable, i.e., the matter variables and their spatial derivatives
were all derived from $\bar{q}$. But with this definition $\bar{q}$ will
only be $C^0$ in case of a piecewise polytropic EOS, even for a single
TOV star. In contrast, $h$, which is given by
\begin{equation}
\frac{\text{d}h}{\text{d}r}= - \frac{h [m(r)+ 4 \pi r^2 p]}{r[r-2m(r)]}
\end{equation}
for a single TOV star, will be at least $C^1$ inside the
star under the assumption that $p,\rho_0 \in C^0$.  [Here $m(r) :=
  4\pi\int_0^r\rho_0(\bar{r})\bar{r}^2\text{d}\bar{r}$, so $m\in
  C^1$.]
For this reason we have switched variables to
\begin{equation}
q := h - 1 ,
\end{equation}
which is at least $C^1$ for single TOV stars. Taking spatial derivatives of
$q$ is thus more accurate than taking them of $p/ \rho_0$.

We can compute the other matter variables in terms of $q$, giving
\begin{subequations}
  \begin{eqnarray}
\rho_0 & = & \left[ \frac{q+1-K_I}{\kappa_I (n_I+1)} \right]^{n_I}, \\ 
     p & = & \rho_0 \frac{q+1-K_I}{n_I+1}, \\ 
\rho_E & = & n_I p + K_I \rho_0 .
  \end{eqnarray}
\end{subequations}

\begin{table}[t]
  \centering
  \caption{\label{tab:listEOS} Properties of the equations of state (EOSs)
    used in this work.  The first seven rows refer to piecewise
    polytropes, where we employ the fits of~\cite{ReaLacOwe09}.
    These EOS use a crust with
    $\kappa_\text{crust}=\kappa_0=8.948185\times10^{-2}$ and
    $\Gamma_\text{crust}=1+1/n_0=1.35692$. The divisions for the
    individual parts are at $\rho_{0,1}={\rho}_\text{crust}$,
    $\rho_{0,2}=8.12123\times 10^{-4}$ and $\rho_{0,3} =
    1.62040 \times 10^{-3}$.  The last two rows refer to simple
    polytropic EOSs.  The columns (for the piecewise polytropes)
    refer to: the name of the EOS, the maximum density in the crust,
    the three polytropic exponents $\Gamma_I = 1 + 1/n_I$ 
    for the individual pieces, and
    the maximum supported gravitational mass $M^\text{max}$, maximum
    baryonic mass $M_\text{b}^\text{max}$, and maximum compactness
    $\mathcal{C}^\text{max}$, respectively, of an isolated
    nonrotating star.  [We define the compactness by $\mathcal{C} :=
      M/R$, where $R$ is the star's radius (in Schwarzschild
      coordinates) and $M$ is its gravitational mass.]  For the
    simple polytropes we present $\Gamma$ and $\kappa$, in addition
    to the same maximum values for an isolated nonrotating star
    given for the piecewise polytropes. }
  \begin{tabular}{l|cccc|ccc}        
    \hline EOS & ${\rho}_\text{crust}\cdot10^{4}$& $\Gamma_1$ &
      $\Gamma_2$ & $\Gamma_3$ & $M^\text{max}$ &
    $M_\text{b}^\text{max}$ & $\mathcal{C}^\text{max}$ \\ \hline SLy
    & 2.36953 & 3.005 & 2.988 & 2.851 & 2.06 & 2.46 & 0.31 \\ ALF2 &
    3.15606 & 4.070 & 2.411 & 1.890 & 1.99 & 2.32 & 0.26 \\ ENG &
    2.99450 & 3.514 & 3.130 & 3.168 & 2.25 & 2.73 & 0.32 \\ H4 &
    1.43830 & 2.909 & 2.246 & 2.144 & 2.03 & 2.33 & 0.26 \\ MPA1 &
    2.71930 & 3.446 & 3.572 & 2.887 & 2.47 & 3.04 & 0.32 \\ MS1 &
    1.52594 & 3.224 & 3.033 & 1.325 & 2.77 & 3.35 & 0.31 \\ MS1b &
    1.84169 & 3.456 & 3.011 & 1.425 & 2.76 & 3.35 & 0.31 \\ \hline
    $\Gamma2$ &
        \multicolumn{4}{l|}{$\Gamma=2, \qquad \ \ \ \kappa =
          123.6489$} & 1.82 & 2.00 & 0.21 \\ $\Gamma2.72$ &
        \multicolumn{4}{l|}{$\Gamma=2.7203, \ \ \kappa = 23841.43$} &
        2.40 & 2.85 & 0.30 \\ \hline
  \end{tabular}
\end{table}

\subsubsection{Updating $\Omega$ and $x_{CM}^1$}
\label{sec:omega_and_xCM}

In order to also solve Eq.~(\ref{eq:h_star}) we need to know the
values of $\Omega$ and $x_{CM}^1$.  We first determine the star
centers $x_{C*{\pm}}^1$ by finding the maximum of the current $h$
along the $x$-axis. Using $\partial_1 h|_{x_{C*{\pm}}^1} =0$ in
Eq.~(\ref{eq:h_star}) we find~\cite{Tic12}
\begin{eqnarray}
\label{eq:forcebalance}
\partial_{1} \ln \left[ \alpha^2 -
  \left(\beta^i+k^i+\frac{w^i}{hu^0}\right)
  \left(\beta_i+k_i+\frac{w_i}{hu^0}\right)
  \right]\Bigg|_{x_{C*{\pm}}^1} && \nonumber \\ =
-2\partial_{1}\ln\Gamma\big|_{x_{C*{\pm}}^1} .  &&
\end{eqnarray}
Note that $\beta^i+k^i$ is a function of $\Omega$ and $x_\text{CM}^1$.
The right hand side of Eq.~(\ref{eq:forcebalance}) is given by
\begin{equation}
\Gamma = \frac{ \alpha u^0 \left[
    1-\left(\beta^i+k^i+\frac{w^i}{hu^0}\right)\frac{D_i\phi}{\alpha^2
      hu^0} - \frac{w_i w^i}{(\alpha hu^0)^2}\right] } { \sqrt{ 1 -
    \left(\beta^i+k^i+\frac{w^i}{hu^0}\right)
    \left(\beta_i+k_i+\frac{w_i}{hu^0}\right)\frac{1}{\alpha^2} } } ,
\end{equation}
which we reproduce here (with $\xi^i$ replaced by $k^i$) because there
were some typos in the original published version in~\cite{Tic12}.
Eq.~(\ref{eq:forcebalance}) is called force balance equation. It gives
one equation for $\Omega$ and $x_\text{CM}^1$ at each star center and
thus can be used to update $\Omega$ and $x_\text{CM}^1$. One
noteworthy caveat is that we evaluate the derivative of $\ln\Gamma$ in
Eq.~(\ref{eq:forcebalance}) for the $\Omega$ and $x_\text{CM}^1$
before the update.

We have found that the force balance equation works well in many
cases.  However, for more massive stars or higher mass ratios the
overall iteration can become unstable.  In this case the center of
mass drifts away and the magnitude of the Arnowitt-Deser-Misner (ADM)
momentum
\begin{equation}
\label{PADM}
P_\text{ADM}^i= \int j^i \psi^{10} d^3x ,
\end{equation}
especially of its $y$-component $P_\text{ADM}^y$ becomes large.  This
problem has also been observed by others~\cite{TanShi10}.  It can be
solved in part by computing $\Omega$ and $x_\text{CM}^1$ in a
different way: Notice first that the matter flux
\begin{equation}
j^i = \alpha (\rho_E + p) (u^0)^2 (V^i + k^i + \beta^i)
\end{equation}
in Eq.~(\ref{PADM}) depends on the Killing vector $k^{\mu}$ and thus
on $\Omega$ and $x_\text{CM}^1$.  Using the $\Omega$ from the previous
iteration we can then solve the equation $P_\text{ADM}^y = 0$ for
$x_\text{CM}^1$. This gives a value for the center of mass such that
the $P_\text{ADM}^y$ will be zero as desired. Once we have determined
$x_\text{CM}^1$ in this way, we next compute $\Omega$ from
Eq.~(\ref{eq:forcebalance}) for each star's center. This will in general
give two different values for $\Omega$. For the final $\Omega$ we simply
use the average of these two values.

\section{Binary neutron stars in Quasi-equilibrium}
\label{sec:Results}

In the following, we discuss the main results on equilibrium
configurations obtained by applying the framework and code
improvements discussed in Sec.~\ref{sec:Method} and
Sec.~\ref{sec:NumericalMethods}.  In particular we analyze the
spin-orbit (SO) interaction for realistic EOSs; we extend the work
of~\cite{MolMarJoh14} on highly eccentric orbits, where we improve
our data by solving Eq.~\eqref{eq:velocity_potential} for the velocity potential;
we investigate inspirals on eccentricity reduced orbits; we also
present significantly unequal mass setups as well as configurations with high compactnesses; we
end with a convergence study.

\subsection{Spins}
\label{sec:spins}

\begin{figure}[t]
  \includegraphics[width=0.48\textwidth]{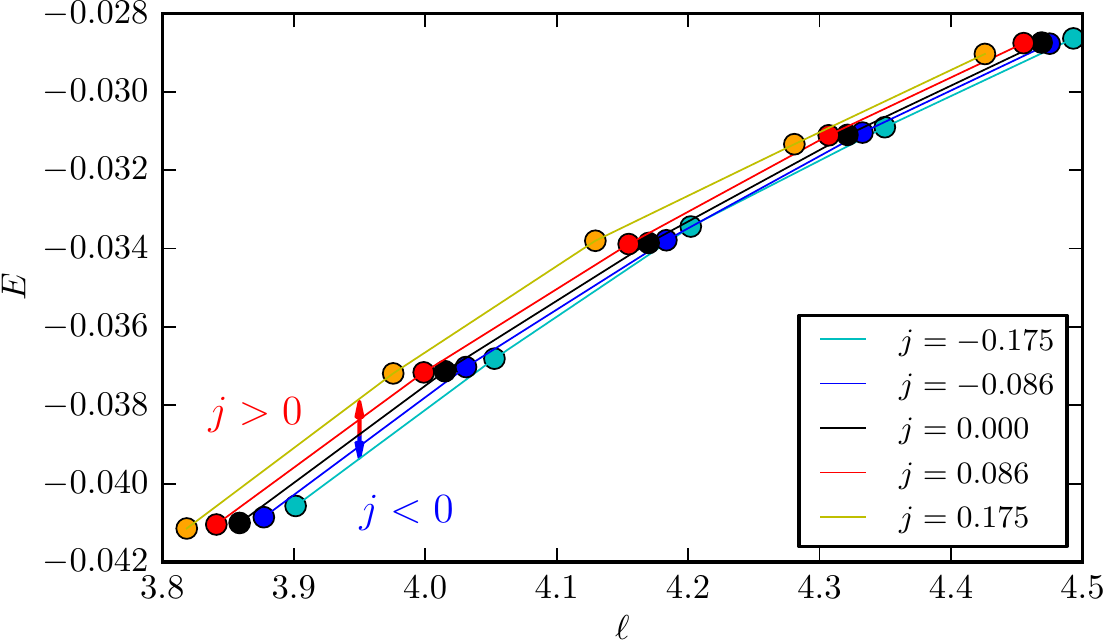}
  \caption{Reduced binding energy vs.~specific total angular
    momentum for a binary system with $M_b^A=M_b^B=1.4895$ and the
    SLy EOS.  The influence of aligned and antialigned spin is shown
    with arrows}
  \label{fig:BNS_spin}
\end{figure}

\begin{figure}[t]
  \includegraphics[width=0.48\textwidth]{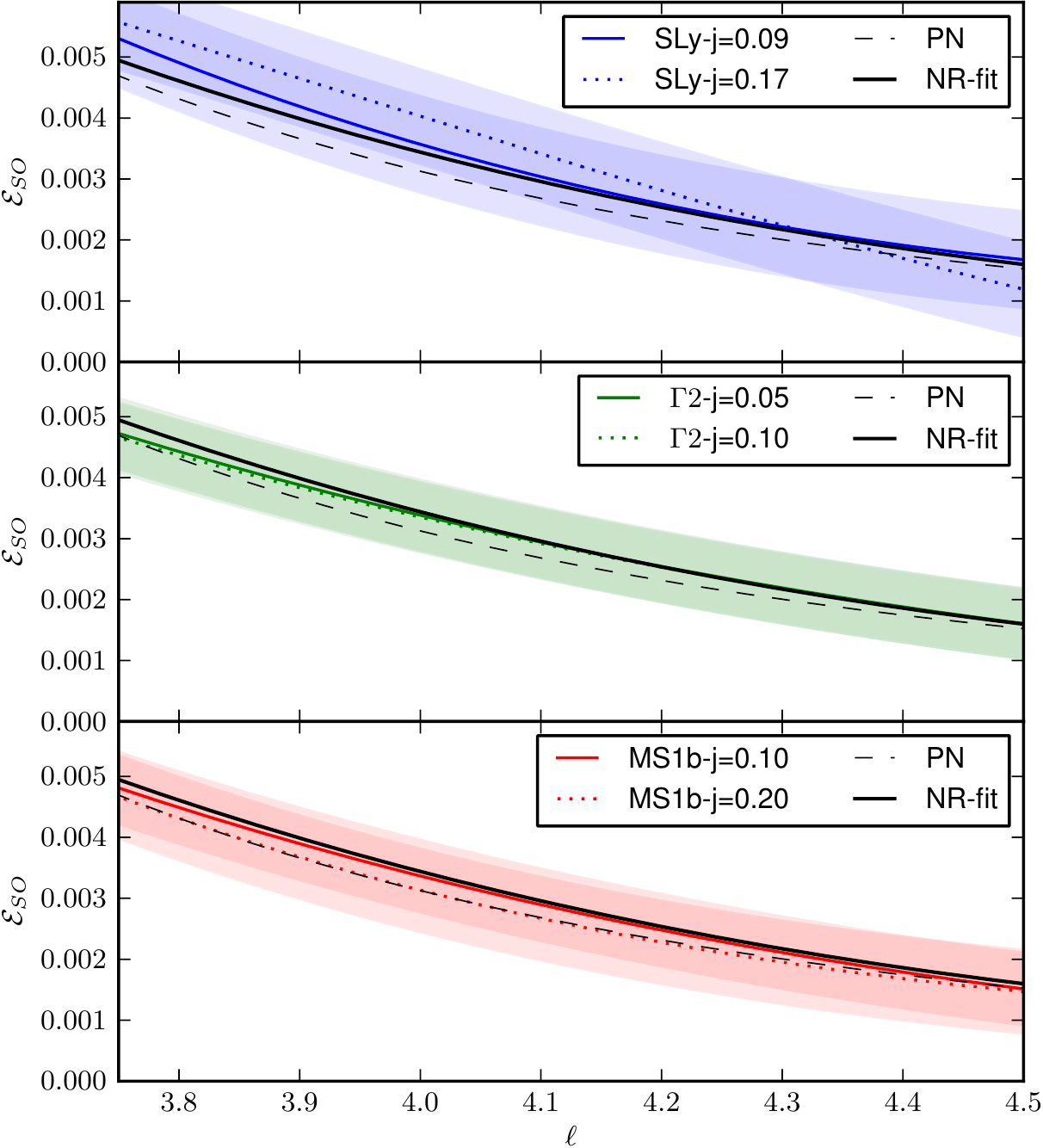}
  \caption{The spin-orbit coefficient $\mathcal{E_{SO}}$ in the
    reduced binding energy of the binary neutron stars.  We compare
    numerical results for the SLy (top panel), $\Gamma2$ (middle
    panel), and MS1b (bottom panel) EOSs with the predictions of the
    post-Newtonian (PN) approximation (black dashed lines). We also
    include an average line for our data in all panels (solid black
    lines).  We estimate the numerical error bars from computations at
    different resolutions and show them as shaded regions.}
  \label{fig:E_SO_EOS} 
\end{figure}

To date, there exist three ways to model the fluid velocity field in
quasi-equilibrium BNS configurations in general relativity.  Neutron
star spins are neglected in the irrotational approach, while they are
treated in an unphysical manner if corotation is assumed. The CRV
approach~\cite{Tic11,Tic12} (see also \cite{BerDieTic13}) is the only
known alternative for the construction of consistent and constraint
solved initial data where spins are included and can be chosen freely.
Alternative approaches have been proposed and employed in dynamical
evolutions~\cite{KasGalAli13,TsaMar13,KasGal14}, but they are based on
constraint-violating data which also violate hydrodynamical
stationarity conditions. Here we discuss some properties of equilibrium sequences of
CRV BNS expanding the work of~\cite{BerDieTic13}, and focusing on the
spin-orbit interaction.

We compute equilibrium sequences for the SLy and MS1b EOSs
(Table~\ref{tab:listEOS}), setting the baryonic masses of the individual
neutron stars to $M_{b}=1.48945$. 
Additionally, we use results of~\cite{BerDieTic13}, where the $\Gamma2$ EOS was
employed and the baryonic mass of the neutron stars was set to
$M_{b}=1.625$. For each of these EOSs we obtain sequences at fixed
baryonic mass and for five different spin magnitudes, two aligned spin
setups ($\uparrow \uparrow$), one irrotational ($00$), and two
anti-aligned setups ($\downarrow \downarrow$). Each sequence
essentially mimics an adiabatic evolution.  (In the aligned and
anti-aligned cases we are considering, the spin directions remain
unchanged during an evolution.)  The resolution employed for the
piecewise polytropes is $n_A=n_B=28$; $n_\varphi=8$,
$n_\text{Cart}=24$, while for the $\Gamma2$ simple polytrope it is
$n_A=n_B=24$, $n_\varphi=8$, $n_\text{Cart}=20$.  We use a higher
resolution for the piecewise polytropic runs to better resolve the
crust region (cf.~Table~\ref{tab:listEOS} and
Sec.~\ref{sec:conv}).

We stress that an unambiguous definition of the individual spins of
the stars in a binary system is in general not possible. It is however
possible to define the spin of single isolated neutron stars within the CRV approach, 
as discussed in Appendix~\ref{sec:single-CRV}. Using this result, one can give an
estimate of the magnitudes of the spins of the stars in a binary by
considering, for each star, an isolated configuration with the same
baryonic mass $M_b$ and the same rotational part of the 4-velocity
$w^\mu$, and assuming that the individual spin magnitudes $S^A$, $S^B$
remain unchanged when we use the same parameters to compute binary
initial data.

We analyze equilibrium sequences in terms of the reduced binding
energy
\begin{equation}
  E = \frac{1}{\nu} \left( \frac{M_{\rm ADM}}{M} -1
  \right), \label{eq:Eb}
\end{equation}
and the specific orbital angular momentum
\begin{equation}
  \ell = \frac{L}{\nu M^2} = \frac{J_\text{ADM} - S^A - S^B}{\nu
    M^2}. \label{eq:ell}
\end{equation}
Here $\nu := M^A M^B / M^2$ is the symmetric mass ratio, $M^A$ and 
$M^B$ are the individual masses of the stars in isolation,
$M := M^A + M^B$,
\footnote{Notice that the individual masses of the stars in
  isolation ($M^A$, $M^B$) are obtained here for spinning neutron
  stars and differ from the results for irrotational stars with the
  same baryonic mass.}
and $M_\text{ADM}$ and $J_\text{ADM}$ are the
binary's ADM mass and angular momentum, respectively.  Here $E(\ell)$
is a gauge-invariant way to characterize the dynamics, which is also
applicable to full numerical relativity 
evolutions; see~\cite{DamNagPol11,BerNagThi12} and Sec.~\ref{sec:precessing}.
The $E(\ell)$ curves are shown in Fig.~\ref{fig:BNS_spin}
for the SLy EOS; other EOSs show qualitatively the same behavior.
From the figure one observes that aligned configurations are less
bound than antialigned configurations. This behavior follows from the
fact that the spin-orbit (SO) interaction, which is the main spin-related
effect, is repulsive for aligned spins and attractive for antialigned
spins (for a fixed $\ell$), see, e.g.,~\cite{Dam01} and below.

In the following we explicitly compute the SO contribution in
our binding energy data, and show its influence on the dynamics as well as
its independence of finite size (EOS) effects. We write the binding
energy as
\begin{equation}
  E(\ell) = E_0 + E_{SO} + E_T + E_{SS} \label{eq:bind_energy_sum} \ ,
\end{equation}
where $E_0$ describes the binding energy of a nonspinning black hole
binary in the conformal flatness approximation, and is therefore
independent of the spin and matter effects; $E_{SO}$ and $E_{SS}$
represent the SO and spin-spin contributions, respectively;
and $E_T$ denotes the tidal contribution.  Assuming for simplicity
that the dimensionless spins are the same (as is the case here)
$\vv{j}:=\vv{j}^A= \vv{S} ^A/ (M^A)^2= \vv{j}^B = \vv{S}^B/(M^B)^2$,
the SO interaction is proportional to
\begin{equation}
  E_{SO} \propto \vv{j} \cdot \vv{L} =
  \|\vv{j}\|\|\vv{L}\|\cos(\angle(\vv{j},\vv{L})). \label{eq:ESO_cos}
\end{equation}  
Thus, the angle between $\vv{j}$ and the orbital angular momentum
$\vv{L}$ defines whether the SO interaction is repulsive or
attractive.  In the cases considered here,
$\cos(\angle(\bf{j},\bf{L}))$ takes the values $1$ and $-1$.  For
these two possibilities we can write
\begin{equation}
  E_{SO} = \mathcal{E}_{SO}(\ell)\; j, \label{eq:ESO}
\end{equation}  
where $j$ denotes the signed magnitude of $\vv{j}$, i.e., 
$j := \|\vv{j}\|\cos(\angle(\bf{j},\bf{L}))$.  
Equation~\eqref{eq:ESO}
allows us to answer two important questions: (i) Do we see an imprint
of the EOS on the SO interaction?  (ii) Does the linear
dependence of $E_{SO}$ on $j$ capture the main dynamics?

According to our spin definition (see above), the spins are constant
during the adiabatic evolution. This is a good approximation at these
separations and also supported by numerical evidence in binary black hole
simulations \cite{LovSchSzi10,SchGieHem14,OssBoyKid15,LouHea15}.  The
$\mathcal{E}_{SO}(\ell)$ term can be computed using
\begin{equation}
  \mathcal{E}_{SO}(\ell)=\frac{E^{(\uparrow \uparrow)}_{j}(\ell) -
    E^{(\downarrow \downarrow)}_{j}(\ell)}{2 j} \label{eq:ESO_2}
\end{equation}
for different (signed) spin magnitudes $j$,
where $E^{(\uparrow \uparrow)}_{j}(\ell)$ and
$E^{(\downarrow \downarrow)}_{j}(\ell)$ is $E(\ell)$ for aligned and
anti-aligned spins. Indeed, for a given spin,
all the terms in~\eqref{eq:bind_energy_sum} except $E_{SO}$ cancel in
the combination~\eqref{eq:ESO_2} because they all have the same sign.

The function $\mathcal{E}_{SO}(\ell)$ is shown in
Fig.~\ref{fig:E_SO_EOS} for all three EOSs considered here: SLy (top
panel), $\Gamma2$ (middle panel), and MS1b (bottom panel).
Additionally, we compute the average for all EOSs (solid line) and
compare our results with the linear-in-spin part of the 4PN energy
from Eq.~(8.23) in \cite{LevSte14},
which is shown as a black dashed line in Fig.~\ref{fig:E_SO_EOS}.
From the figure we observe: (i) $\mathcal{E}_{SO}(\ell)$ is positive,
therefore the SO-interaction is repulsive/attractive
(positive/negative) according to the sign of $j$ [in general, this
  depends on $\angle(\bf{j},\bf{L})$]; (ii) all the curves agree
within their errors, i.e., there is no significant dependence on
finite size (EOS) effects.  (iii) the PN expression from~\cite{LevSte14}
captures the behavior of our conformally flat data for all employed
EOSs.

We notice that the $E_{SS}$ and the $E_T$ terms in
\eqref{eq:bind_energy_sum} can be extracted in a similar way.
However, for the spin magnitudes and orbital separations considered
here they lie within the uncertainty of our data.
  
\subsection{Highly eccentric configurations}
\label{sec:eccentricity_high}

In~\cite{MolMarJoh14} we described a method to produce
hydrodynamically consistent initial data for relativistic stars on orbits with
arbitrary eccentricities for the first time. In our initial implementation of the method,
we used an elliptic solver based on a Cartesian multigrid method, for which
it is technically difficult to solve an equation to determine the
velocity potential, since this would require boundaries at the
star's surface. Instead of introducing surface-fitted coordinates,
we employed a constant three-velocity approximation, which could
be motivated by the restriction to irrotational binaries.  This
means that we assumed the instantaneous (at apoastron)
three-velocity $v^y$ of a fluid element measured by a coordinate
observer to be constant throughout the star, so the four-velocity
could be written as
\begin{equation}
  u^\alpha = u^t(t^\alpha+v^y y^\alpha).
\end{equation}
\begin{figure}[t]
  \includegraphics[width=0.48\textwidth]{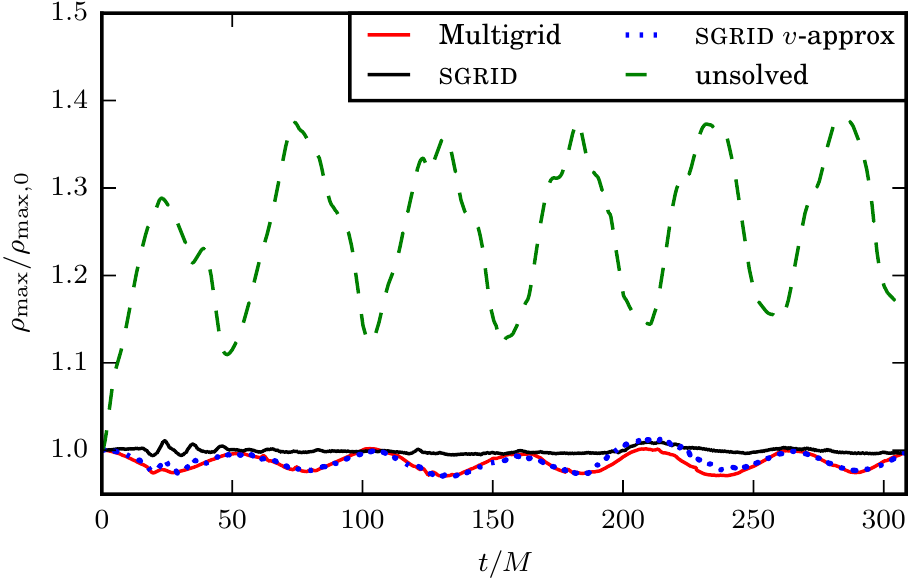}
  \caption{Star's maximum density scaled by its initial value:
    Simple superposition of TOV stars without solving the
    constraints (dashed green), Cartesian multigrid method solver
    of~\cite{MolMarJoh14} (solid red), \SGRID data assuming
    constant three-velocity (dotted blue) and \SGRID data solving
    the fluid potential equation~\eqref{eq:velocity_potential}
    (solid black).}
  \label{fig:VwAo_comparison}
\end{figure}
However, \SGRID provides surface-fitted coordinates and allows us
to solve easily for the velocity potential. First, we want to briefly
compare our old results to the newly obtained \SGRID results and
show the improvement of the initial data gained by solving the additional
equation for the velocity potential.
Fig.~\ref{fig:VwAo_comparison} compares the oscillations of the
central density throughout the evolution for the previous
multigrid solver and \SGRID for two equal mass stars with baryonic
masses $M_{b}^{A,B}=1.620$ on a quasicircular orbit with an initial $(2,2)$ mode
GW-frequency of $M\omega_{22}^0 = 0.053$.

This simple test case with two polytropic stars ($\Gamma2$)
clearly shows the influence of the fully solved velocity
potential: While we observe strong oscillations of $30\%$ in the
central density for superimposed TOV stars (which gives constraint
violating initial data), we only observe roughly $4\%$ oscillations for the
constraint solved data using the constant $3$-velocity approximation.
In this case, \SGRID and the Cartesian multigrid data give a good
agreement. If we drop the approximation and solve for $\phi$, we
can obtain even lower oscillations, improved by a factor of five, i.e., less than $1\%$
(solid black line in Fig.~\ref{fig:VwAo_comparison}).  Here we use
the same evolution setup in BAM as in \cite{MolMarJoh14}, in order
to make a direct comparison: We use the
Baumgarte-Shapiro-Shibata-Nakamura~\cite{NakOohKoj87,ShiNak95,BauSha98}
formulation, and 98 points in each direction in each of $5$
refinement levels, with a grid spacing of $h_5=0.1875$ on
the finest level. We do not use the conservative mesh refinement
introduced in~\cite{DieBerUje15}.  The \SGRID initial data use
$n_A=n_B=24$, $n_\varphi=8$ and $n_{\rm{Cart}}=20$ points.  The
reduced density oscillation with our new setup will allow us to study
orbit induced oscillations as in~\cite{GolBerThi11} in more detail
and disentangle the orbital effect from oscillations due to the
initial data construction, which were present in previous
attempts.

\begin{figure*}[t]
  \includegraphics[width=0.98\textwidth]{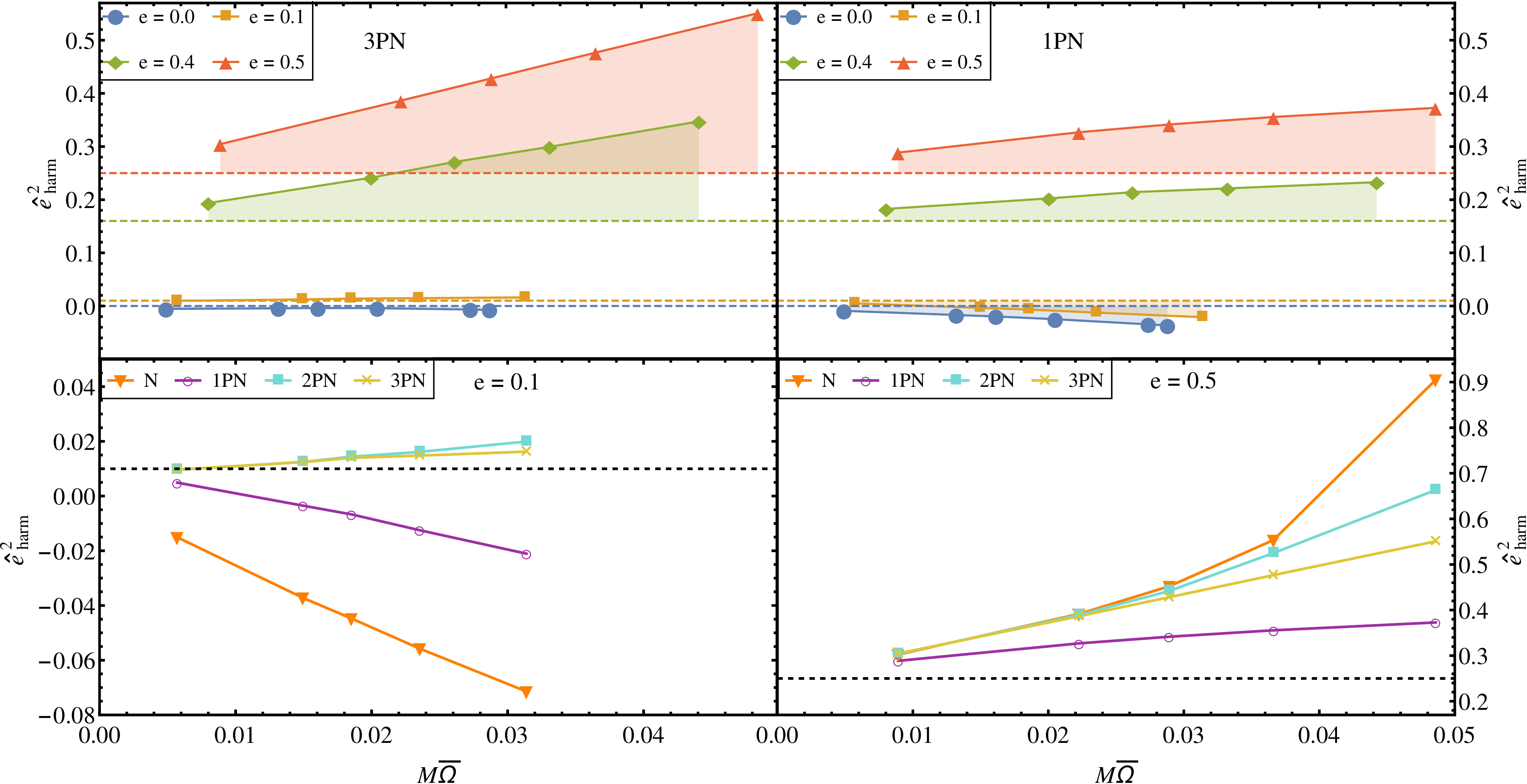}
  \caption{The two plots in the top panel show the squared
    eccentricity $\hat{e}^2_\text{harm}$ calculated from the ADM
    energy $E$ and angular momentum $J$ using harmonic coordinate 
    PN expressions.
    We show the results at $3$PN (top left) and $1$PN (top right) for
    runs with different input eccentricity $e$ (given as
    horizontal dashed lines).  (We plot the square of the PN
    eccentricity since this quantity becomes negative for small
    $e$.) While the results for small eccentricities are
    calculated more accurately with increasing PN orders, the
    error is no longer monotonic with PN order for larger
    eccentricities.  The two plots on the bottom panel show
    convergence of the eccentricity estimate with increasing PN
    order for an input eccentricity $e=0.1$ on the left and how
    this convergence becomes more erratic for higher
    eccentricities (using $e = 0.5$ as an example) on the right side. }
  \label{fig:eccPN_comparison}
\end{figure*}

We now consider quasi-equilibrium sequences for different eccentricities.
We compare our eccentric sequences to post-Newtonian results for the
eccentricity in Fig.~\ref{fig:eccPN_comparison}.  Here we compute the
PN eccentricity from the ADM expressions for the energy and angular
momentum; we subtract the gravitational masses of the stars in
isolation from the total ADM energy to obtain the binding energy that
enters the PN calculation. (We did not perform such a comparison
in~\cite{MolMarJoh14}, since the BAM implementation does not use a
compactified grid including spatial infinity and was thus unable to
compute the ADM quantities sufficiently accurately for this comparison.)
In this figure, we see that the PN eccentricity indeed converges
nicely to the input eccentricity as the binary's separation increases,
but it only converges well to the input eccentricity as one increases
the PN order for smaller eccentricities ($0$ and $0.1$).
For the two larger eccentricities we consider ($0.4$ and $0.5$), the $1$PN results
are closer to the input eccentricity than the $3$PN results (though
the $3$PN results are closer than the Newtonian or $2$PN results).

In order to obtain the PN expressions we used to create this figure,
we, in essence, derived the $3$PN extension of the $1$PN expression
for the eccentricity given in Eq.~(2.36) of Mora and
Will~\cite{MorWil03}, using the general $3$PN results they give.  (The
only difference in the derivations is that since we solve for the
square of the eccentricity, we do not perform the PN expansion of the
square root given in their expression.)  We start from the expressions
for $\tilde{E} := E_b/(M\nu)$ and $\tilde{J} := J/(M\nu^2)$ (the scaled binding energy and angular
momentum)
that Mora and Will give in Eqs.~(2.35) and~(2.40) for both
harmonic and ADM coordinates. (Note that $\tilde{E}$ and $\tilde{J}$ are the same as the $E$ and
$\ell$ used in Sec.~\ref{sec:spins}; the latter agreement only holds for the zero spin case we consider here.) We then invert the series for
$\tilde{E}$ to express the PN parameter $\zeta$ in terms of
$\tilde{E}$; we now use $\tilde{E}$ as our expansion parameter. We can
thus calculate the series for $\tilde{E}\tilde{J}^2$ by substituting
the series for $\zeta$ in terms of $\tilde{E}$ into the PN series for
$\tilde{J}$ and expanding consistently.

We then use the resulting harmonic coordinate expression to obtain a PN series for the
square of the eccentricity parameter, $\hat{e}^2_\text{harm}$, in
terms of $\tilde{E}$ and $\tilde{J}$. We experimented with various
treatments of the expansion (e.g., not expanding after substituting
the expression for $\zeta$ in terms of $\tilde{E}$ and/or solving for
the eccentricity numerically instead of by series inversion) and found
that a consistent expansion to a given PN order produced the most
reasonable-looking results. Specifically, this gives
\begin{widetext}
\begin{equation}
\label{eq:eharm2}
\begin{split}
\hat{e}^2_\text{harm} &= 1 - 2\xi + [-4 - 2\eta + (-1 + 3\eta)\xi]\tilde{E} + \left[\frac{20 - 23\eta}{\xi} -22 + 60\eta + 3\eta^2 - (31\eta + 4\eta^2)\xi\right]\tilde{E}^2\\
&\quad + \biggl[\frac{-2016 + (5644 - 123\pi^2)\eta - 252\eta^2}{12\xi^2} + \frac{4848+(-21128+369\pi^2)\eta+2988\eta^2}{24\xi} -20 + 298\eta - 186\eta^2 - 4\eta^3\\
&\quad  +\left(-30\eta + \frac{283}{4}\eta^2 + 5\eta^3\right)\xi\biggr]\tilde{E}^3,
\end{split}
\end{equation}
\end{widetext}
where $\xi := -\tilde{E}\tilde{J}^2$.
We also performed the same calculation
with the ADM coordinate expressions, which differ from the harmonic
coordinate ones starting at $2$PN, and found the expected small
differences in the results ($\lesssim5\%$, with smaller differences
for larger eccentricities).

Our original motivation for using these particular PN results was to
avoid using expressions which are not well-behaved in the limit of a
head-on collision from rest, which is what our eccentric data approach
in the limit $e\nearrow1$, e.g., expressions that have factors of $\tilde{J}$ in
the denominator. However, even though we start from expressions that are
well-behaved in this limit (by considering $\tilde{E}\tilde{J}^2$ instead of $\tilde{J}$
by itself), we still obtain factors of $1/\xi$ in the final expression for $\hat{e}^2_\text{harm}$
[Eq.~\eqref{eq:eharm2}] starting at $2$PN. This is not a significant problem since $J_\text{ADM}$ is
not very small for the systems we are considering. We thus also investigated
the expressions given by Memmesheimer, Gopakumar, and Sch{\"a}fer
(MGS)~\cite{MemGopSch04} which are not well-behaved in the limit of
constant energy and vanishing angular momentum. Here one can compute
the three eccentricities defined in the post-Keplerian
parametrization of an eccentric orbit ($e_r$, $e_t$, and $e_\phi$)
from Eqs.~(20) and~(25) in MGS (these give the expressions in ADM and
harmonic coordinates, respectively).

We find that $e_t$ (particularly in harmonic coordinates) agrees quite
well with the input eccentricities, especially at large separations,
with fractional errors of $<2.4\%$ for all separations we consider for
$e = 0.5$. In addition, we can compute the coordinate separation $r$
of the stars from the ADM energy and angular momentum by noting that
the binary is at apoapsis, so we can take $\dot{r} = 0$ in Eqs.~(A1)
and~(A3) in MGS (again, these give the expressions in ADM and harmonic
coordinates, respectively) and then solve for $r$ numerically. Here we
find that the value for $r$ we obtain by solving the harmonic
coordinate equation agrees quite closely with the coordinate
separation from \SGRID for all the separations we consider,
particularly for the two larger eccentricities we consider (fractional
errors of $< 2\%$). Such close agreement may simply be fortuitous,
since there is no reason \emph{a priori} to expect the coordinate
systems used to agree so closely.

\subsection{Orbits with reduced eccentricity} 
\label{sec:eccentricity_reduction}

The \textit{helliptical approximate} symmetry vector (or, more
formally, the \textit{instantaneous helical vector}) introduced in
Eq.~\eqref{eq:ellinspiralKV} allows one to vary the binary's
initial radial velocity, in addition to the control on the initial
tangential velocity provided by the orbital frequency $\Omega$.
If we do not make use of this freedom, i.e., if we set the radial
velocity and the eccentricity parameter to zero, $v_r = e = 0$, we
obtain the well known limit of standard quasicircular initial
data. However, when we evolve such data, the separation between
the stars oscillates. From these oscillations (or their imprint on the
gravitational wave signal) we deduce actual measured eccentricities
of $\hat{e}\sim 10^{-2}$. For the vast majority of
astrophysical scenarios, this remaining eccentricity is purely
artificial, because GWs efficiently circularise the orbit during
the inspiral, leading to almost vanishing eccentricities during
the last minutes before merger. In particular, the remaining
eccentricity at merger $\hat{e}$ of the six observed double
neutron star systems that will coalesce within a Hubble time will
be $\hat{e}\lesssim 10^{-5}$~\cite{KowBulBel11} and thus several
orders of magnitude smaller than the eccentricity obtained when
evolving standard quasicircular initial data.

The effect of this artificial eccentricity can be observed in
various quantities, most notably the gravitational
waveform. Therefore, we apply an iterative method to reduce the
eccentricity, as outlined in \cite{MolMarJoh14}. This method is
similar to the standard eccentricity reduction procedure for
binary black holes~\cite{PfeBroKid07,HusHanGon07,TicMar10}, and
the recent work on eccentricity reduction for binary neutron
stars~\cite{KyuShiTan14}.

The basic idea is to find a measure for the eccentricity and
determine which corrections have to be applied in order to remove
the measured eccentricity from a Keplerian orbit. In this work, we use the
proper distance $d$ inside the hypersurface (measured along the
coordinate line connecting the two local minima of the lapse,
corresponding to the centers of the two stars) as well as the GW
frequency $\omega_{22}$ to estimate the remaining eccentricity
$\hat{e}$.  The coordinate distance $d_\text{coord}$ can also be
used to estimate $\hat{e}$.  However, $d_\text{coord}$ depends
much more strongly on the particular gauge choice than the proper
distance does and thus gives reliable results only in certain
cases (cf.\ \cite{KyuShiTan14}), which is why we choose not to use
it.  Using $d$ for the two neutron stars, we track this quantity
throughout the evolution and fit it to the model
\begin{equation}
\label{eq:d_fit}
  d(t) = S_0 +A_0 t + \frac{1}{2} A_1 t^2 - \frac{B}{\omega_\text{f}}
  \cos(\omega_\text{f} t+ \phi)
\end{equation}
Note that it is also possible to fit the time derivative $\dot
d(t)$ and get rid of one fitting parameter. However, taking the
derivative of $d$ introduces noise, especially for lower
eccentricities.  Applying some low-pass filters can help to
improve the results, but $d$ is generally better suited for
obtaining the eccentricity measure than $\dot{d}$.  The
following results, however, apply to both methods.  A comparison
of the fitting model with the expected Keplerian orbits under the
assumption that the eccentricity $e$ is small for quasicircular
orbits (so we can neglect higher orders of a series expansion),
yields an eccentricity
\begin{equation}\label{eq:estimate_e}
  e = \frac{B}{d_0\omega_\text{f}} =: \hat{e}_{d},
\end{equation}
where $d_0$ is the initial separation.

The ellipticity part of the model 
[$-\frac{B}{\omega_\text{f}} \cos(\omega_\text{f} t+ \phi)$]
leads to an initial radial velocity of $B \sin\phi$
for the stars, which means that the initial radial velocity $v_r$
in Eq.~\eqref{eq:ellinspiralKV} has to be corrected by
\begin{equation}\label{eq:correction_vx}
  \delta v_r = - B \sin\phi
\end{equation}
in order to remove the eccentricity generated by the radial
velocity.  Similarly, we can adjust the eccentricity parameter $e$
(and thus the initial tangential velocity) in order to
remove the residual eccentricity induced by radial acceleration.
The orbital angular frequency at apoastron $\Omega_\text{apo}$ is
related to the $\Omega$ of the symmetry vector in
Eq.~(\ref{eq:ellinspiralKV}) by
$\Omega_\text{apo} \approx (1-e)\Omega$. 
In Newtonian physics we have $\Omega_\text{apo}^2 = (1-e) G M/d_0^3$.
So if we change $e$ by a small $\delta e$, this will lead to a change 
in $\Omega_\text{apo}^2$ by 
$\delta \Omega_\text{apo}^2 \approx -\frac{G M}{d_0^3}\delta e 
= \frac{\Omega_\text{apo}^2}{e-1}\delta e$.
Here all contributions of $O(e^2)$ have been dropped, since we are only
considering situations where $e^2$ is quite small ($\lesssim10^{-4}$).

The radial acceleration in Eq.~(\ref{eq:d_fit}) is given by
$B\omega_\text{f}\cos\phi$. Considering that a change of 
$\delta\Omega_\text{apo}^2$ yields a change of the acceleration
of $d_0\delta\Omega_\text{apo}^2$, we obtain a necessary correction of
\begin{equation}\label{eq:correction_e}
  \delta e = \frac{B\omega_\text{f} \cos\phi}{\Omega_\text{apo}^2 d_0}(1-e),
\end{equation}
where $\Omega_\text{apo}$ can be approximated by either 
$\omega_\text{f}$ or $\Omega$.
We obtain the correction parameters \eqref{eq:correction_vx},
\eqref{eq:correction_e} from our fit to the data from the
evolution, where we need at least one to two orbits for the fit to
be accurate enough. Then we iterate this process until the
eccentricity is sufficiently small, i.e., in most cases two or
three iterations.

\begin{figure}[t]
  \includegraphics[width=0.46\textwidth]{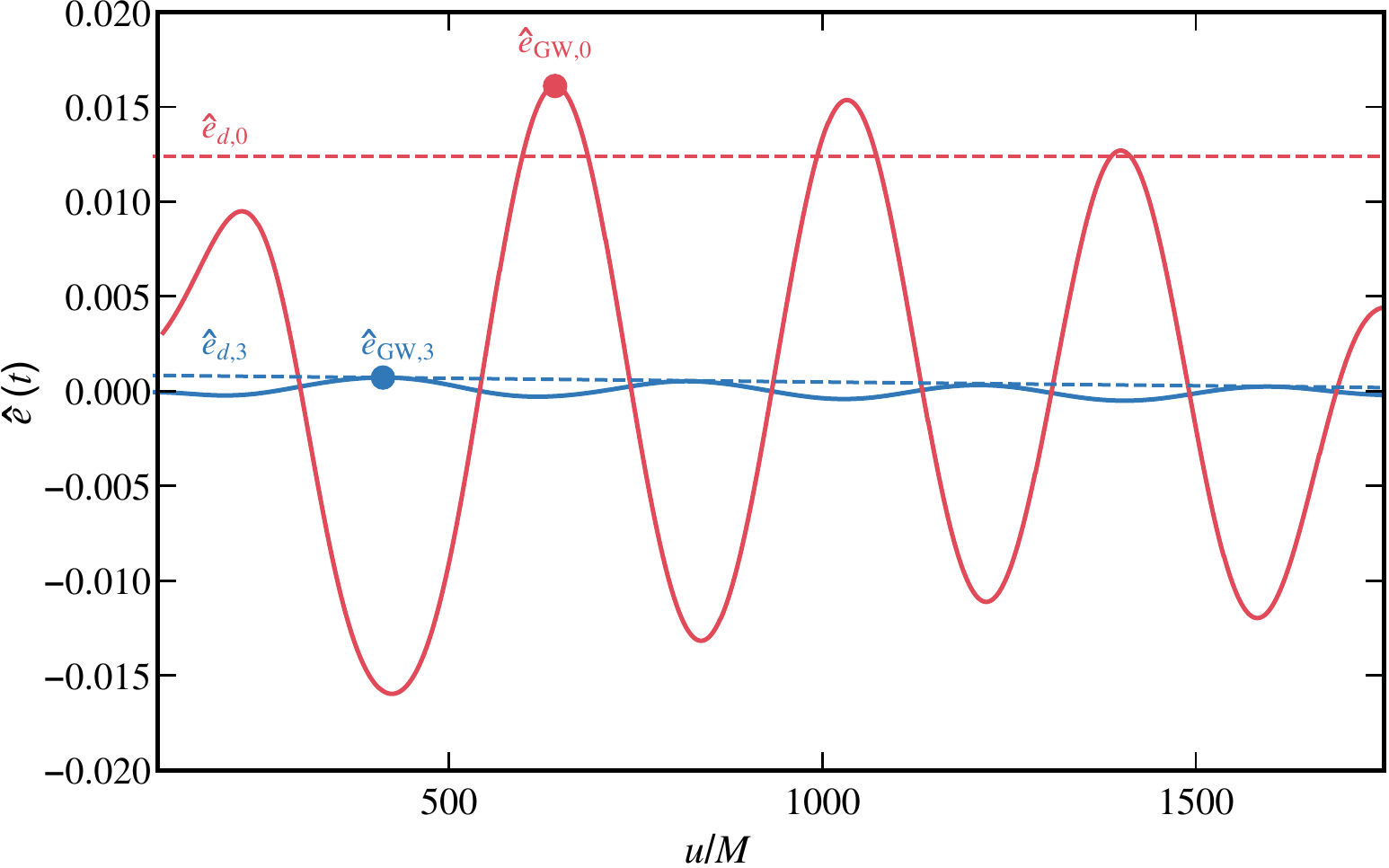}
  \caption{The eccentricity as a function of retarded time $u$
    computed from the residual of the GW frequency for SLy initial
    data with an initial [$(2,2$) mode] GW frequency of $M\omega_{22}^0=0.0365$. 
    We shown the quasicircular data (red) and the third iteration of the
    eccentricity reduction (blue).  The global extremum, marked by the points, gives
    eccentricities of $\hat{e}_{\rm{GW,0}}=0.0156$ and $\hat{e}_{\rm{GW,3}}=0.00084$
    for these interations, which is a factor of 20 improvement. The horizontal dashed lines mark the 
    eccentricities $\hat{e}_d$ calculated from the proper distance.}
    \label{fig:Ecc_omega_estimation}
\end{figure}

Besides the already mentioned methods, there exist several other
techniques to determine the eccentricity. To check the reliability
of our results, we want to make use of the GWs to give an
additional estimate of the eccentricity $\hat{e}_\text{GW}$. We
follow the procedure given in \cite{WalBruMue09}, based on
\cite{BakMetMcW06a}, and model the GW frequency motivated by
post-Newtonian calculations as (cf.\ \cite{HanHusOhm10})
\begin{equation}
  \omega_{\rm{fit}} = \frac{1}{4}\tau^{-3/8}\left(1+c_1
  \tau^{-1/4}+c_2 \tau ^{-3/8}\right),
\end{equation}
where $c_1$, $c_2$ are determined by fitting and
\begin{equation}
  \tau^2 = \frac{\nu^2(t_c-t)^2}{25M^2}+\tau_0^2,
\end{equation}
where $t_c$ and $\tau_0$ are again fitting parameters. We proceed by looking at
the quantity
\begin{equation}
  \hat e_\omega (t) = \frac{\omega_{22}(t)-\omega_{\rm{fit}}(t)}{2
    \omega_{\rm{fit}}(t)}.
\end{equation}
This measure of eccentricity is time dependent and strongly
oscillatory, but the global extremum
\begin{equation}
  \hat{e}_{\rm{GW}} = \displaystyle\max_{t} \left|\hat e_\omega
  (t)\right|
\end{equation}
can be seen as a time-independent measure of
eccentricity; see Fig.~\ref{fig:Ecc_omega_estimation}.  As it
is to be expected intuitively, this value can mostly be found in
the beginning of the evolution, just after the initial gauge
noise has decayed. Note that the initial noise has to be cut off
in order to obtain consistent data; this is also necessary for
the estimate based on the proper distance.  Comparing the
eccentricities obtained this way with the eccentricities
computed from the proper distance, we observe an agreement
within roughly five percent (comparing
$\hat{e}_{\text{GW,3}}=8.4 \times 10^{-4} $ to
$\hat{e}_{d\text{,3}}=8.7 \times10^{-4}$ for the third
eccentricity reduction iteration step of an SLy EOS run;
cf.\ Table \ref{tab:listEccRedPars}).  Considering the decrease
in signal-to-(numerical)-noise of the proper distance
oscillations at higher iterations, and thus the reduced accuracy
in the parameters one obtains from the fitting procedure, this
agreement is satisfying and shows the consistency of the
measures. In order to minimize the computational effort for
finding eccentricity reduced initial data, it makes sense to
evaluate $\hat{e}_{d}$ to find the next iteration step's
parameters, since in this case we have to evolve for fewer time
steps.  Using the gravitational wave signal necessitates longer
evolutions to ensure that the wave has reached the extraction
radius.

\begin{figure}[t]
  \includegraphics[width=0.48\textwidth]{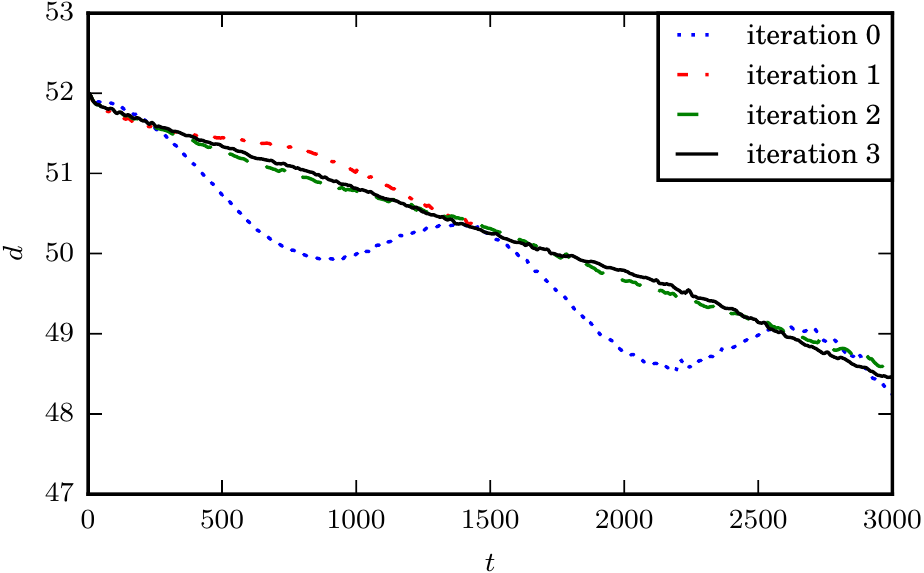}
  \caption{Comparison of the eccentricity of the simple polytropic
    $\Gamma2$ setup by looking at the proper distance as a
    function of time.  The blue dotted line is from an evolution
    of the original initial data set, while the solid black line
    uses the third iteration step of the eccentricity reduction
    procedure.}
  \label{fig:EccRed_propdist}
\end{figure}

\begin{table}[t]
  \centering
  \caption{ \label{tab:listEccRedPars} Iteration procedure for two
    binary setups starting at $M\omega_{22}^0=0.0365$.
    The first
    configuration is a $\Gamma2$ binary with individual masses of
    $M^A = M^B = 1.515$ and a total ADM mass of
    $M_{\rm{ADM}}=3.006$. The second one is an equal mass SLy
    configuration with $M^A = M^B = 1.350$ and
    $M_{\rm{ADM}}=2.6782$.  
    The columns give the
    eccentricity $e$ and radial velocity $v_r$ input to the code
    [cf.\ Eq.~\eqref{eq:ellinspiralKV}], along with $\hat{e}_d$,
    the remaining eccentricity in the binary evolution measured
    using the proper distance $d$ in the hypersurface. We also
    list the values of the binding energy $E_b=M_{\rm{ADM}}-M$ and
    the angular momentum $J_{\rm{ADM}}$, which we normalize by $M$
    and $M^2$, respectively.}
  \begin{tabular}{@{}c|cccc|cc@{}}        
    \toprule[0.1em] EOS &Iter & $e$ $[10^{-3}]$ & $v_r$
    $[10^{-3}]$ & $\hat{e}_d$ $[10^{-3}]$ & $E_b/M$ $[10^{-3}]$ &
    $J_{\rm{ADM}}/M^2$\\ \hline \multirow{4}{*}{$\Gamma2$} &0 & 0
    & 0 & 9.77 & -7.984 & 1.0700 \\ &1 & -6.8 & -1.63 & 1.38 &
    -7.922 & 1.0729 \\ &2 & -5.7 & -1.14 & 0.91 & -7.920 & 1.0738
    \\ &3 & -6.3 & -1.16 & 0.56 & -7.920 & 1.0734 \\ \hline
    \multirow{4}{*}{SLy } &0 & 0 & 0 & 12.41 & -8.115 & 1.0541
    \\ &1 & -6.0 & -1.13 & 7.80 & -8.103 & 1.0580 \\ &2 & -12.1 &
    -1.91 & 3.97 & -8.088 & 1.0615 \\ &3 & -13.7 & -1.09 & 0.87 &
    -8.085 & 1.0625 \\
    \bottomrule[0.1em]
  \end{tabular}
\end{table}
Table~\ref{tab:listEccRedPars} shows the numerical values for the
eccentricity reduction iteration for two runs with different
setups, viz., two equal-mass binaries with different total masses, one
with the SLy EOS and the other with the $\Gamma2$ EOS.
Additionally, we show the proper distance $d$ for the
latter setup in Fig.~\ref{fig:EccRed_propdist}.  We used \SGRID
with $n_A=n_B=26$, $n_\varphi=8$, $n_{\rm Cart}=22$ points. The
evolution was done in BAM with a constraint damping Z4c evolution
scheme, as described in Sec.~\ref{sec:Dynamical_Evolutions}. We
used a total of 7 refinement levels, where the two inner levels
contained 96 points in each direction in each box with a grid spacing of 0.15 in the
finest. The other outer boxes all contain 192 points in each direction; the grid
spacing is doubled on each successive level moving outwards.  Finally, the
outermost level is given by a cubed sphere
(cf.~\cite{RonIacPao96,Tho04,PolReiSch11,HilBerThi12}; this is also
used for the evolutions in Sec.~\ref{sec:Dynamical_Evolutions}) with
192 and 84 points in the radial and azimuthal directions, respectively.
Over the course of three
iteration steps, using the method outlined above to iteratively
determine the correction parameters $\hat{e}_d$ and $v_r$, we
were able to decrease the eccentricity in the $\Gamma 2$ case from
$\hat{e}_{d\text{,0}} = 9.8\times 10^{-3}$ to
$\hat{e}_{d\text{,3}} = 5.6\times 10^{-4}$, which can be seen in
a significant improvement of the oscillations of the proper
distance (Fig.~\ref{fig:EccRed_propdist}).  Similar improvements
in the gravitational waves can be seen in
Fig.~\ref{fig:Ecc_omega_estimation} for the SLy case.  
  
Furthermore, we can try to give some PN estimates of the
improvement of the data by looking again at the PN expressions
from Mora and Will~\cite{MorWil03}, which we already utilized in
Sec.~\ref{sec:eccentricity_high}. We can even expand our
comparisons to fourth post-Newtonian order (as summarized in
\cite{KyuShiTan14}) without any essential changes in the results,
since the 4PN contribution is small. 

Given the fairly large initial separation of the binaries, we
expect the fourth post-Newtonian order results for the ADM energy
and angular momentum in terms of the binary's angular velocity to
be quite accurate and we indeed find that the eccentricity reduced
data gives a better match to the values for these quantities for a
circular orbit than do the original data:
Table~\ref{tab:listEccRedPars} gives the binding energy $E_b$ and
the angular momentum $J_{\rm{ADM}}$ and if we compare these to PN
for the SLy run (with $J_{\text{ADM, PN}}=1.0637M^2$ and $E_{b
  \text{, PN}}=-0.008081M$), we find a relative error of $\sim
1\%$ for original data and an error of $\sim 0.2\%$ for the data
at the third iteration.  Similarly, for the simple polytropic
$\Gamma 2$ run, we compare our results to $J_{\text{ADM,
    PN}}=1.0731 M^2$ and $E_{b \text{, PN}}=-0.007918M$ and find
an improved agreement of the eccentricity reduced data with a
deviation of $\sim 0.04\%$, compared with a deviation of $\sim
0.4\%$ for the original data.  We thus see that the eccentricity
reduced data are closer to quasi-equilibrium than the standard
initial data, as was noted by~\cite{KyuShiTan14}.

\subsection{Unequal masses}
\label{sec:unequal}

\begin{figure}[t]
  \includegraphics[width=0.5\textwidth]{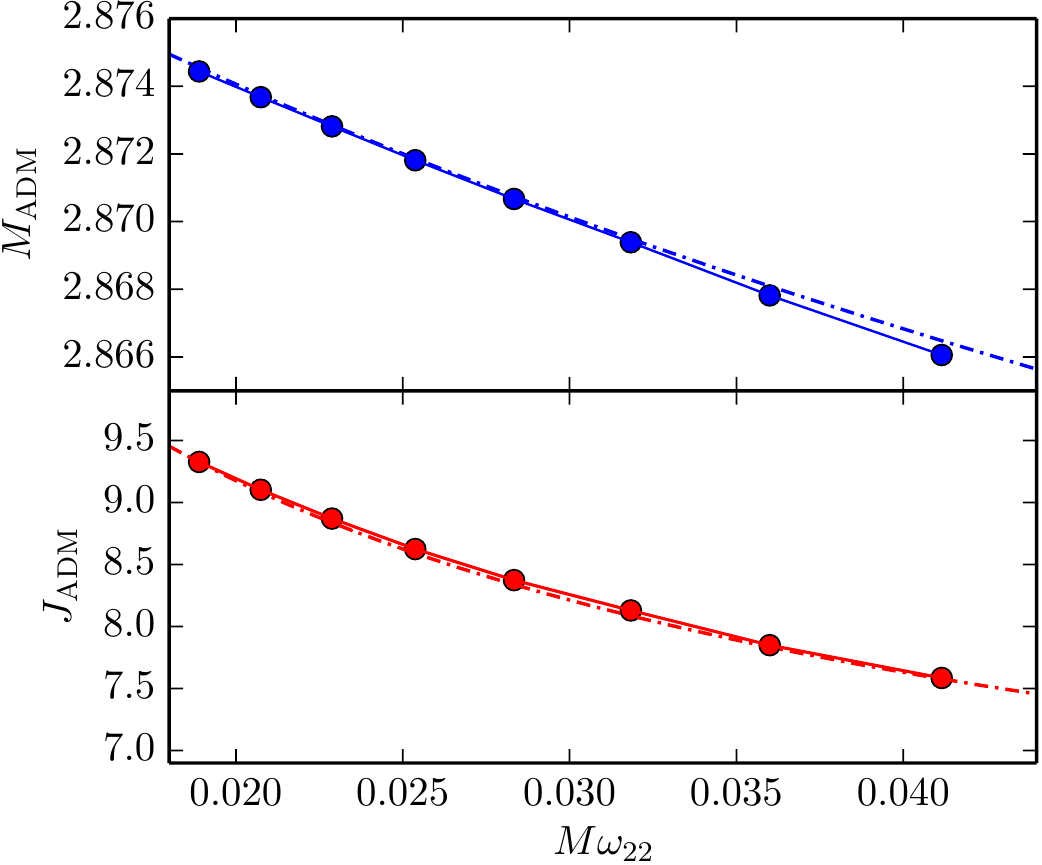}
  \caption{$q=2.06$ sequence.  Top panel: ADM mass compared with
    4PN tidal results (dot dashed lines).  Bottom panel: ADM
    angular momentum compared with 4PN tidal results (dot dashed
    lines).}
  \label{fig:Unequal_mass_sequence}
\end{figure}

As outlined in Sec.~\ref{sec:Introduction} and explained in more
detail in Appendix~\ref{sec:Population_Synthesis_M}, some population
synthesis models predict relatively high mass-ratio BNS systems.  In
this section we want to illustrate \SGRID's ability to generate
quasiequilibrium configurations for such systems.  We model the
systems by the stiff EOS MS1b. This EOS allows rather large baryonic
masses up to $M_b = 3.35$ and gravitational masses up to $M = 2.76$
for single neutron stars in isolation. As is seen in the next
subsection, it is one of the EOSs for which we achieved high neutron
star masses in equal-mass binaries with \SGRID.

The particular configuration constructed below consists of
$M_b^A=1.00 \ (M^A=0.94)$ and $M_b^B=2.20 \ (M^B=1.94)$ neutron
stars resulting in a mass ratio of $q=2.06$.  We want to highlight
that this is (i) a slightly larger mass ratio than the largest
predicted by the population synthesis models discussed in
Appendix~\ref{sec:Population_Synthesis_M}, (ii) the largest mass ratio
computed for a realistic (irrotational) binary configuration in
quasi-equilibrium,\footnote{In~\cite{Tic09a} (see that paper's
  Table~1), one of us computed corotating binary neutron star
  initial data with baryonic mass ratios up to $3$ for large
  separations, using a polytropic EOS and small baryonic masses.} and (iii) the largest mass
ratio evolved in full general relativity;\footnote{Up to now the
  highest mass ratio evolved in full general relativity was
  $q=1.5$~\cite{ShiTan06a,DieBerUje15}.}  see
Sec.~\ref{sec:dyn_highq}.  We employ a grid with $n_A=n_B=28$,
$n_\varphi=8$, $n_{\rm Cart}=24$ and construct a sequence varying
the distance parameter $b\in[16,30]$. This results in GW frequencies
of $M \omega_{22} \in [0.019,0.041]$.

The ADM mass and angular momentum for this sequence are shown in
Fig.~\ref{fig:Unequal_mass_sequence}.  As a comparison, we show the
$4$PN results including tidal components given in Appendix~A
of~\cite{KyuShiTan14} (obtained from~\cite{Bla14}
and~\cite{VinFla10}).  However, the influence of the tidal contributions and
higher-PN terms (above 2PN) is negligible at the scale
shown here.
Additionally, we emphasize that a direct comparison of these results
is not really warranted, since \SGRID uses the conformal flatness
approximation, which is known to be violated at $2$PN (see, e.g.,
the discussion in Sec.~III~A of~\cite{MolMarJoh14}).

Due to the newly implemented mechanism in \SGRID which adjusts the
center of mass so that $P^y_{\rm ADM}$ is kept small (ideally
vanishing), the linear momentum of the configuration stays within
$\sim 10^{-5}$ to $\sim 10^{-4}$. In particular, for the
configuration with $M \omega_{22} = 0.0359$ that we evolve in
Sec~\ref{sec:dyn_highq}, the linear momentum is $\vv{P}_{\rm
  ADM}=(4.23\times 10^{-7},-1.10\times 10^{-5},-1.96\times
10^{-6})$.
    
  \subsection{High compactness}

\begin{table*}[t]
  \setlength{\tabcolsep}{4pt} \centering
  \caption{ Overview of high compactness configurations without fine
    tuning the iteration procedure.  The columns refer to: the \SGRID
    distance parameter $b$, the dimensionless gravitational wave
    frequency $M\omega_{22}$, the maximum baryonic mass $M_{b\text{,
        max}}$, the corresponding gravitational mass $M_\text{max}$, and
    compactness $\mathcal{C}_\text{max}$.
    \label{tab:highcompact}   }
  \begin{tabular}{c|ccccc|ccccc|ccccc}        
    \hline EOS & $b$ & $M \omega_{22}$ & $M_{b\text{, max}}$ & $M_\text{max}$ &
    $\mathcal{C}_\text{max}$& $b$ & $M \omega_{22}$ & $M_{b\text{, max}}$
    & $M_\text{max}$ & $\mathcal{C}_\text{max}$& $b$ & $M \omega_{22}$ &
    $M_{b\text{, max}}$ & $M_\text{max}$ & $\mathcal{C}_\text{max}$
    \\ \hline SLy & 22 & 0.032 & 1.727 & 1.534 & 0.199 & 26 & 0.026 &
    1.744 & 1.547 & 0.200 & 32 & 0.020 & 1.780 & 1.575 & 0.204 \\ ENG & 22
    & 0.033 & 1.788 & 1.583 & 0.199 & 26 & 0.024 & 1.824 & 1.611 & 0.203 &
    32 & 0.021 & 1.852 & 1.632 & 0.206 \\ MPA1 & 22 & 0.035 & 1.871 &
    1.661 & 0.200 & 26 & 0.029 & 1.899 & 1.676 & 0.202 & 32 & 0.022 &
    1.927 & 1.683 & 0.203 \\ ALF2 & 22 & 0.036 & 1.899 & 1.678 & 0.200 &
    26 & 0.029 & 1.937 & 1.708 & 0.204 & 32 & 0.023 & 1.976 & 1.738 &
    0.208 \\ H4 & 22 & 0.038 & 1.976 & 1.762 & 0.197 & 26 & 0.031 & 1.996
    & 1.777 & 0.199 & 32 & 0.024 & 2.026 & 1.800 & 0.203 \\ MS1b & 22 &
    0.043 & 2.183 & 1.931 & 0.198 & 26 & 0.036 & 2.216 & 1.963 & 0.201 &
    32 & 0.027 & 2.250 & 1.982 & 0.203 \\ MS1 & 22 & 0.043 & 2.205 & 1.955
    & 0.197 & 26 & 0.036 & 2.238 & 1.981 & 0.200 & 32 & 0.028 & 2.273 &
    2.007 & 0.203 \\ $\Gamma2.72$ & 22 & 0.042 & 2.162 & 1.908 & 0.205 &
    26 & 0.034 & 2.183 & 1.924 & 0.207 & 32 & 0.027 & 2.238 & 1.966 &
    0.212 \\
    \hline
\end{tabular}
\end{table*}

While population synthesis models predict a number of binary neutron
star systems containing high-mass (and thus high-compactness) neutron
stars (see Appendix~\ref{sec:Population_Synthesis_M}), computing initial
data for binary neutron stars with high compactnesses is a challenging
task for most codes. The highest compactness achieved for a neutron
star in a relativistic binary is: $\mathcal{C}\simeq0.26$ for a
neutron star-black hole system~\cite{HenFouKid14},
$\mathcal{C}\simeq0.25$ for corotating binary neutron
stars~\cite{UsuUryEri00}, $\mathcal{C}\simeq0.26$ for irrotational
binary neutron stars~\cite{TanShi10}, and $\mathcal{C}\simeq0.22$ for
binary neutron stars with relatively low spins~\cite{TsoUryRez15}.
  
In the following we present results for most EOSs listed in
Table~\ref{tab:listEOS} (except for the $\Gamma2$ EOS, which does not
allow high compactness neutron stars) and provide an estimate of the
highest compactness easily reachable with our numerical method. In
physical terms, the maximum feasible compactness depends mostly on two
different properties: (i) the chosen EOS and (ii) the binary
separation. However, in \SGRID the resolution and iteration procedure
(e.g., the softening parameter and mass increment) also play important
roles.
  
As a starting point, we choose a simple procedure to estimate the
maximum compactness $\mathcal{C}_{\rm max}$ reachable with \SGRID. We
consider equal-mass nonspinning binaries, fix the separation parameter
$b$, and increase the baryonic masses of the two stars in steps of
$\Delta M_b=0.005$, starting from $M_b^A = M_b^B =1.2$.  We stop
increasing the mass when the elliptic solve of
Eqs.~\eqref{eq:final_eqs}, in particular the Hamiltonian constraint
equation, does not reduce the residual and no solution can be found.
For our test we use a resolution of $n_A=n_B=24$, $n_\varphi=8$,
$n_{\rm Cart}=20$.
  
Table~\ref{tab:highcompact} summarizes the maximum compactness
achieved for different EOSs and separations. In all cases we were able
to achieve higher compactnesses at larger separations. The maximum
compactness for the EOSs and separations considered lies within
$(0.197,0.212)$ for this simple iteration procedure. We obtain the
highest compactness for this test (with a constant number of grid
points) with the simple polytropic EOS $\Gamma2.72$. This is to be
expected, since runs with piecewise polytropic EOSs require higher
resolutions to obtain the same accuracy as runs with simple polytropic
EOSs, as discussed in Sec.~\ref{sec:conv}.
   
Even higher compactnesses can be achieved by reducing $\Delta M_b$,
decreasing the softening parameter to 0.2, and using higher
resolutions. We now give an explicit example of this procedure.

In order to achieve high compactness with the SLy EOS, we have
considered a non-spinning equal mass system with a separation
parameter of $b=31$. We started with baryonic masses $M_b^A = M_b^B
=1.5$ and increased the mass in each iteration by a factor of 1.07
until we reached $M_b^A = M_b^B =1.82$. The highest resolution used
for this configuration was $n_A=n_B=28$, $n_\varphi=8$, $n_{\rm
  Cart}=24$. This results in a binary with ADM mass
$M_\text{ADM}=3.193$, angular momentum $J_\text{ADM}=12.759$, orbital
angular velocity $\Omega=0.00336$, dimensionless GW-frequency $M
\omega_{22}=0.0216$, and a coordinate separation of $63.270$ between
the star centers.  A single TOV star with the same baryonic mass would
have an ADM mass of 1.60564 and a radius of 7.69608 (in standard
Schwarzschild areal radius coordinates), which implies a compactness
of 0.2086.

Increasing the baryonic masses beyond $M_b^A = M_b^B =1.82$ has proven
very difficult. For masses above this value we observe that the
elliptic solver cannot always solve Eq.~\eqref{eq:metric1} for $\psi$
within the main iteration, because the values from the previous
iteration are not good enough as an initial guess for the
Newton-Raphson scheme we use. We were able to address this problem by
making two changes to our Newton-Raphson scheme. First, we take
smaller Newton steps if this results in a smaller residual error than
a full step (this procedure is known as backtracking). And second, if
backtracking also fails we simply skip the elliptic solve for
$\psi$. The overall iteration still succeeds if we do not skip the
elliptic solve for $\psi$ too often. Using this trick we started from
the initial data for $M_b^A = M_b^B =1.82$ and increased the masses by
a factor of 1.02 in the main iteration until we reached $M_b^A = M_b^B
=2.0$.  The resulting initial data were again computed with
$n_A=n_B=28$, $n_\varphi=8$, $n_{\rm Cart}=24$ and result in a binary
with ADM mass $M_\text{ADM}=3.459$, angular momentum
$J_\text{ADM}=14.444$, orbital angular velocity $\Omega=0.00347$,
dimensionless GW-frequency $M \omega_{22}=0.0242$, and a coordinate
separation of $63.262$ between the star centers. The Hamiltonian
constraint violation for this data set is about twice as big as for
the lower mass data, because we do not always solve
Eq.~\eqref{eq:metric1}, which is the Hamiltonian constraint.  A single
TOV star with the same baryonic mass would have an ADM mass of 1.74067
and a radius of 7.6104 (in standard Schwarzschild area radial
coordinates), which implies a compactness of 0.2287.

If we try to increase the baryonic masses beyond $M_b^A = M_b^B =2.0$
we can never solve Eq.~\eqref{eq:metric1} for $\psi$, so we obtain
constraint violating initial data.

\subsection{Convergence of SGRID}
\label{sec:conv}

\begin{table}[t]
  \caption{ \label{tab:convergence} Convergence test
    setups. Columns refer to: Name of the configuration, EOS,
    distance parameter $b$, eccentricity parameter $e$
    [from~\eqref{eq:x_circlecenters}], the baryonic masses $M_b^A$
    and $M_b^B$, and the angular velocity vector $\omega^i$ as
    specified in Eq.~\eqref{eq:matterasm0.5}.  }
  \begin{tabular}{c|ccccccc}        
   \hline 
   Name & EOS & $b$ & $e$ & $M_b^A$ & $M_b^B$ & $\omega^i_{A,B}$ \\ 
   \hline 
   $\Gamma$2q100w000e00 & $\Gamma 2$ & 20 & 0.0 & 1.4336 & 1.4336 & $ (0,0,0) $\\ 
   $\Gamma$2q100w005e00 & $\Gamma 2$ & 20 & 0.0 & 1.4336 & 1.4336 & $ 0.005 \cdot (1,1,1)$ \\ 
   $\Gamma$2q100w000e03 & $\Gamma 2$ & 20 & 0.3 & 1.4336 & 1.4336 & $ (0,0,0) $ \\ 
   $\Gamma$2q100w005e03 & $\Gamma 2$ & 20 & 0.3 & 1.4336 & 1.4336 & $ 0.005 \cdot (1,1,1)$ \\ 
   $\Gamma$2q116w000e00 & $\Gamma 2$ & 20 & 0.0 & 1.5491 & 1.3199 & $ (0,0,0) $ \\ 
   \hline 
   H4q100w000e00 & H4 & 20 & 0.0 & 1.4687 & 1.4687 & $ (0,0,0) $  \\ 
   H4q100w005e00 & H4 & 20 & 0.0 & 1.4687 & 1.4687 & $0.005 \cdot (1,1,1)$ \\ 
   H4q100w000e03 & H4 & 20 & 0.3 & 1.4687 & 1.4687 & $ (0,0,0) $ \\ 
   H4q100w005e03 & H4 & 20 & 0.3 & 1.4687 & 1.4687 & $0.005 \cdot (1,1,1)$ \\ 
   H4q116w000e00 & H4 & 20 & 0. & 1.5887 & 1.3506 & $(0,0,0) $ \\ 
   \hline
  \end{tabular}
\end{table}
    
\begin{figure}[t]
  \includegraphics[width=0.49\textwidth]{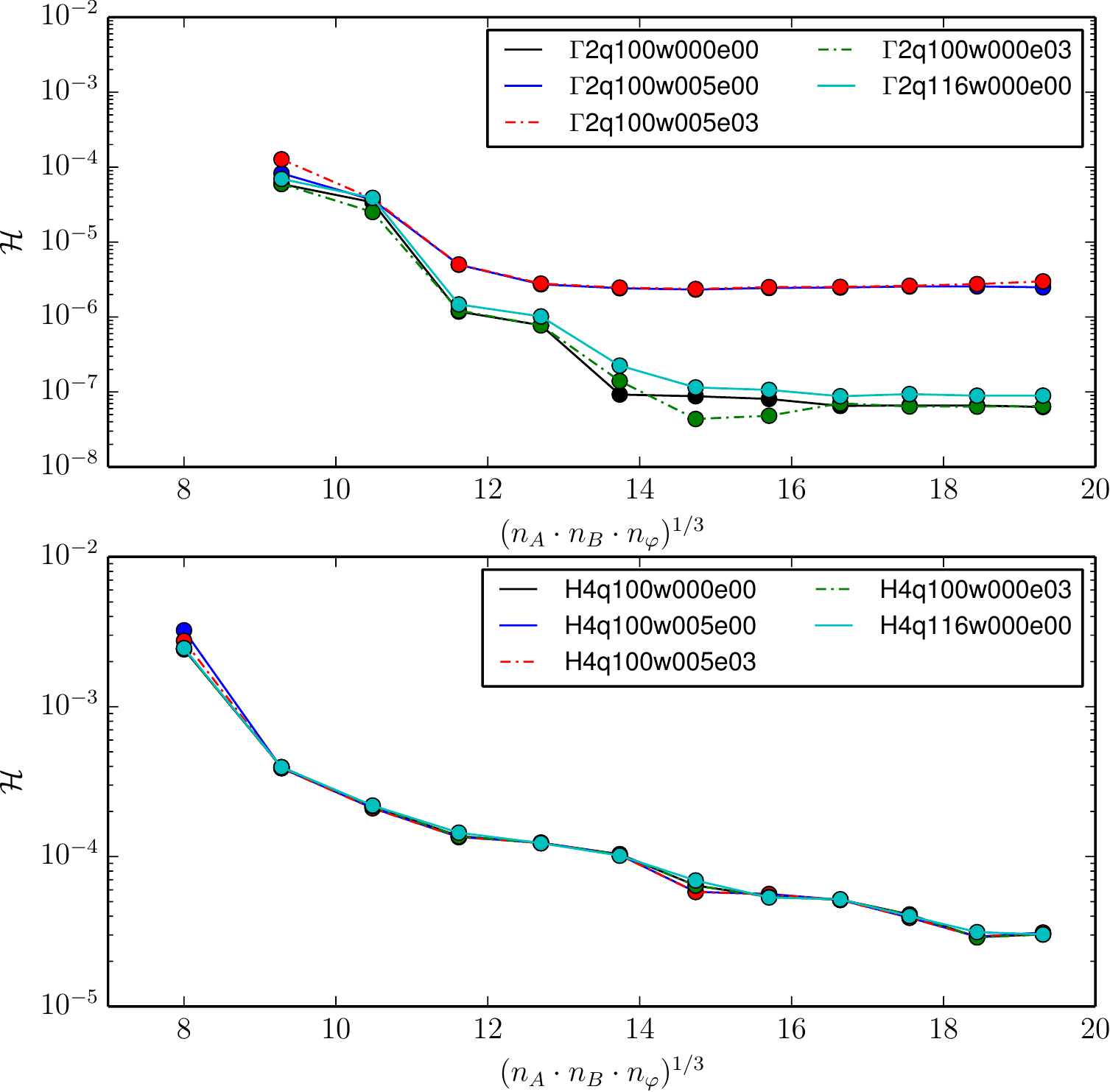}
  \caption{Convergence analysis for representative configurations.
    Simple polytropic $\Gamma2$ configurations (top panel) and
    piecewise polytropic (H4) configurations (bottom panel).  The plot
    shows the average Hamiltonian constraint inside the two stars in
    the regions $A \in [0,A_\text{max}]$ and thus includes the stars'
    surfaces, which are the most problematic regions. We have fixed
    $n_\varphi=8$ and use $n_A = n_B$, $n_\text{Cart}=n_A-4$.  }
  \label{Fig:conv}
\end{figure}

The convergence of \SGRID was already presented in Fig.~3
of~\cite{Tic09a} for a corotating equal-mass quasicircular binary
neutron star system with a simple polytropic EOS. However, we want
to show here how well the code converges in more complicated
situations.  To illustrate this, we consider the configurations
presented in Tab.~\ref{tab:convergence}.  In particular, we choose
configurations with $q=1.0$ where each star has a gravitational mass of $1.35$. We
either investigate nonspinning configurations [i.e.,
  $\omega^i_A=\omega^i_B=(0,0,0)$] or choose an angular velocity of
$\omega^i_A=\omega^i_B=(0.005,0.005,0.005)$. We also use different
eccentricities of $e=0.0$ and $e=0.3$.  Finally, we consider an
unequal mass configuration with gravitational masses $1.45$ and $1.25$
($q=1.16$).  We compute initial data for these systems for both a
simple polytropic EOS as well as a piecewise polytropic EOS (the fit
to the realistic H4 EOS).

Figure~\ref{Fig:conv} summarizes our findings. In the simple
polytropic runs (upper panel), we encounter a higher Hamiltonian
constraint for $\Gamma2$q100w005e00 and $\Gamma2$q100w005e03 than
for the other configurations. This clearly shows that the addition
of spin increases the complexity of the system, while adding
eccentricity has no noteworthy effect.  In the cases where no spin
is present, the Hamiltonian constraint is one to two orders of
magnitude smaller. For unequal masses we see a slightly larger
constraint violation.  The momentum constraints converge in a
similar manner, but their magnitude is roughly one order of
magnitude smaller.

When piecewise polytropes are employed, the constraint violations
increase substantially.  This is to be expected, since the solution
is only $C^1$ at the interfaces between the different pieces, so one
will no longer obtain exponential convergence from the spectral
method. In this case, however, the addition of spin and eccentricity
or the consideration of unequal masses has no noteworthy additional
effect. Since spin most likely affects the convergence primarily
through the deformation of the star's surface, it is not surprising
that it does not affect the convergence so much for piecewise
polytropes at these resolutions, where the portion near the surface
is already quite troublesome, due to the necessity of resolving the
crust. We expect that one would see a difference due to the addition
of spin at much higher resolutions or for higher spin magnitudes.
This reduction in convergence when using piecewise
polytropes means that in most cases we use higher resolutions when 
computing piecewise polytropic setups, compared to previous work
with just simple polytropes.

\section{Dynamical Evolutions}
\label{sec:Dynamical_Evolutions}
  
In this section we present evolutions of our new initial data for two new
configurations: (i) a high-mass ratio simulation with $q =
2.06$, which is the \textit{highest mass ratio binary neutron star}
ever evolved in full numerical relativity; (ii) an unequal-mass
configuration where the spins are not aligned with the orbital
angular momentum, which is the \textit{first precessing
binary neutron star} simulation performed so far.
The new dynamical evolutions presented here show,
as a proof of principle, that the BNS phenomenology can be very rich
in these newly accessible regions of parameter space, where we are now able 
to study mass exchange during the inspiral or the precession and nutation 
of the orbital plane during a binary neutron star simulation. 
Additionally, we discuss (iii) the effect of eccentricity reduction on the
waveform phasing. \\

We perform the evolutions with the newest version of the BAM
code~\cite{BruGonHan06,ThiBerBru11,DieBerUje15} including the
recently implemented conservative mesh refinement algorithm 
employed in~\cite{DieBer15,DieBerUje15}. We
also use the Z4c scheme~\cite{BerHil09,HilBerThi12} with
constraint preserving boundary
conditions~\cite{RuiHilBer10,HilBerThi12}.  The BAM grid setup
consists of 7 refinement levels. The outermost level ($l=0$) uses a
multipatch ``cubed-sphere''
grid~\cite{RonIacPao96,Tho04,PolReiSch11,HilBerThi12}.

The setups of Sec.~\ref{sec:precessing} do not employ any symmetry
condition. For the other simulations we employ reflection symmetry
across the orbital plane, letting us use half the number of grid points
in the $z$-direction.  Further information about the particular grid
and initial conditions is given in Tab.~\ref{tab:simu-ID}.

\begin{table*}[t]
  \centering
  \caption{ \label{tab:simu-ID} Initial data and grid details for the
    dynamical evolutions. The columns refer to: the simulation name,
    the gravitational and baryonic masses of star A and B, the stars'
    dimensionless angular momenta $j^{A,B}$, the number of grid points
    employed in each direction in the fixed and moving boxes, the number of radial and azimuthal
    points, the finest grid spacing, and
    the outer boundary location.}
  \begin{tabular}{l|cccccc|cccccc} 
  \hline
    name & $M^A$& $M_b^A$ &$M^B$& $M_b^B$ & $\bm{j}^A$ & $\bm{j}^B$ & $n$ & $n^{mv} $& $n_r$ & $n_\theta$ & $h_6$ & $r_b$\\ 
    \hline
    MS1b-q206 & 1.944 & 2.200 & 0.944 & 1.000 & (0,0,0)  & (0,0,0)  & 128 & 72 & 144 & 63 & 0.250 & 1692 \\ 
    \hline
    SLy$^{(\nearrow \nearrow)}$ & 1.3553 &  1.500 & 1.1072 & 1.200 & $(0.13/\sqrt{3}) (1,1,1)$ & $(0.16/\sqrt{3})(1,1,1)$ & 128 & 64 & 128 & 56 & 0.245 & 1532 \\
    SLy$^{(\uparrow \uparrow)}$ & 1.3547 & 1.500 & 1.1067 & 1.200 & $0.077 (0,0,1)$ & $0.089 (0,0,1)$ & 128 & 64 & 128 & 56 & 0.245 & 1532 \\ 
    SLy$^{(00)}$ & 1.3544 & 1.500 & 1.1065 & 1.200 & (0,0,0)  & (0,0,0)  & 128 & 64 & 128 & 56 & 0.245 & 1532 \\ 
    \hline
    SLy-eccred &1.350& 1.495 &1.350& 1.495 &(0,0,0) &(0,0,0) & 192 & 96 & 192 &84 & 0.15 & 1384 \\ \hline
  \end{tabular}
\end{table*}
    
\subsection{A $q=2.06$ binary} 
\label{sec:dyn_highq}

\begin{figure}[t]
  \includegraphics[width=0.48\textwidth]{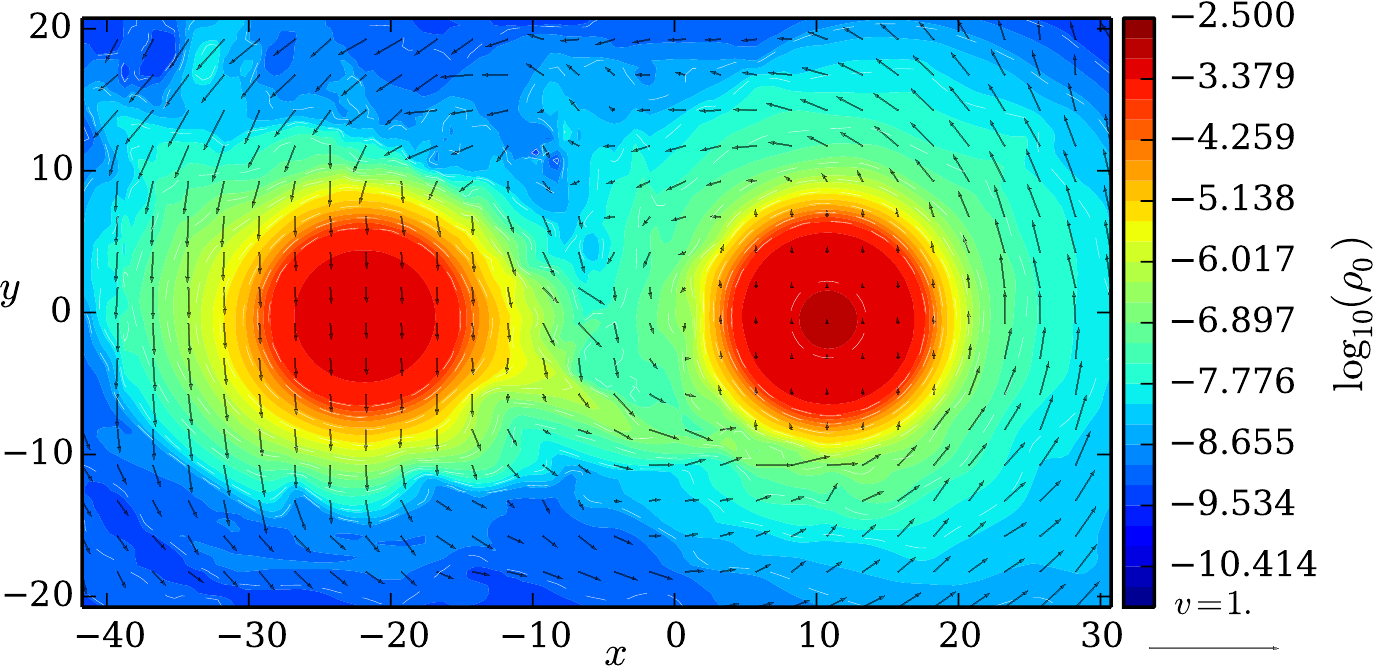}
  \includegraphics[width=0.48\textwidth]{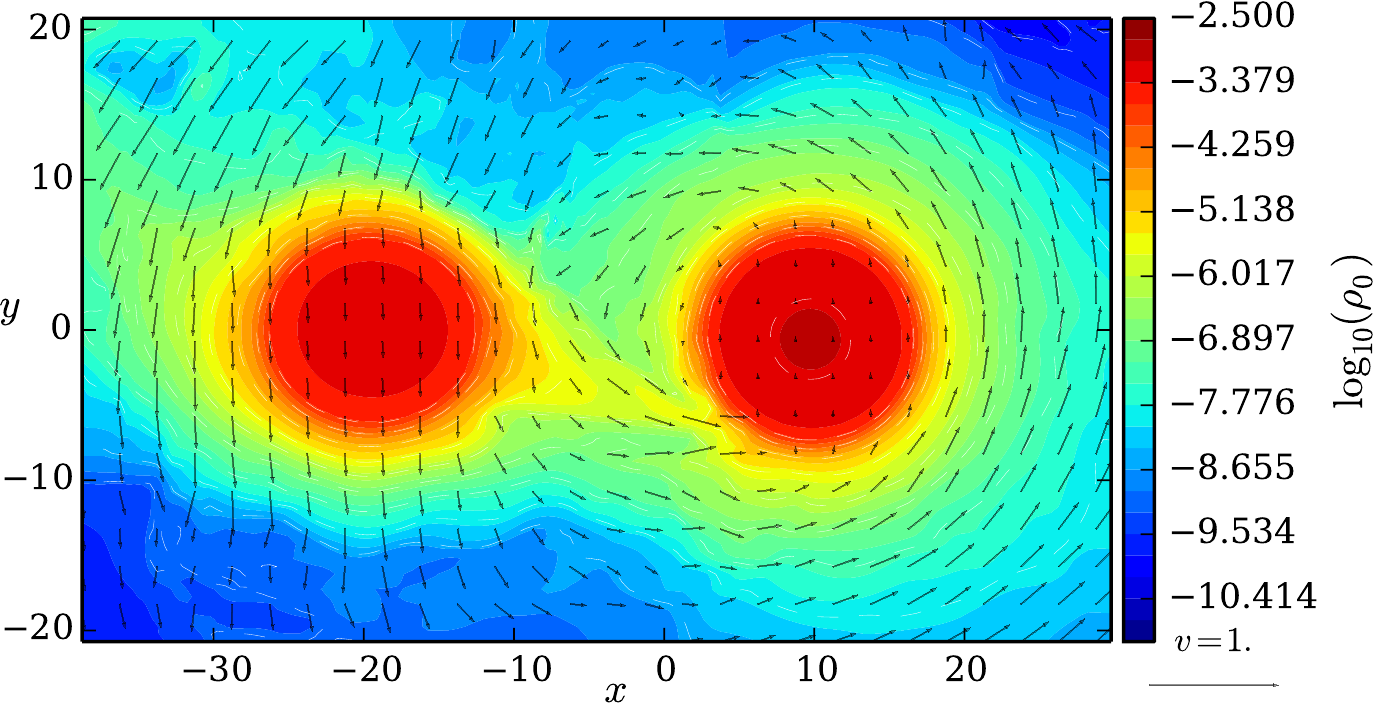}
  \includegraphics[width=0.48\textwidth]{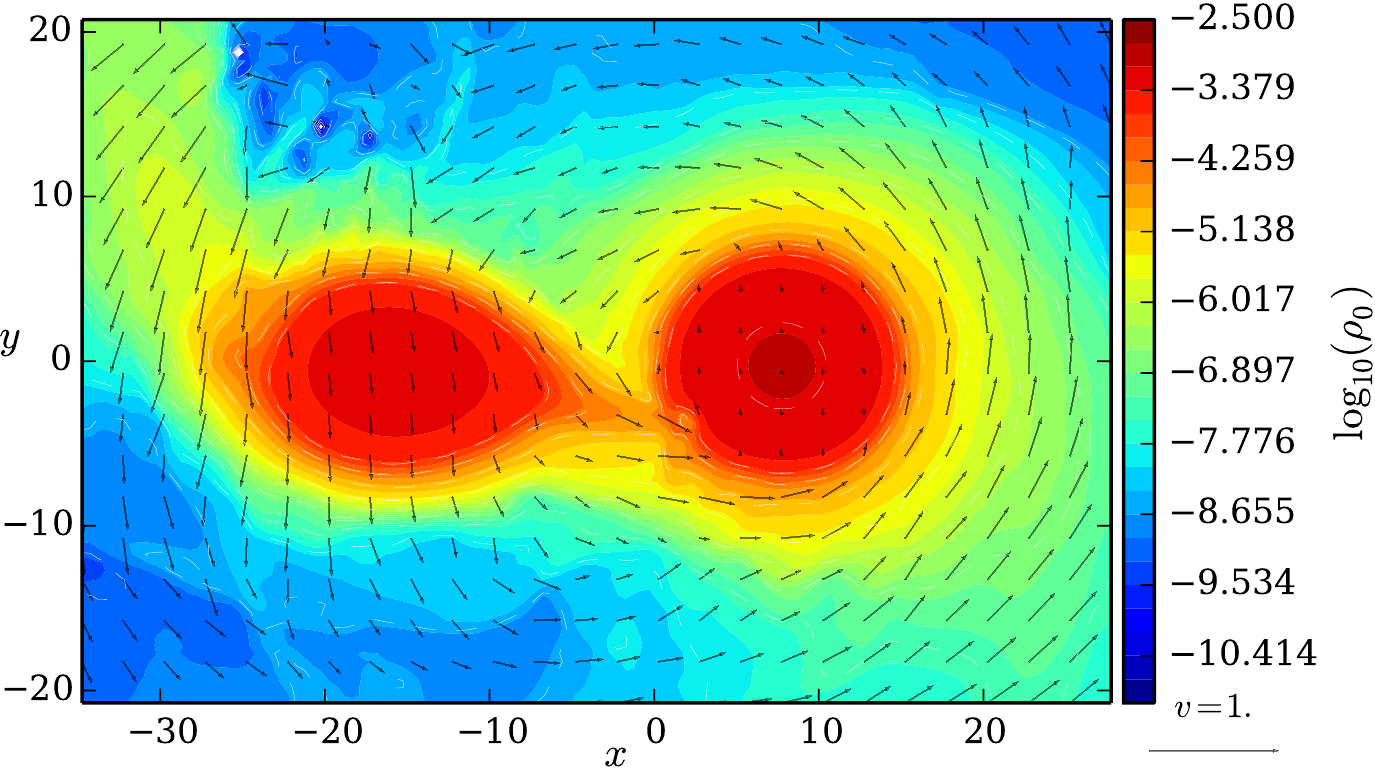}
  \caption{Plot of the matter density $\rho_0$ and the velocity $v^i$ in
    the orbital plane for the $q = 2.06$ simulation at times $t=1726M$
    (upper panel), two revolutions later at $t=2227M$ (middle panel),
    and yet another two revolutions later at $t=2644M$ (lower
    panel). Over the 4 revolutions shown, one can see a clear mass
    transfer between the two stars. Note that each plot has a somewhat
    different scale.}
  \label{fig:q206_dyn_1}
\end{figure}
    
In this section we evolve one of the models computed in
Sec.~\ref{sec:unequal}. We have set the rest mass of the primary
star to $M_b^A= 2.200$, corresponding to a gravitational mass of
$M^{A}= 1.944$ and compactness $\mathcal{C}^A=0.199$ in isolation,
while the companion is characterized by $M_b^B=1.000$, $M^B= 0.944$,
$\mathcal{C}^B=0.103$. We have not tried to add spin/eccentricity
or performed eccentricity reduction since we want to focus solely on
the high mass ratio effects.  Nevertheless, the setup shows a rather
small eccentricity of $\hat{e}_d=4.2 \times 10^{-3}$,
$\hat{e}_\text{GW} = 5.1 \times 10^{-3}$ and despite the high mass
ratio a small linear momentum of $\vv{P}_{\rm ADM}=(4.23 \times
10^{-7},-1.10 \times 10^{-5},-1.96 \times 10^{-6})$.  The initial
gravitational wave frequency is $M\omega_{22}^0=0.0359$. Star B is
already deformed at this separation indicated by a mass shedding
parameter of $\chi\simeq0.89$; cf.~Eq.~\eqref{eq:mass_shedding}.

This configuration, with a gravitational mass ratio of $q=2.06$, is
the highest mass ratio binary neutron star evolved in full general
relativity, and well above the mass ratio of $q=1.5$ considered
before~\cite{ShiTan06a,DieBerUje15}. Note that~\cite{FryBelRam15}
reports results of Newtonian smooth particle hydrodynamics (SPH)
evolutions (including radiation reaction) for a similar $q = 2$
setup [gravitational masses of $(1.0 + 2.0)$], but does not mention any
 mass transfer, which we find in our simulation. 
However, mass transfer is observed in white dwarf
binary simulations with $q=2$ \cite{DanRosGui11}.

\subsubsection{Mass transfer} 

In Fig.~\ref{fig:q206_dyn_1} we show snapshots of the density in the
orbital plane during the inspiral. The upper panel of
Fig.~\ref{fig:q206_dyn_1} shows the binary at $t=1726M$. Although
the stars are clearly separated, mass transfer from the companion
($M^B$) to the primary star ($M^A$) has already set in. Two
revolutions later [$t=2227M$ (middle panel)] and four revolutions
later [$t=2644M$ (lower panel)] the mass transfer becomes more
dramatic. During this period, $\sim (2$--$3) \times10^{-2} M_\odot$
of material was transferred between the two stars, i.e., $\sim
(2$--$3) \%$ of the rest mass of the less massive star. The estimate
is based on the rest mass leaving the finest refinement level around
star B and entering the refinement level of star A. The uncertainty
is mainly related to mass loss due to the artificial atmosphere
treatment~\cite{DieBerUje15}. (Note that the overall rest mass is conserved
to better than $0.12\%$ in this simulation until very late times, post-merger,
when matter starts leaving the grid.) We observe that for this system the
mass transfer happens continuously until the companion is tidally
disrupted.

The average rate of mass transfer is $\dot{M}_{AB} \sim 10^{-5} \sim
1 M_\odot \text{ s}^{-1}$, taking place for $\sim 10^{-2}\text{ s}$,
from which one can estimate the accretion power. Here we just make a
simple, order-of-magnitude estimate, since this is likely all that
is warranted by the accuracy of our estimate of the mass transfer.
If one just considers the change in energy of this matter going down
the more massive star's Newtonian potential well, we have an average
accretion power of $\sim\dot{M}_{AB}\mathcal{C}^A \sim 10^{-6} \sim
10^{53}\text{ erg s}^{-1}$, comparable to the neutrino luminosities
found in simulations of BNS
mergers~\cite{SekKiuKyu11b,SekKiuKyu15,PalLieNei15}.  (Recall that
$\mathcal{C}^A$ is the compactness of $M^A$ in isolation.)  This
gives a total accretion energy of $\sim 10^{51}\text{ ergs}$, which
is comparable to the energy released in a supernova.  But of course,
one cannot say anything definite about the amount of this energy
that would be released in photons or neutrinos, since these are not
present in our simulation. While one might expect to be able to see
the effects of this amount of mass transfer on the phase of the
gravitational waves, the present simulation does not appear to be
accurate enough to make such a comparison.

The merger happens at $t=2692M$. We classify the merger remnant as a
supramassive neutron star (SMNS), since it is below the maximum
supported gravitational mass of a rigidly rotating star for the MS1b
EOS, which is $\sim 3.2$--$3.3M_\odot$, i.e., roughly $\sim
15$--$20\%$ larger than the corresponding maximum gravitational mass
of a nonrotating neutron star (see, e.g., Sec.~2.9.1
in~\cite{Ste03}).  We thus do not expect the SMNS to collapse on
dynamical timescales. This is supported by our simulation, where no
indication for a collapse is present through the end of the
simulation at $t=6500M\simeq 0.09$~s. The central density reaches a
constant value $\rho_c~\simeq 9.8 \times 10^{-4} \simeq 6 \times
10^{14} \text{ g cm}^{-3}$ and the final merger product settles into
a stable configuration.

\begin{figure}[t]
  \includegraphics[width=0.45\textwidth]{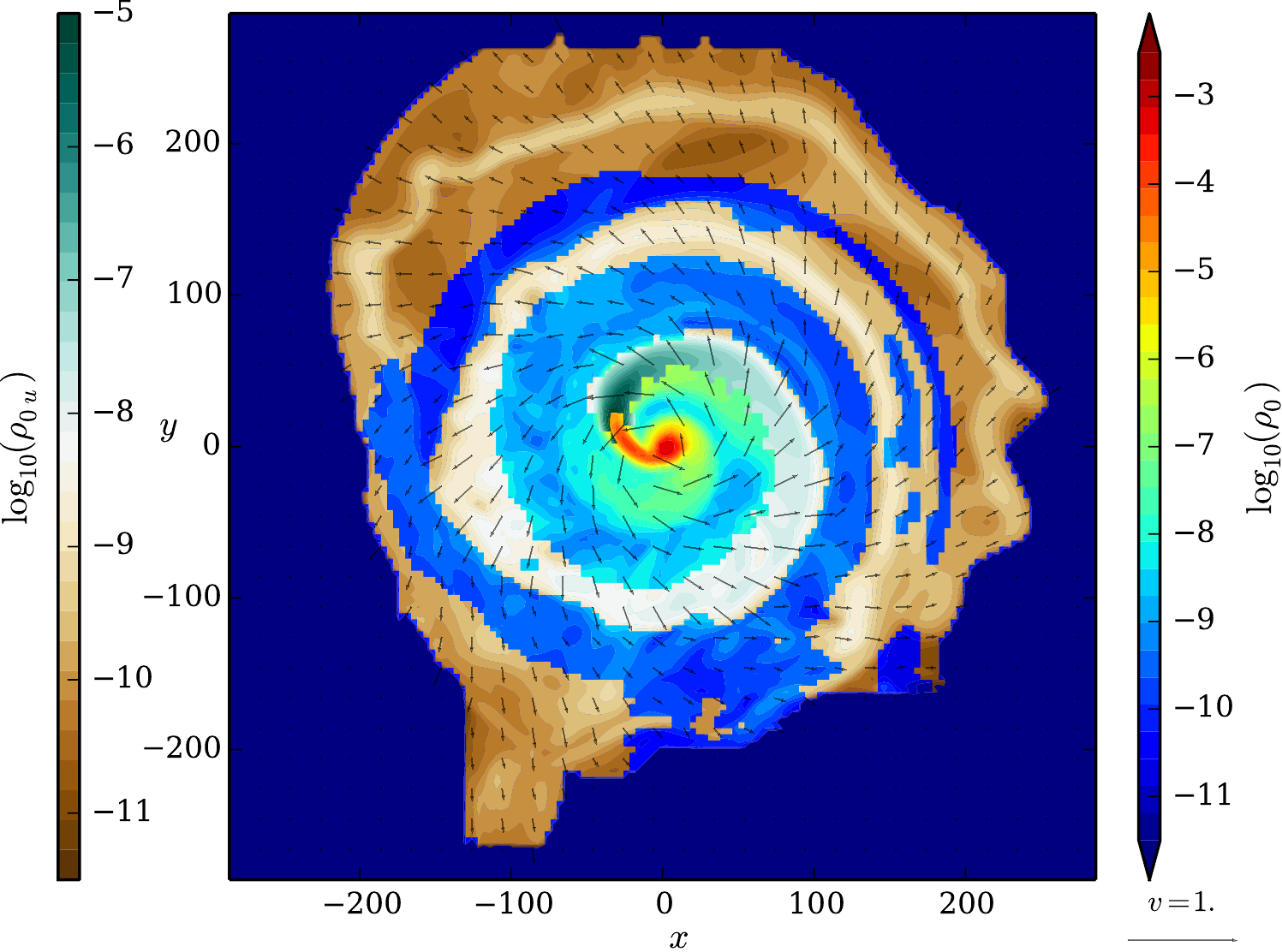}\\ 
  \includegraphics[width=0.45\textwidth]{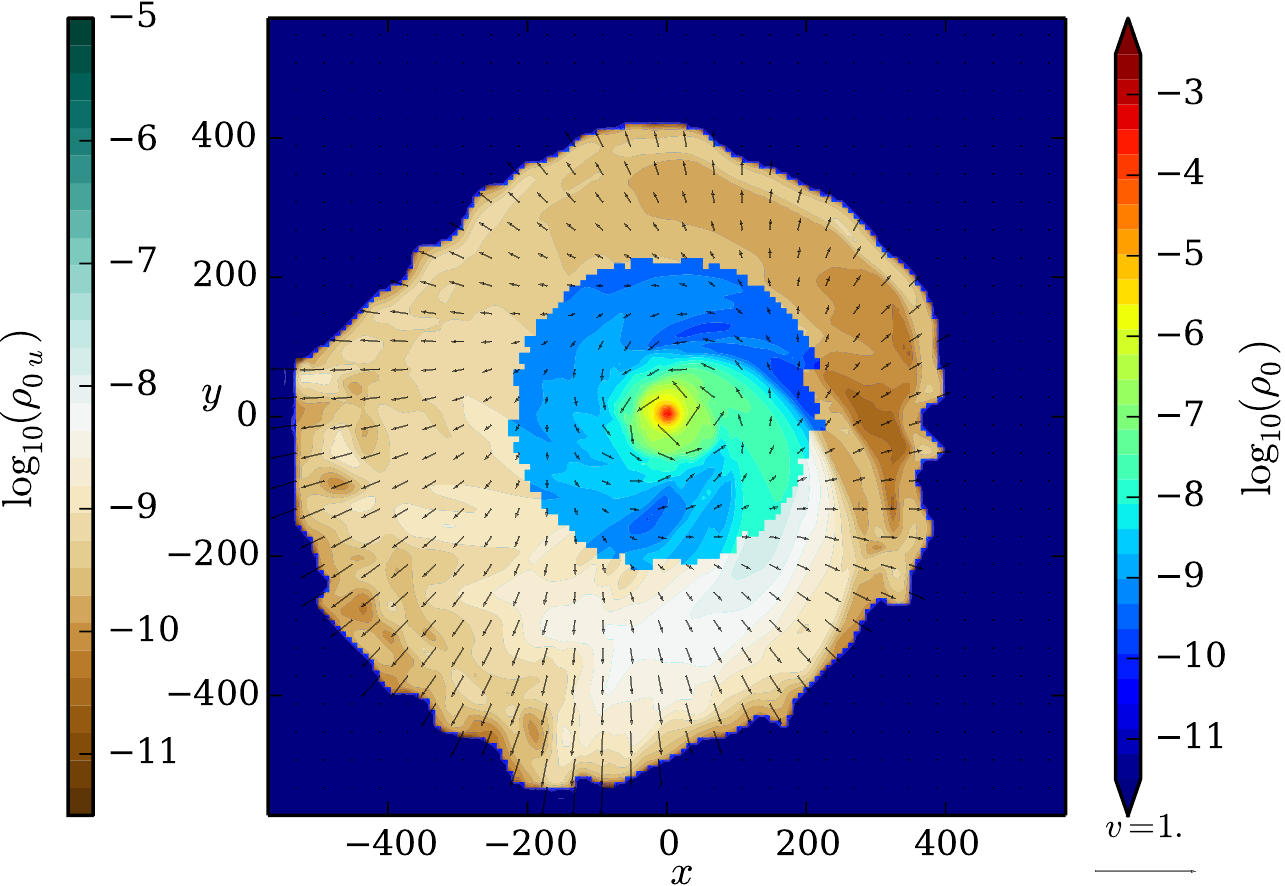}
  \caption{Plot of the matter density $\rho_0$, the density of unbound
    matter $\rho_{0u}$, and the velocity field $v^i$ in the orbital
    plane for the $q = 2.06$ simulation after merger. 
    In the upper panel ($t=2798M$) a
    clear spiral like pattern in the ejecta is visible, where material
    is expelled due to torque in the tidal tail of the companion star.
    The lower panel ($3421M$) shows that material is ejected over the
    entire grid anisotropically. Note that the two panels have
    different scales.}
  \label{fig:q206_dyn_3}
\end{figure} 

\subsubsection{Ejecta and kick}   

In this simulation we observe a significant mass ejection, $M_{\rm
  ejecta} \simeq 7.6 \times 10^{-2}M_\odot$, which is among the
largest found for full general relativistic simulations of binary
neutron star mergers, including the case of eccentric
binaries~\cite{EasPre12}, and much larger than any of the ejecta
masses found for the quasicircular case in the studies
in~\cite{HotKiuKyu12,DieBerUje15}. 

Figure~\ref{fig:q206_dyn_3} visualizes the ejected material,
distinguishing it from the bound material by using a different color
bar. Here we compute the unbound material using the method given
in~\cite{DieBerUje15}. In our simulation most of the material is ejected into the 
orbital plane by torque on the tidal tail of the companion star; 
cf.~the discussion in~\cite{HotKiuKyu12,DieBerUje15}. We can also see this in the
spiral-like pattern in the upper panel of Fig.~\ref{fig:q206_dyn_3},
at a time of $t=2798M$. In the lower pattern we see also clearly
that the ejection happens anisotropically, where the density inside
of the ejected material at a given radial distance from the SMNS
differs by several orders of magnitude in different directions.  The kinetic energy of the
ejecta is $\sim 2 \times 10^{-3} \simeq 4 \times 10^{50} \text{ erg}$. 
Note that the ejecta mass we
find is a factor of $\sim 2$ larger than that found in the Newtonian
calculation of~\cite{FryBelRam15} (Table~1 there). This is
significant, since, as
discussed in~\cite{DieBerUje15}, the uncertainty on the ejecta mass
is about $\lesssim 20\%$, and mainly due to the resolution.

The anisotropic mass ejection causes the merger remnant to
recoil. We approximate the ejecta's linear momentum by computing
\begin{equation}
  \vv{P}_\text{ej}=M_\text{ej} \mean{\vv{v}_\text{plane}} = M_\text{ej}  \frac{\int D
      \ \vv{v}_\text{plane} dx dy}{\int D dx dy} ,
\end{equation}

where the integrals here are restricted to the orbital plane (while $M_\text{ej}$ is
computed from the integral of the rest-mass density $D$ of the unbound matter over all three
dimensions) and $\vv{v}_\text{plane}$ denotes the 
ejecta velocity in the orbital plane (i.e., just the $x$ and $y$ components). We find that 
$v^\text{ej}_\text{kick}=\|\vv{P}_\text{ej}\|/M\sim 100-1000 \text{ km
  s}^{-1}$. 
This number is also consistent with the one obtained from the
coordinate position of the SMNS.
However, the value should just be seen as an order of magnitude
estimate, where the main difficulties here are the low
resolution, the long simulation time, and the gauge dependence of the measurement. 
Since this is an unequal-mass system, we also expect some contribution to the kick from
anisotropic gravitational wave emission. We compute this from the
GW linear momentum flux, 
\be
\vv{v}_\text{kick}^\text{GW} = -\frac{1}{M} \int\dot{\vv{P}}_\text{GW}dt 
\ee
where the linear momentum flux $\dot{\vv{P}}_\text{GW}$ is computed as in
\cite{BruGonHan06}. We find $v_\text{kick}^\text{GW}\simeq100\text{
  km s}^{-1}$ at merger. 
The GW kick is smaller than the ejecta
kick, as is found in black hole-neutron star
simulations~\cite{KyuIokOka15,KawKyuNak15}.

\begin{figure}[t]
  \includegraphics[width=0.5\textwidth]{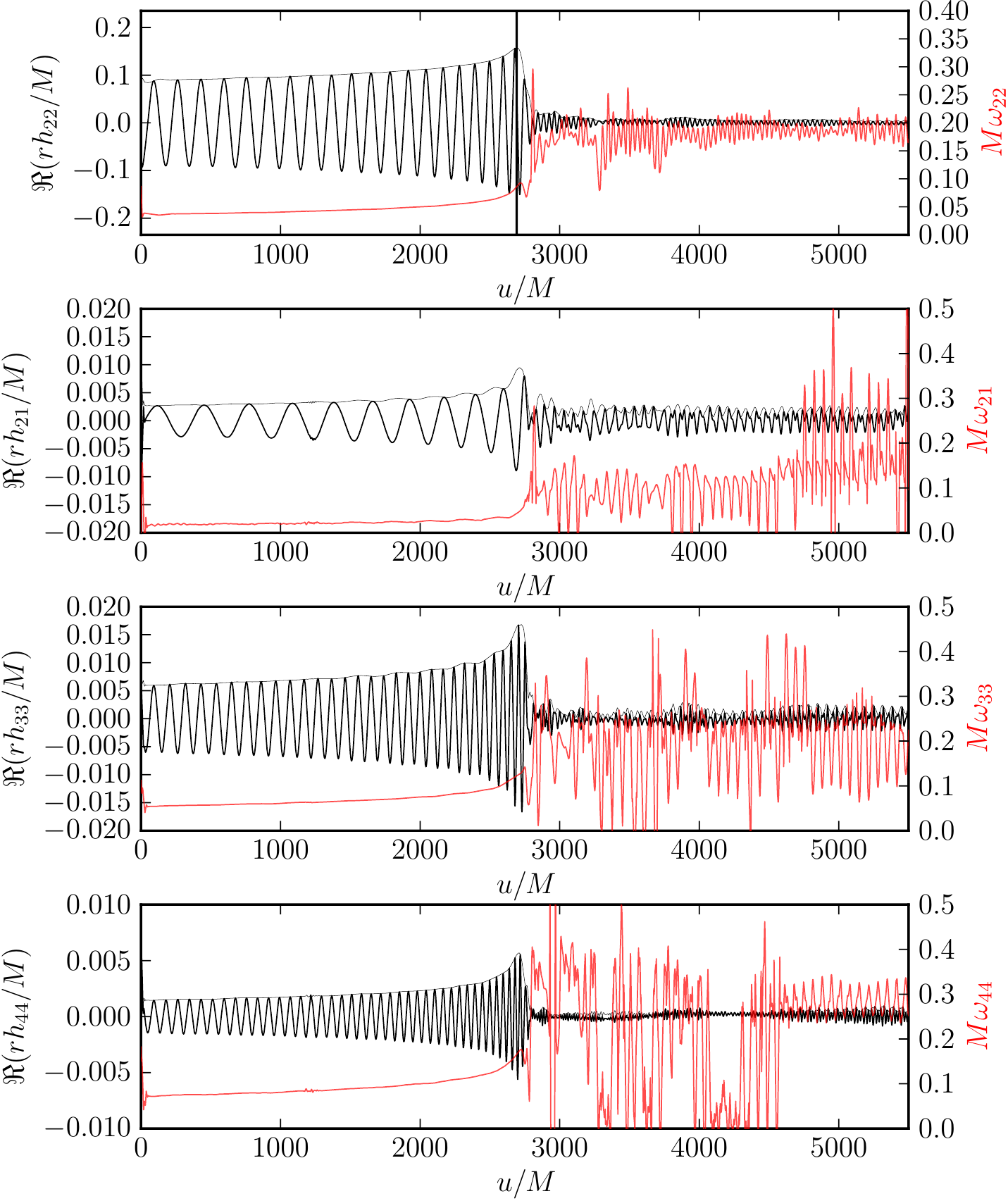}
  \caption{The four dominant multipoles of the GWs from the $q = 2.06$ simulation,
    viz., $(l,m)=(2,2)$, $(2,1)$, $(3,3)$, and $(4,4)$.  We plot
    the real part of the modes in black and show the dimensionless GW
    frequency for all modes in red. In the top plot, the solid
    vertical line marks the merger. The frequency oscillations are in
    part unphysical and resolution dependent. The large frequency spikes 
    in the post-merger phase are caused by zeros of
    the amplitude.}  
  \label{fig:q206_wave}
\end{figure}

\begin{figure}[t]
  \includegraphics[width=0.5\textwidth]{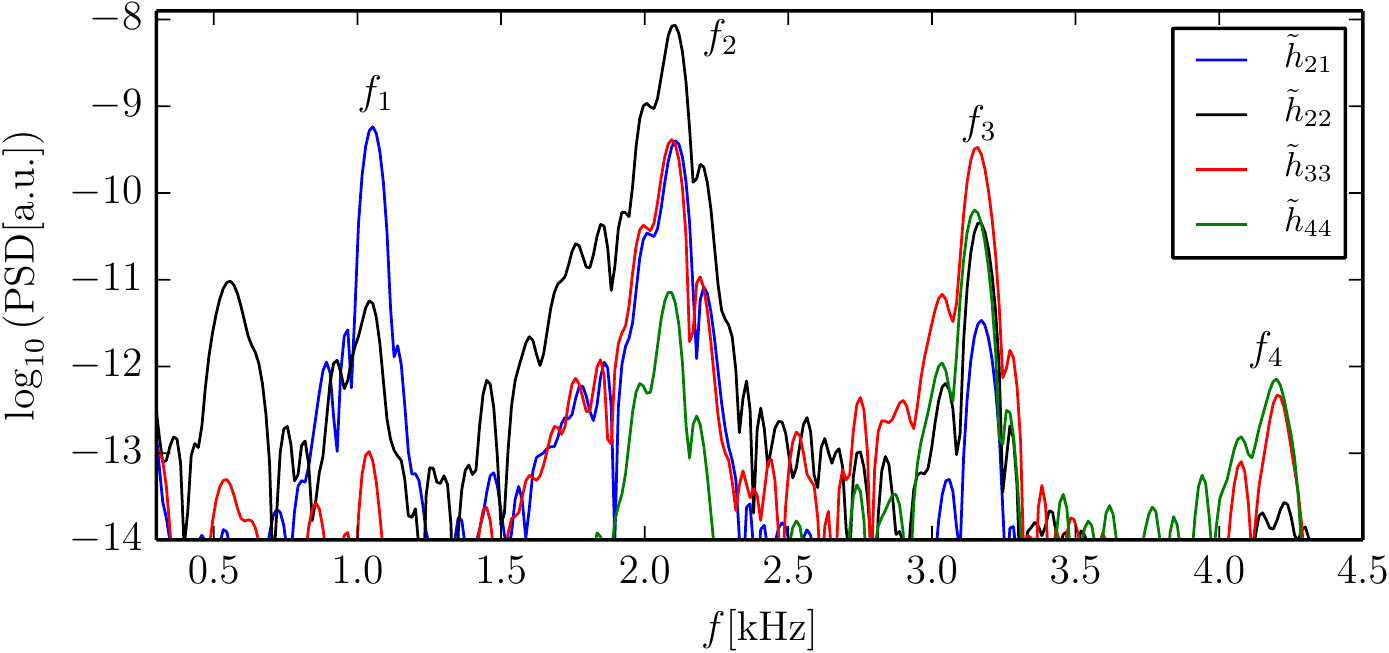}
  \caption{Power spectrum density for the dominant postmerger modes
    of the $q = 2.06$ simulation.  We used a time interval $u\in
    [3800M,5700M]$ for the computation, and therefore focus only
    on the SMNS spectrum. The harmonicity of the
    frequencies $f_1$, $f_2$, $f_3$, $f_4$ is clearly visible.}
  \label{fig:q206_spectra}
\end{figure}

\subsubsection{Gravitational waves} 

Let us discuss the gravitational waveform. The four
dominant modes of the gravitational waveform are presented in
Fig.~\ref{fig:q206_wave}. 
The inspiral-merger signal ends at $u_{\rm mrg} = 2692M$, and we find no
evidence for an obvious GW signature of the mass transfer described above.
We study the effect of the mass-ratio on
the GW multipolar structure by computing the relative GW energy
contribution of each dominant mode over the total energy released up to merger
($E_{lm}/E$) and comparing with the $q=1$ case with the same EOS~\cite{BerNagDie15}
(though with a slightly different total mass: $2.70$, as opposed to $2.89$ for the $q = 2.06$
simulation; the $q = 1$ system also radiates almost $40\%$ more energy per total mass than
the $q = 2.06$ system). 
We find that for the $q=1$ simulation $\sim99.6\%$ of the energy is 
released in the $(2,2)$-mode, while $\sim98\%$ is released for MS1b-q206.
For $q=1$ the $(2,1)$ and the $(3,3)$ modes are zero by symmetry; for $q=2.06$
the $(2,1)$ mode releases about $\sim0.08\%$ of the energy and the $(3,3)$ mode $\sim1.4\%$.
The next dominant mode, $(4,4)$, contributes to $\sim0.19\%$ of the
total energy for $q=1$ and $0.22\%$ for $q=2.06$.

Finally, we study the postmerger GW frequencies for the dominant
modes.  We perform a Fourier analysis on the interval $u\in
[3800M,5700M]$ in Fig.~\ref{fig:q206_spectra}.  The $f_2$ frequency
[the dominant frequency of the $(2,2)$ mode] is clearly visible in
all modes (though it may not be dominant) and is close to $f_2 =
2.09{\rm kHz}$ ($M f_2 = 0.0298$).  $f_2$ agrees within $2.5 \%$
with the relation presented in~\cite{BerDieNag15}, which indicates the
robustness of the fit given there even for high mass ratios. Comparing
with the $q=1$ case, we do not find a significant difference in the
$f_2$ value. Inspection of the peaks of the other modes,
reveal that the $f_k$ frequencies are harmonic to a very high
accuracy, i.e., $f_1 = f_2/2 =f_3/3 =f_4/4$. In particular we have
$f_1=1.05$~kHz ($M f_1 = 0.0150$), $f_3=3.16$~kHz ($M f_3
= 0.0449$), $f_4=4.20$~kHz ($M f_4 = 0.0598$).  The agreement
is better than $1\%$, though the uncertainties in the frequencies
are larger than this, about $0.15$~kHz and are mainly caused by a
shift of the frequencies over time due to the changing compactness
of the neutron star.  This harmonicity was also found
in~\cite{SteBauZag11}, though there they obtained the mode
frequencies from a Fourier transform of the pressure and used the
spatial conformal flatness approximation in their simulation. Here
we have verified that the same harmonicity is present in an
observable quantity (the modes of the gravitational radiation) in a
fully relativistic simulation.

\subsection{A precessing unequal mass binary}
\label{sec:precessing}

\begin{figure*}[t]
  \includegraphics[width=0.95\textwidth]{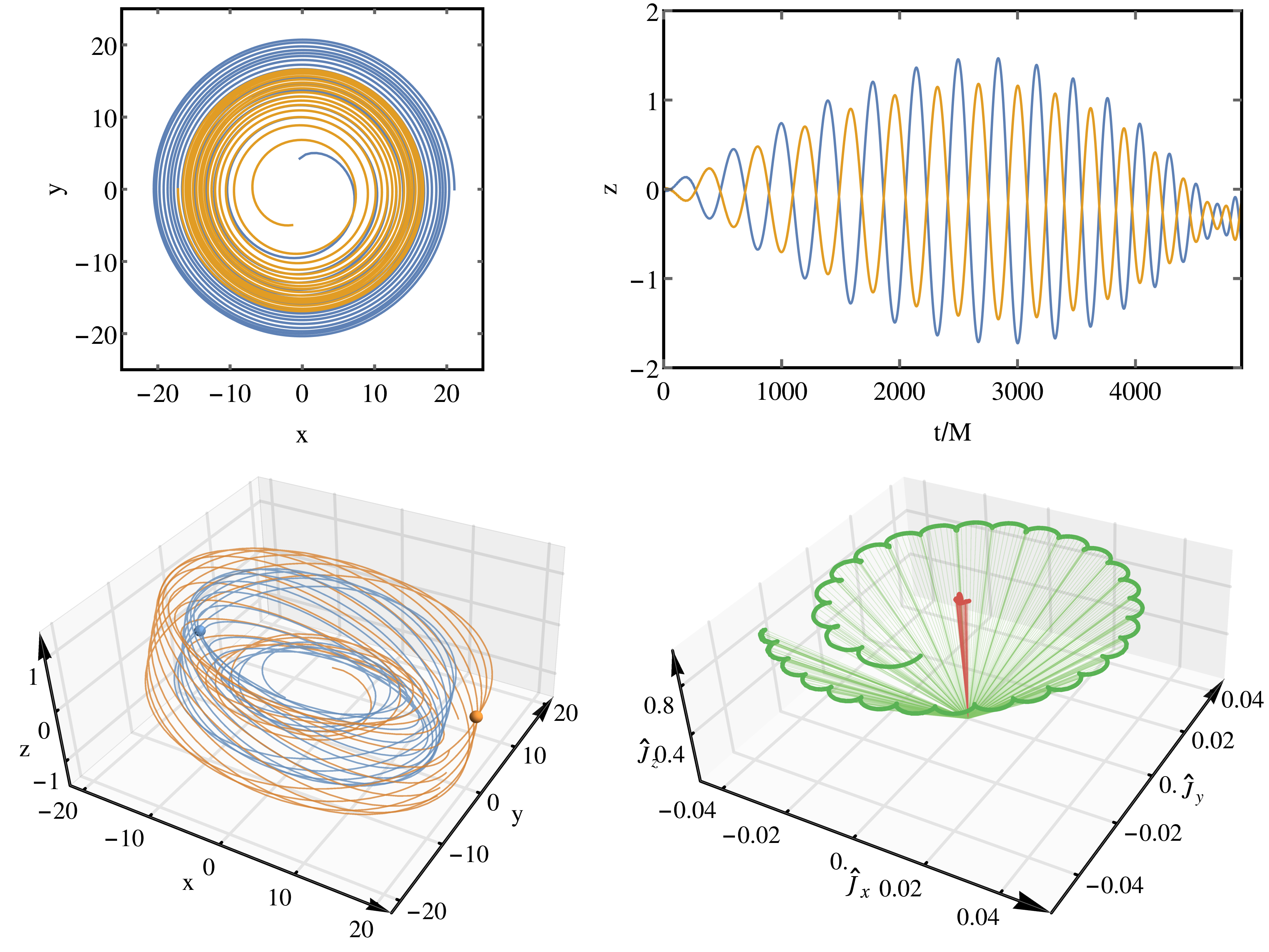}
  \caption{Orbital dynamics of the precessing binary neutron star
    inspiral. Upper left: neutron star tracks in the $xy$-plane.  Upper
    right: $z$-coordinates of the stars' centers as a function of the coordinate time $t$.
    Lower left: 3D neutron star tracks visualizing the precession of
    the orbital plane, where the spheres denote the original positions of 
    the stars. Lower right: Precession and nutation of the
    orbital angular momentum $\hat{\vv{L}}$ (green lines).
    The coordinate system is rotated such that $\hat{\textbf{J}}(t=0)$ lies
    along the z-axis. The orbital
    angular momentum performs slightly more than one precession cycle with a
    period of $T_\text{precess} \simeq 4720M$.  The opening angle
    $\lambda_L$ decreases from $\sim 0.05 \simeq 3^\circ$ to $\sim
    0.035\simeq 2 ^\circ$ over the inspiral.  The direction of the
    total angular momentum also precesses at the same period with a
    considerably smaller opening angle of $\lambda_J \simeq 1.5\times
    10^{-3} \simeq 0.086^\circ$ (red lines).
    }
  \label{fig:precessionCone}
\end{figure*} 
  
In this section we investigate the evolution of a BNS system
with a rather generic spin configuration and $q=1.22$. The CRV
angular velocity $\omega^i$ for star A and star B is set as
$\vv{\omega}=0.005 (1,1,1)$, i.e., their spins both point at an angle of
$45^\circ$ to the orbital angular momentum. 
This evolution is the first precessing BNS merger simulated in
numerical relativity. We indicate this precessing spin
configuration with SLy$^{(\nearrow \nearrow)}$ as we use the SLy 
EOS (Tab.~\ref{tab:listEOS}).  The dimensionless
spin magnitudes are $j^A \simeq 0.13$ and $j^B
\simeq 0.16$.  Although such high spins are not observed in double
neutron star systems so far, there is no physical reason to exclude
such a scenario. In particular, binary neutron star systems with
spins misaligned from the orbital angular momentum, as here, are
more likely to form dynamically in dense stellar regions, such as
globular clusters, where there are many rapidly spinning neutron
stars, as is discussed in Appendix~\ref{sec:BNS_spins}. The stars have
baryonic masses of $M_b^A= 1.5$ and $M_b^B=1.2$;
the gravitational masses are $M^A=1.3553$ and $M^B=1.1072$.

Together with SLy$^{(\nearrow \nearrow)}$ we evolve for comparison
two other configurations with the same baryonic masses (and $q$) but with the CRV angular
velocity set to $\vv{\omega}=0.005 (0,0,1)$ (SLy$^{(\uparrow \uparrow)}$), 
and $\vv{\omega}=0.0$ (SLy$^{(00)}$). Thus SLy$^{(\uparrow \uparrow)}$ has 
no precession, and SLy$^{(00)}$ no spin interactions at all.

All the \SGRID data are computed with $n_A=n_B=28$, $n_\varphi=8$,
$n_{\rm Cart}=24$.  We have not tried to reduce the eccentricity
for this particular setup to save computational resources.  The
residual eccentricity of the precessing system is $\hat e_{d}\simeq
4\times10^{-3}$, $\hat e_{\rm GW}\simeq 5\times10^{-3} $ for the
two eccentricity measures considered in
Sec.~\ref{sec:eccentricity_reduction}. The resolution in the
finest level covering the neutron star is $h_6=0.245$, 
similar to the low resolution setups of ~\cite{BerNagBal14,BerNagDie15,BerNagBal15}, but no
symmetries are applied to the grid (full 3D). 
Thus, although the principal dynamics are properly modeled at this
resolution, quantitative statements come with quite large uncertainties.

\subsubsection{Dynamics}

\begin{figure}[t]
  \includegraphics[width=0.49\textwidth]{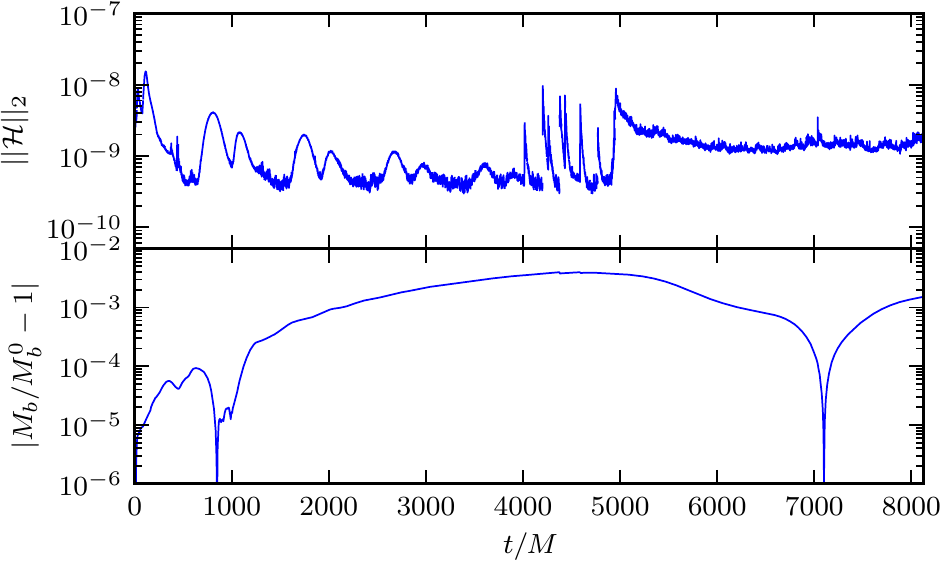}
  \caption{$L^2$ norm of the Hamiltonian constraint (upper panel) and
    baryonic mass 
    conservation (lower panel). Quantities are computed in level
    $l=1$, i.e., the outermost Cartesian box of the numerical domain.}
  \label{fig:precessiondyn}
\end{figure}   
  
Figure~\ref{fig:precessionCone} (upper and lower left) shows the
coordinate tracks (positions of the local minimum of the lapse)
of the neutron stars in the SLy$^{(\nearrow\nearrow)}$ simulation. The change of the
orbital plane due to the misaligned spin of the binary neutron
stars is clearly visible. This effect can also be seen in the change of the
$z$-coordinate over time (upper right panel). During the inspiral
approximately one precession cycle is finished, i.e., the orbital
plane again coincides (approximately) with the $xy$-plane during merger.

We present the evolution of the $L^2$ norm of the Hamiltonian constraint in the upper panel of
Fig.~\ref{fig:precessiondyn} and of the rest mass 
conservation in the lower panel. Because of the constraint propagation
and damping properties of the Z4c scheme, the constraint violations stay
at or below the value of the initial data. The rest mass
conservation over the entire simulation is up to $\sim0.3\%$. Overall,
these diagnostics indicate the errors in the simulation are under control. The
small violation of rest-mass conservation is related to the artificial atmosphere and the
relatively low resolution we employed; see the discussion in~\cite{DieBerUje15}.

The initial angular momentum of the system is
$\vv{J}_\text{ADM}(t=0)= (0.251,0.239,6.951)$ or normalized
$\hat{\vv{J}}=(0.0361,0.0343,0.9989)$. We find that $\hat{\vv{J}}$
changes slightly over the entire simulation, which is to be
expected, as the total angular momentum precesses slightly in
post-Newtonian calculations, as discussed in,
e.g.,~\cite{ApoCutSus94}. We can estimate the opening angle of the
precession cone (neglecting spin-spin effects) using Eq.~(51)
from~\cite{ApoCutSus94} and evaluating everything at $t = 0$, using
the initial orbital frequency to evaluate $M/r$ to $1$PN order.  We
calculate the orbital angular momentum by subtracting the spin
angular momenta of the two stars in isolation from the ADM angular
momentum. Using this method, we obtain (as a first approximation)
an opening angle of the precession cone of $\lambda_J \simeq
1.3\times10^{-3}$. Due to the decreasing orbital separation over
the evolution, we expect that the opening angle should be slightly
larger in our full GR simulation. In fact, this can be observed and
we find an angle of $\sim 1.5 \times 10^{-3}\simeq 0.086^\circ$,
where the numerical uncertainty is $\lesssim10^{-4}$ based on
comparison with simulations without precession. 
The opening angle of the precession cone for the
orbital angular momentum is $\lambda_L \sim 0.05$, from a simple
calculation of the initial angle between the orbital and total
angular momenta, which agrees with the initial opening angle found
in our simulation. 
The opening angle
$\lambda_L$ decreases from $\sim 0.05 \simeq 3^\circ$ to
$\sim0.035\simeq 2 ^\circ$ over the inspiral. 

The lower right panel of Fig.~\ref{fig:precessionCone} presents the
precession of the orbital plane, as given by the direction of the
orbital angular momentum. The effects of precession and nutation
are clearly visible. The normalized orbital angular momentum
$\hat{\vv{L}}$ we plot in this figure is constructed as the
vector orthogonal to the orbital plane, which we estimate using the
coordinate line between the two star centers at two adjacent
timesteps. To minimize high frequency noise we apply a low-pass
filter. The precession period is $T_{\rm precess} \simeq 4720 M$,
which agrees within $\lesssim 10 \%$ with the PN estimates, based on
a PN evolution similar to the discussion
in~\cite{OssBoyKid15}.\footnote{The formulation
  of~\cite{OssBoyKid15} uses non-spinning terms at the 3.5PN level,
  SO terms up to 4PN, spin-spin terms up to 2PN, and for the
  precession SO contributions up to the
  next-to-next-to-leading order are included.}

\begin{figure}[t]
\includegraphics[width=0.44\textwidth]{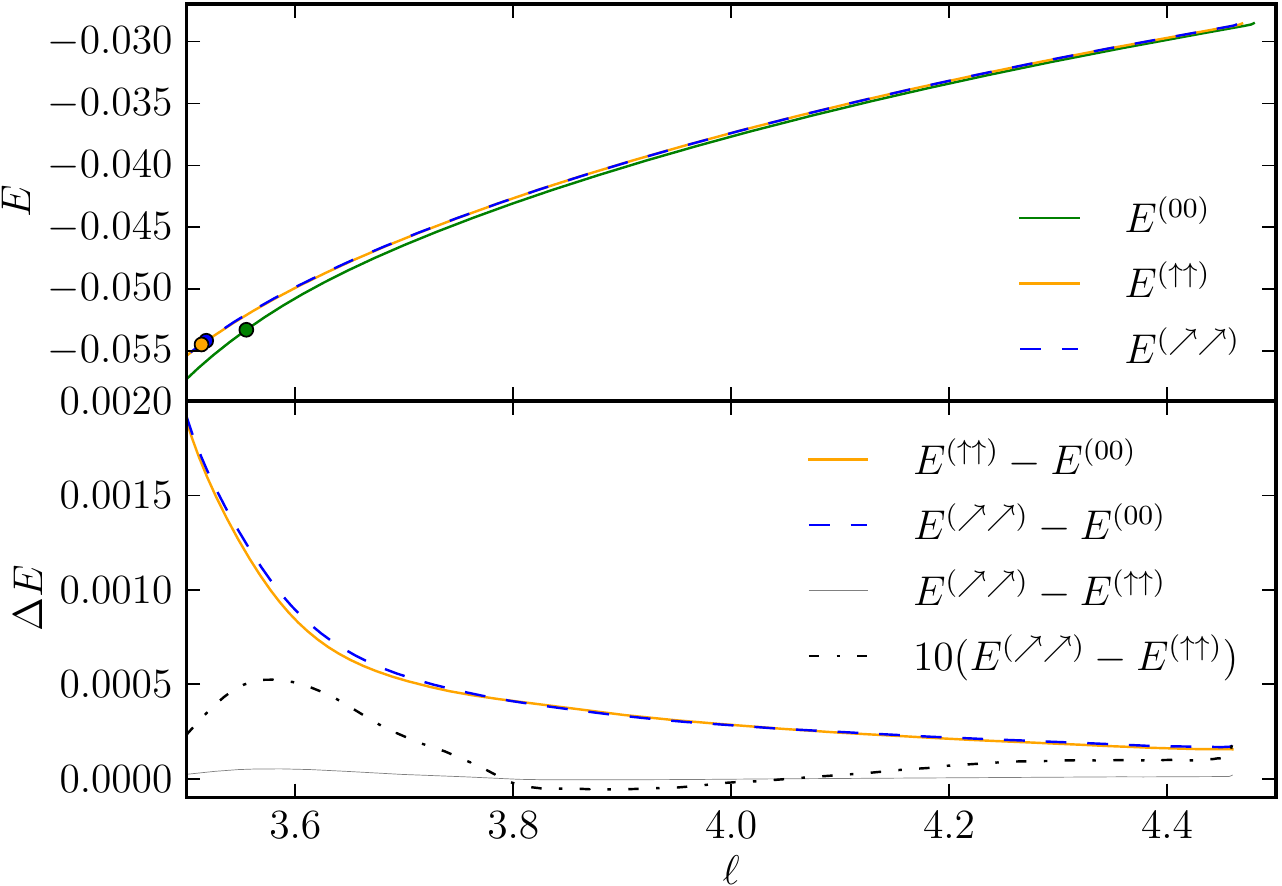}  
\caption{Binding energy as a function of the reduced orbital angular
  momentum for the precessing simulation SLy$^{(\nearrow \nearrow)}$ 
  (blue dashed), SLy$^{(00)}$ (solid orange), and SLy$^{(\uparrow
  \uparrow)}$ (solid green). 
  The bottom panel reports differences between the curves.}
\label{fig:precessionEb}
\end{figure}

Binding energy vs.~reduced orbital angular momentum
curves provide a gauge-invariant way of characterizing the binaries'
dynamics~\cite{BerNagThi12,BerDieTic13,BerNagDie15}. In order to compute
such curves, we 
modify Eqs.~\eqref{eq:Eb} and~\eqref{eq:ell} to take the emitted
energy and total angular momentum of the GWs into account; see,
e.g., Eqs.~(12) and (13) in~\cite{BerDieTic13}. Energy curves for
SLy$^{(\nearrow \nearrow)}$, SLy$^{(\uparrow \uparrow)}$, and
SLy$^{(00)}$ are shown in Fig.~\ref{fig:precessionEb}, together with
their pairwise differences (bottom panel).

The differences between the spinning configurations and SLy$^{(00)}$
essentially quantify the repulsive spin-orbit (SO) interaction contribution
to the binding energy, which is the dominant one for the dynamics. The
energetics of SLy$^{(\nearrow\nearrow)}$ are quite close to those of
SLy$^{(\uparrow\uparrow)}$. This happens because the leading order SO terms
$(\propto\vv{L}\cdot \vv{S}_i/r^3)$ are identical in the initial data for
SLy$^{(\nearrow \nearrow)}$ and SLy$^{(\uparrow \uparrow)}$.\footnote{The
following discussion is based on leading-order PN expressions for the 
binding energy that can be found in, e.g., Eq.~(2.7) of~\cite{Kid95}.}
During the evolution, the difference between the SO interactions of
SLy$^{(\uparrow \uparrow)}$ and SLy$^{(\nearrow \nearrow)}$ is solely due to
the slight changes in the projection of the spins onto the angular momentum
as they precess. For distances $d\sim16-8$ this corresponds to corrections
on the order of $10^{-4}$ to the binding energy. The leading order spin-spin
(SS) contribution ($E_{SS}=[3(\vv{n}\cdot\vv{S_1})(\vv{n}\cdot\vv{S}_2)-
(\vv{S}_1\cdot\vv{S}_2)]/r^3$, where $\vv{n}$ denotes the unit vector pointing from
one star to the other) is exactly zero in the SLy$^{(\nearrow \nearrow)}$
initial data and order $10^{-6}$ in the SLy$^{(\uparrow \uparrow)}$ initial
data. During evolution the SS contribution of SLy$^{(\nearrow \nearrow)}$ is
of the order $\sim10^{-5}$. This explains the differences between
SLy$^{(\nearrow \nearrow)}$ and SLy$^{(\uparrow \uparrow)}$ at the level of
$\sim10^{-4}$--$10^{-5}$ due to these SO and SS corrections, which can be observed
in Fig.~\ref{fig:precessionEb}. However, significant differences between
SLy$^{(\nearrow \nearrow)}$ and SLy$^{(\uparrow \uparrow)}$ can be observed
in some of the higher modes, notably the $(2,1)$ mode, as we shall see in
the following.

\subsubsection{Gravitational Waves}

\begin{figure*}[t]
  \includegraphics[width=1\textwidth]{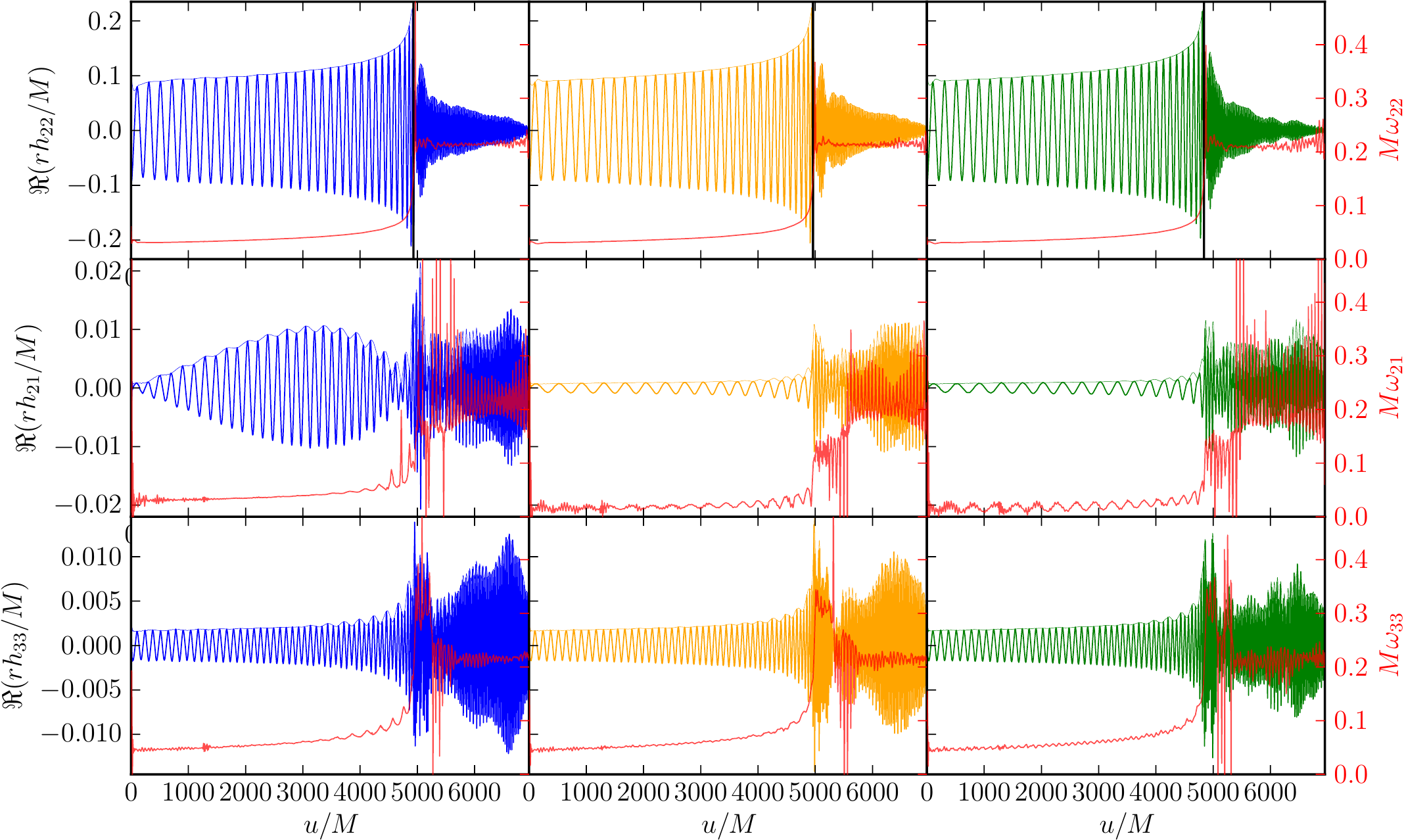}
  \caption{The three most dominant multipoles of the GW $(l,m)=(2,2)$,
    $(2,1)$, $(3,3)$ for SLy$^{(\nearrow \nearrow)}$ (left panels),
    SLy$^{(\uparrow \uparrow)}$ (middle panels), and
    SLy$^{(00)}$(right panels).  We plot the real part of the modes
    and the dimensionless GW frequency in red.  In the upper plots,
    the solid vertical lines mark the moment of merger (i.e., the
    maximum of $|h_{22}|$), where a later merger in the two cases with
    spin is seen, due to the spin-orbit interaction.  We observe that
    the $(2,1)$ mode in the precessing case is dominated by the large
    contribution from the mass quadrupole at twice the orbital
    frequency due to the precession of the orbital plane, while it
    only has the much smaller contribution from the current quadrupole
    at the orbital frequency in the two nonprecessing cases.  We
    extracted the wave at a radius of $r=446M$. Some very small
    noise in the frequency at early times is visible and due to
    reflections at the boundary. After merger, the frequency
    calculation is less accurate and affected by larger oscillations,
    mostly unphysical.}
    \label{fig:precessionGW}
\end{figure*}  

We present the three largest amplitude modes
of the GW signal in Fig.~\ref{fig:precessionGW}. As in the
nonprecessing case, the dominant
emitter of GWs is the $(2,2)$-mode.  However, interesting physical
aspects are present in the subdominant modes.  The amplitude of the
$(2,1)$-mode is modulated by the precession period, giving a second
possibility to extract the value of that period, which agrees with
the estimate given above. 
Additionally, the amplitude of the
modulation we observe is consistent with the expected contribution
to the $(2,1)$-mode from the binary's dominant mass quadrupole
radiation [which only appears in the $(2,2)$-mode for nonprecessing
systems] due to the precession of the orbital plane. In particular,
since the initial angle between the total and orbital angular
momenta is small in this case, $\lambda_L :=
\angle(\hat{\vv{L}},\hat{\vv{J}}) \simeq 0.05$, decreasing slightly
over the evolution, as discussed in the previous section (see also 
Fig.~\ref{fig:precessionCone}), one can work to linear order in
$\lambda_L$, where one finds that the maximum amplitude of the mass
quadrupole's contribution to the $(2,1)$-mode is $2\lambda_L$ times
the amplitude of the Newtonian mass quadrupole
radiation.\footnote{%
  Similarly, the amplitude of the $(2,2)$-mode is the
  same as the amplitude of the Newtonian mass quadrupole radiation up
  to corrections that are suppressed by factors of $v\lambda_L\delta$,
  $v^2$, or $\lambda_L^2$, where $v$ is the binary's orbital velocity
  and $\delta := (M^A-M^B)/M \simeq 0.1$.}
One obtains the expression for
the maximum mass quadrupole contribution to the $(2,1)$ mode by
noting that $2\lambda_L$ is the largest angle the orbital angular
momentum (originally aligned with the $z$-axis to a very good
approximation) makes with the $z$-axis; cf.\ the PN expressions for
the modes expanded in $\iota$ in Eqs.~(4.17)
of~\cite{AruBuoFay09}. (Note that $\iota$ is defined as the angle
between the orbital angular momentum and the initial total angular
momentum, which they take to be along the $z$-axis in the coordinate
system they use to define the mode decomposition.)
Of course, there are contributions to the $(2,1)$ mode from the
current quadrupole, as well, but these are much smaller than the
contribution due to precession, as is seen in
Fig.~\ref{fig:precessionGW}, since the current quadrupole
contributions are suppressed by a factor of $v\delta$ compared to
the Newtonian mass quadrupole radiation.  

The systems finally merges after $\sim 30$ GW cycles [in the $(2,2)$
mode] at a frequency of $M\omega_{22}^{\rm mrg}\sim0.128$ (solid vertical
line in upper left panel; for consistency with our previous work, we
define the merger as the maximum of $|h_{22}|$) at $t=4927M$.

Comparing SLy$^{(\nearrow \nearrow)}$ with
the other two simulations, we observe two main differences during
the inspiral. First, without spin the merger happens earlier at
$t=4839M$, due to the fact that no repulsive spin-orbit interaction
is present. In case of the SLy$^{(\uparrow \uparrow)}$ simulation,
the merger happens at $t=4960M$, which agrees within $\Delta t =
33M$ with the precessing simulation. The second observation is that
the amplitude of the $rh_{21}$ is much smaller than for the
precessing simulation. The non-zero amplitude is caused by the
unequal masses of the two stars, but no clear imprint of the spin is
visible (cf.\ the middle and right panels). Due to the small
amplitude of the $(2,1)$ mode in the nonprecessing simulations it is not
very well resolved, which is clearly visible in unphysical
sinusoidal oscillations of the frequency. The same holds to a lesser
extent
for the $(3,3)$-mode. Additionally, the large frequency spikes present
in the post-merger phase of the subdominant modes are caused by zeros of
the amplitude.

\begin{figure}[t]
  \includegraphics[width=0.5\textwidth]{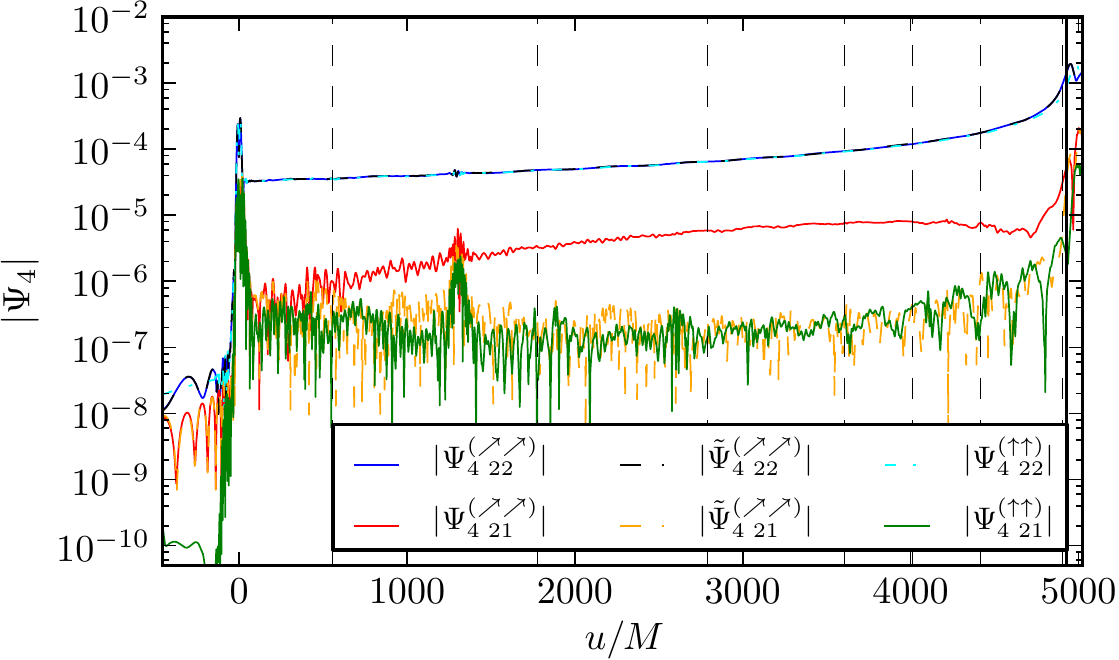}
  \caption{Amplitude of the $(2,2)$ and $(2,1)$ modes of $\Psi_4$ for the
           SLy$^{(\nearrow \nearrow)}$ and Sly$^{(\uparrow \uparrow)}$
           simulations, along with the same amplitudes for
           SLy$^{(\nearrow \nearrow)}$ in the nonprecessing frame
           (denoted by $\tilde{\Psi}_{4}$) obtained
           as in~\cite{SchHanHus10}. Note that we only consider the inspiral
           here. The transformation is done in chunks (separated by vertical
           dashed lines). The agreement between $|\tilde{\Psi}_{4 \
           21}^{(\nearrow \nearrow)}|$ and $|\Psi^{(\uparrow \uparrow)}_{4 \
           21}|$ after transformation to the non-precessing frame is clearly
           visible. The noise around $u\approx 1500M$ is caused by
           reflections of the outer boundary.}
    \label{fig:prec_copre}
\end{figure}

In order to further assess the differences between the 
SLy$^{(\nearrow \nearrow)}$ and SLy$^{(\uparrow \uparrow)}$ waveforms,
we compare the nonprecessing data to the precessing ones after 
transformation to the precessing frame. The transformation is performed
with a method similar to the one used in binary black hole
simulations in~\cite{SchHanHus10}, and the main result is shown in
Fig.~\ref{fig:prec_copre}. In particular, the rotation 
of the $\Psi_{4 \ lm}$ multipoles reads (Eq.~(A9) in~\cite{SchHanHus10})
\be
\tilde{\Psi}_{4 \ lm} = \sum_{m'=-l}^{l} e^{im' \gamma}
d^l_{m'm}(-\beta) e^{im\alpha} \Psi_{4 \ lm'} \ ,
\ee
where $d_{m' m}^{l}$ are the Wigner $d$-matrices.
We focus on the $(2,1)$ and $(2,2)$ modes and perform the rotation as follows:
(i) we look for the Euler angles $\beta,\gamma$ for which $|\tilde{\Psi}_{4
  \ 21}|$ is minimal (Ref.~\cite{SchHanHus10} maximizes $|\tilde{\Psi}_{4
  \ 22}|^2+|\tilde{\Psi}_{4 \ 2-2}|^2$);  
(ii) we split the dataset in several chunks (vertical dashed lines
in Fig.~\ref{fig:prec_copre}) and fit the obtained Euler angles with
low-order polynomials. This step optimizes
the fits and minimizes numerical oscillations. We do not need to worry about the
third Euler angle $\alpha$ here, as it is irrelevant in the case we consider, where we
only look at the magnitude of the modes of $\tilde{\Psi}_4$.
From Fig.~\ref{fig:prec_copre} one sees that $|\tilde{\Psi}_{4 \ 2m}^{(\nearrow \nearrow)}|$
(i.e., the version in the nonprecessing frame) is almost equivalent to
$|\Psi_{4 \ 2m}^{(\uparrow \uparrow)}|$.
The effect is very clear in the $(2,1)$ mode. This preliminary result suggests it
will be possible to model precessing BNS waveform using aligned spin
BNS models for moderate spin magnitudes (cf.~\cite{SchHanHus10}).

The merger remnant is a hypermassive neutron star (HMNS), which mostly
emits in the $(2,2)$ channel at early times. After $t\approx 5900M$
the amplitude of the $(2,2)$ mode decreases until the $(2,1)$ and
$(3,3)$ modes have the same 
amplitude as the $(2,2)$ mode. Note that at this time the $(2,0)$
and $(4,4)$ modes also have comparable amplitudes. 
We find a frequency shift of the $f_2$-frequency [the dominant
frequency in the $(2,2)$ mode] of $\sim 60$~Hz due to the
additional angular momentum of the HMNS formed by the spinning
configurations; the origin of such frequency shifts was discussed in detail
in~\cite{BerDieTic13}. The estimated $f_2$-frequencies are $2.75$,
$2.79$, and $2.81\text{ kHz}$, for SLy$^{(00)}$, SLy$^{(\uparrow
  \uparrow)}$, and SLy$^{(\nearrow \nearrow)}$, respectively.

\subsection{Effect of eccentricity reduction on waveform phasing}
\label{sec:dyn_eccred}

Although an eccentricity reduction procedure for BNS quasicircular
initial data has been already presented in~\cite{KyuShiTan14}, its
performance on the relevant observable quantity, i.e., the GW
phase and amplitude, has not been evaluated directly.  Here, we
investigate the effect of eccentricity reduced data on the GW
phasing and amplitude by a direct comparison with data in which
eccentricity reduction has not been performed. Such a comparison
is particularly important since the eccentricity reduction
procedure is computationally expensive and might be not necessary
for certain applications or when the data are affected by larger
uncertainties due, for example, to truncation errors.

We compare the GW phase of two SLy $q=1$ runs: In one case we
evolve initial data with $\hat{e}_d=1.241\times 10^{-2}$ (Iter~0)
and in the other data with $\hat{e}_d=8.7\times10^{-4}$ (Iter~3);
see Table~\ref{tab:listEccRedPars}.  

We focus on the $\ell=m=2$ multipole and omit the subscript in the
waveform quantities. Waveforms are aligned on the interval
$[t_1,t_2]=[1000,6000]\simeq[370M,2222M]$ 
shifting by constant time and phase
offsets $T,\Phi$. The latter are determined by minimizing the
function~\cite{BerThiBru11} 
\be G(T,\Phi) = \int_{t_1}^{t_2} 
|\phi_1(t) - \phi_2(t+ T) - \Phi|^2dt , 
\ee 
where $\phi_{1,2}$ denotes the GW phase of the two datasets. 
A more robust alignment procedure based on a frequency interval can also be
used~\cite{BerNagDie15}, but the current procedure is sufficient for our
purposes.

\begin{figure}[t]
  \includegraphics[width=0.48\textwidth]{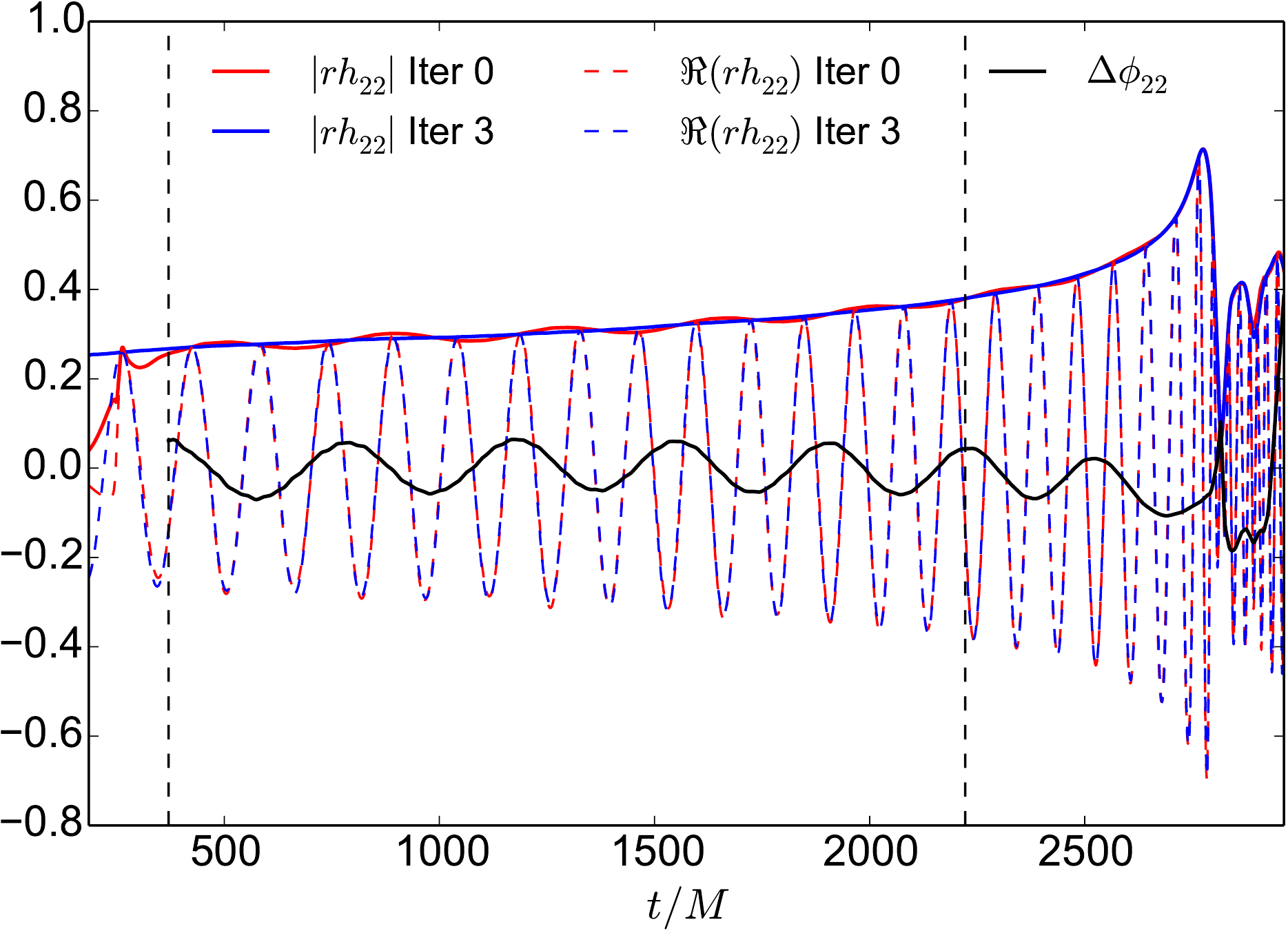}
  \caption{GW phasing of the evolved initial data without (Iter~0)
    and with (Iter~3) eccentricity reduction for the SLy EOS $q=1$
    configuration. Vertical dotted lines denote the region over which
    we align the waveforms.}
      \label{fig:EccRed_phasing}
\end{figure}

We present the results in Fig.~\ref{fig:EccRed_phasing}. The phase
difference $\Delta\phi_{22}$
oscillates between $[-0.06,0.06]$~rad during the
$\sim21$ GW cycles, and it is essentially flat up to merger,
$t\simeq2780M$.  Similarly, the amplitude of the non-eccentricity
reduced data (Iter~0) oscillates around the eccentricity reduced
ones (Iter~3); the amplitude oscillations are $\sim5\%$ at early
times $t\sim370M$ and decrease as the system approaches
merger.

Overall, these results show that the use of eccentricity reduced
data with $\hat{e}_d\sim10^{-4}$ (about 3 iterations of our
procedure) improves the waveform quality for GW modeling
purposes, and should be employed in future precision studies of
the gravitational waveform.  However, eccentricity reduction is
likely to be effective only if combined together with an
improvement of other source of errors, notably truncation
errors. We notice in this respect that $\Delta\phi\sim0.12$~rad is
at least a factor two smaller than the typical uncertainty
introduced by truncation errors at the resolutions employed
here~\cite{BerNagDie15}.
  
\section{Conclusions}
\label{sec:Conclusion}

Due to advances in the construction of constraint solved and
consistent initial data, simulations of binary neutron stars in full
general relativity are now able to cover more of the binary neutron
star parameter space accurately. In particular, it is now possible to
study different EOS~\cite{ShiTanUry05,SekKiuKyu11b,SekKiuKyu11}, 
large mass ratios~\cite{TanShi10,DieBerUje15}, spinning neutron
star configurations~\cite{BerDieTic13}, highly eccentric
setups~\cite{MolMarJoh14}, and neutron stars on orbits
with less residual eccentricity than the
standard ones~\cite{KyuShiTan14,HotKyuOka15}.  
We have recently upgraded the \SGRID code to be
able to generate consistent, constraint-solved initial data with the ability to
vary all these parameters.\footnote{Note that here we only
  consider the ``inspiral parameter space,'' consisting of the purely
  relativistic hydrodynamic parameters (eccentricity, masses, spins,
  and EOS) that can be measured with an inspiral gravitational wave
  signal with current or proposed gravitational wave detectors in
  physically expected scenarios.} The most noteworthy improvement was
the combination of the constant rotational velocity approach
(from~\cite{Tic11}) with a generic specification of the symmetry
vector (from~\cite{MolMarJoh14}) allowing arbitrary eccentricities
(including eccentricity reduction).

In this paper, we have exhibited \SGRID's ability to generate binary
neutron star initial data for many situations of interest and included 
dynamical simulations of some generic configurations evolved with the
BAM code.\\

\subsection{Quasi-equilibrium sequences}

We have constructed the following quasi-equilibrium setups:

{\it Binaries in the constant rotational velocity approach.} 
With the help of quasi-equilibrium sequences we studied spinning
neutron star configurations for different equations of state and
characterized the spin-orbit contribution to the binding energy. Our
results show that the spin-orbit interaction can be well approximated
by post-Newtonian theory within the uncertainty of our numerical
method.

{\it Highly eccentric binaries.}
We also constructed highly eccentric sequences following the
description of~\cite{MolMarJoh14}, but with the advantage of solving
the elliptic equation for the velocity potential, which results in
smaller artificial density oscillations by a factor of $\sim5$.  This
will allow a more detailed analysis of tidally-induced oscillations
in the neutron stars than was performed in~\cite{GolBerThi11} using
superposed initial data, which led to much larger initial
oscillations.  Additionally, we compared our eccentric sequences with
PN calculations in various ways, where our analysis showed that, as
expected, close agreement is only obtained in the limit of large
separations or small eccentricities.

{\it Eccentricity reduced binaries.}
We can use the same technology that allows us to create highly
eccentric orbits to reduce the eccentricity present in standard binary
neutron star initial data constructed using a helical Killing
vector. Here we iteratively adjust our eccentricity and radial
velocity parameters, similar to the procedure presented
in~\cite{KyuShiTan14}.

{\it Varying compactnesses and mass-ratios.}
As additional parameters, we considered different mass ratios and
compactnesses.  We were able to construct a sequence for a mass ratio
of $q=2.06$ and equal-mass binary neutron stars with compactness up to
$\mathcal{C}=0.23$. \\

This shows that with its recent upgrades, \SGRID allows one to
construct binary neutron star configurations in a substantial
portion of the possible inspiral parameter space (and one can vary
all the relevant parameters independently).  Of course, there is a
considerable amount of physics that we do not include here. However,
the physics we have neglected does not affect the inspiral at a level
that can be detected via gravitational waves for the parameter values expected
to be present in neutron star binaries. Nevertheless, this
missing physics can still play an important role in the merger or
potentially produce other interesting effects, and includes magnetic
fields, elasticity (e.g., in the solid crust), and composition, all of
which would need to be appropriately incorporated in the initial data.
Of this additional physics, the inclusion of magnetic fields is likely
the most pressing. Unfortunately, there is no known method for including
magnetic fields consistently in constraint-solved initial data: All simulations
of magnetized binary neutron stars
(e.g.,~\cite{AndHirLeh08a,LiuShaEti08,GiaRezBai09,KiuKyuSek14,DioAliRez15})
add the magnetic field by hand after constraint solving. Developing
such a method would be a useful advance in binary neutron star initial
data construction.\\

\subsection{Dynamical Evolutions}

To ensure that the constructed data are suitable for dynamical
simulations, we have evolved three configurations. 

{\it $q=2.06$ run.} 
As a first example, we evolved the highest mass ratio ever considered 
in a full general relativistic binary neutron star configuration. 
The configuration consisted of a $q=2.06$ setup with the MS1b EOS. 
Because of the high mass ratio and the
rather stiff EOS, we observed mass transfer between the two stars
several revolutions before merger. During this process material with a
rest-mass of $\sim(2-3)\times10^{-2}M_\odot$ is transferred, with an average
accretion power of $\sim10^{53}\text{ erg
  s}^{-1}$. During the merger process, $\sim
7.6\times10^{-2}M_\odot$ get unbound and are released from the system
with a kinetic energy of $\sim 4 \times 10^{50} \text{ erg}$. The
ejecta process happens primarily due to torque in the tidal tail of
the lower massive star and forms a spiral like pattern.  Due to this
anisotropic ejection of material, the merger remnant receives a large
kick of $O(100)\text{ km s}^{-1}$. The final merger remnant can be
characterized as a supramassive neutron star, which is not expected to
collapse on dynamical timescales. An investigation of the merger remnant's GW spectrum,
including more than just the dominant $(2,2)$-mode, reveals that many
of the peak frequencies are harmonically related to high accuracy.

{\it Precessing and unequal masses run.} 
As a second example, we considered the first precessing binary neutron
star merger simulation. Contrary to most BNS investigations (except,
e.g.,~\cite{BerNagThi12,ReiHaaOtt12}) we present more than just the
dominant $(2,2)$ mode and find a clear imprint of the precession in
the subdominant $(2,1)$ mode, where the amplitude is modulated by the
precession frequency. Considering for comparison a simulation with
the same leading-order spin-orbit interaction, we show
that the relation between the gauge-invariant binding energy vs. reduced
orbital angular momentum up to the merger exhibits only a minor imprint of the
precession for the spin magnitudes we consider. Regarding the post-merger GW spectrum, we observed a
frequency shift of the dominant $f_2$ mode due to the spins of the
binary's components (cf.~\cite{BerDieTic13}).

{\it Reduced eccentricity run.} 
Finally, we have performed simulations of an equal-mass configuration
with and without eccentricity reduced initial data and we have
quantified the differences in the waveform's amplitude and phase. We
found that although the eccentricity reduction improves the waveform
quality, one also needs to reduce other errors in the waveforms,
notably truncation errors, in order for the improvement due to
eccentricity reduction to be effective.

\begin{acknowledgments}
  It is a pleasure to thank Roland Haas, Michael Kramer, Alessandro Nagar, 
  Jan Steinhoff, Maximiliano Ujevic for helpful discussions 
  and valuable comments. We are particularly indebted to Patricia Schmidt
  for her help understanding the precession effects on the waveform. 
  This work was supported in part by DFG grant
  SFB/Transregio~7 ``Gravitational Wave Astronomy,'' the
  Graduierten-Akademie Jena, and the DFG Research Training Group
  1523/1 ``Quantum and Gravitational Fields.'' N.K.J.-M. acknowledges
  support from the AIRBUS Group Corporate Foundation through a chair
  in ``Mathematics of Complex Systems'' at the International Centre
  for Theoretical Sciences.  S.B. acknowledges partial support from
  the National Science Foundation under grant numbers NSF AST-1333520,
  PHY-1404569, and AST-1205732.
  C.M. was supported  by   the
STFC grant PP
/
E001025
/
1. W.T. was supported by the National Science Foundation under
  grant PHY-1305387.
  The authors also gratefully acknowledge the Gauss Centre for Supercomputing e.V.~for funding
  this project by providing computing time on the GCS Supercomputer
  SuperMUC at Leibniz Supercomputing Centre and the computing time
  granted by the John von Neumann Institute for Computing provided on
  the supercomputer JUROPA at J\"ulich Supercomputing Centre.  We also
  acknowledge usage of computer time on the Fermi CINECA machine
  allocated through the ISCRA initiative.  Additionally, this work
  used the Extreme Science and Engineering Discovery Environment,
  which is supported by National Science Foundation grant number
  ACI-1053575, computer resources at the Institute of Theoretical
  Physics of the University of Jena, and the HPC cluster KOKO at
  Florida Atlantic University.
\end{acknowledgments}

\appendix

\section{Astrophysical predictions for more extreme masses, mass ratios, and spins for binary neutron stars} 
\label{sec:BNS_par_space}

Here we assess the prospects for binary neutron star mergers with more
extreme masses and spins---such as those we simulated in Sec.~\ref{sec:Dynamical_Evolutions}---actually occurring in nature. 
(Note that this assessment only covers the masses and spin magnitudes, not the equation of state
or spin misalignment.) 

\subsection{Large and small masses and larger mass ratios} 
\label{sec:Population_Synthesis_M}

There is good evidence that the population of neutron stars in the
universe extends at least from $\sim1M_\odot$ up to $2M_\odot$. In
particular, there are two precise measurements of high-mass ($\sim
2M_\odot$) neutron stars in binaries with white
dwarfs~\cite{DemPenRan10,AntFreWex13} as well as some less precise
measurements of low-mass ($1M_\odot$) neutron stars in X-ray
binaries~\cite{RawOroMcC11,OezPsaNar12}.
There are also recent measurements of low compactnesses for isolated
neutron stars~\cite{HamNeuSul14}, which would imply quite small masses
$\sim1M_\odot$ for many of the equations of state we
consider.\footnote{Note that these authors quote compactnesses in
  units of $M_\odot/\text{km}$, not the dimensionless compactnesses we
  use in this work, as is mentioned explicitly in the caption to
  Fig.~1 of~\cite{NeuHamHoh12}.}  Additionally, there is recent work
that suggests that the initial (pre-accretion) mass of the low-mass
millisecond pulsar J$0751$+$1807$ (whose present mass is
$1.26\pm0.12M_\odot$) could have been as low as
$1.1M_\odot$~\cite{ForBejHae14}.

The theoretical bounds on the neutron star mass allow for an even
larger mass range (and thus, in principle, large mass ratios, up to
$\sim 3$), as some EOSs have maximum masses of $\sim 3M_\odot$ (e.g.,
the $\sim 2.8M_\odot$ maximum masses for the MS1 and MS1b EOSs we
consider in this work; see Table~\ref{tab:listEOS}). However, the
minimum mass of neutron stars in the universe is likely around the
minimum observed mass of $\sim 1M_\odot$. While the minimum mass of a
star constructed from a cold dense matter equation of state is quite
small ($< 0.1M_\odot$), the minimum mass of a hot protoneutron star is
considerably larger, $0.89$--$1.13M_\odot$ for the models considered
in~\cite{StrSchWei99}. This minimum mass provides a practical lower
bound on neutron star masses formed from supernovae, barring formation
of lower-mass stars by fragmentation (see, e.g.,~\cite{PopBlaGri07}),
which is quite speculative. Moreover, the high
kicks that neutron stars formed by fragmentation would be expected to receive
make them unlikely components of binaries. Additionally, there is a further restriction from the
baryonic mass of the iron core of the supernova progenitor, which
gives a minimum mass of $1.15$--$1.2M_\odot$, as discussed in Sec.~3.3
of~\cite{Lat12}, though there are uncertainties in both these bounds
due to uncertainties in supernova physics. (Note that Tauris, Langer,
and Podsiadlowski~\cite{TauLanPod15} estimate that the minimum mass of
a neutron star formed in an ultra-stripped supernova is $1.1M_\odot$.)
See~\cite{Lat12,LatXX} for a general review of neutron star masses.

As mentioned in Sec.~\ref{sec:Introduction}, the mass range of the
observed binary neutron star systems is much smaller, particularly if
one only considers the six systems that will merge within a Hubble
time, where the minimum and maximum masses are $1.25$ and
$1.44M_\odot$ (PSRs J0737$-$3039B and B1913$+$16, the less massive
star in the Double Pulsar and the Hulse-Taylor pulsar, respectively),
and the largest observed mass ratio is $1.07$, for the Double Pulsar
(see, e.g., Table~1 in~\cite{Lat12,LatXX} and Table~3
in~\cite{PosYun14}). Thus, one might na{\"\i}vely not expect to see a
very wide range of masses (and thus mass ratios) in many binary
neutron star coalescences. However, for certain values of poorly
constrained parameters, population synthesis calculations (e.g., the
Synthetic Universe models from Dominik~\emph{et
  al.}~\cite{DomBelFry12}) predict the existence of binary neutron
stars (formed ``\emph{in situ,}'' i.e., not by dynamical capture) with
masses over the entire observed range of neutron star masses, and some systems with
reasonably large mass ratios.

Let us now consider the predictions of the Synthetic Universe
population synthesis data available online~\cite{SynthUniv}; these are
the standard model and Variations $1$--$15$ of Dominik~\emph{et
  al.}~\cite{DomBelFry12}, where each of the variations varies one of
the poorly constrained parameters in the calculation---see Table~1
in~\cite{DomBelFry12}. (See~\cite{DomBerOsh14} for a study of the
predicted gravitational wave detection rates using a few of these
models and~\cite{MinBel15} for a study of the effects of varying
certain initial conditions.) Additionally, each of these models has
four further variants, given by the four combinations of two choices
for the metallicity (solar and $0.1$ solar)\footnote{Note that the
  metallicity of objects in the universe varies, generally increasing
  for more recent formation times, as discussed in,
  e.g.,~\cite{DomBelFry13}. These two choices of metallicities are
  intended to give an indication of the effects of metallicity on
  these calculations.} and two treatments of the common envelope phase
of the binary's evolution (submodels A and B, which correspond to the
optimistic and pessimistic predictions, respectively, for the fate of
binaries which enter the common envelope phase when the donor is in
the Hertzsprung gap). These models all assume a minimum neutron star
mass of $1M_\odot$ (as mentioned in~\cite{DomBerOsh14}) and a maximum
mass of $2.5M_\odot$ (except for Variations 5 and 6, which assume a
maximum mass of $3$ and $2M_\odot$, respectively).

All of these models predict a galactic binary neutron star merger rate
above the estimated lower bound of $2.1\text{ Myr}^{-1}$ inferred from
observations (see the discussion in Sec~4.1 of~\cite{BelDomRep14}) at
solar metallicity, except for Variation~$1$ (both submodels), and
submodel B in Variations~$2$, $4$, and $12$ (see Table~2
in~\cite{DomBelFry12}). We still show results for these models, for
comparison, particularly since that lower bound is not particularly
firm.

\begin{figure}[htp]
  \includegraphics[width=0.5\textwidth]{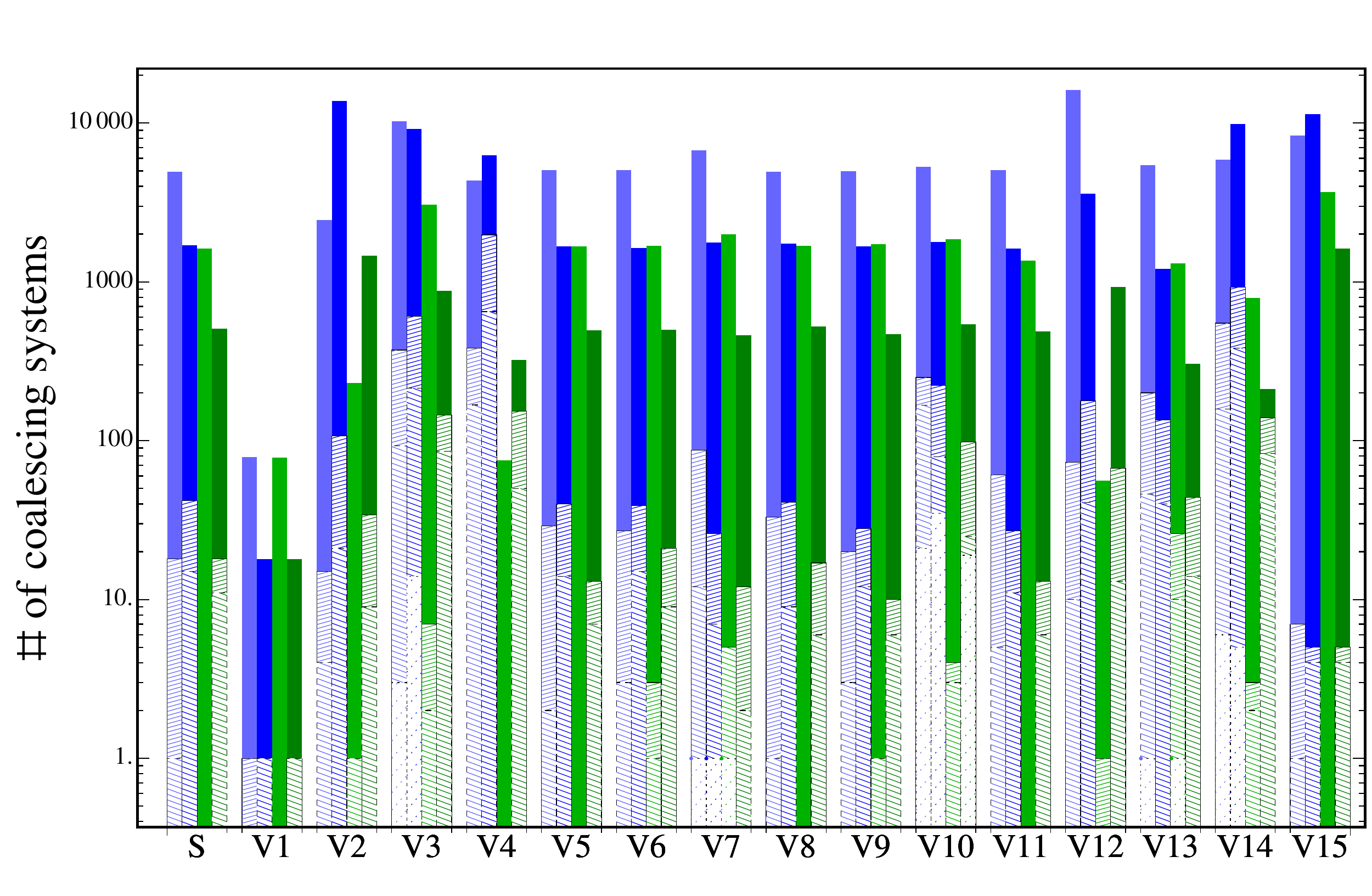}
  \includegraphics[width=0.5\textwidth]{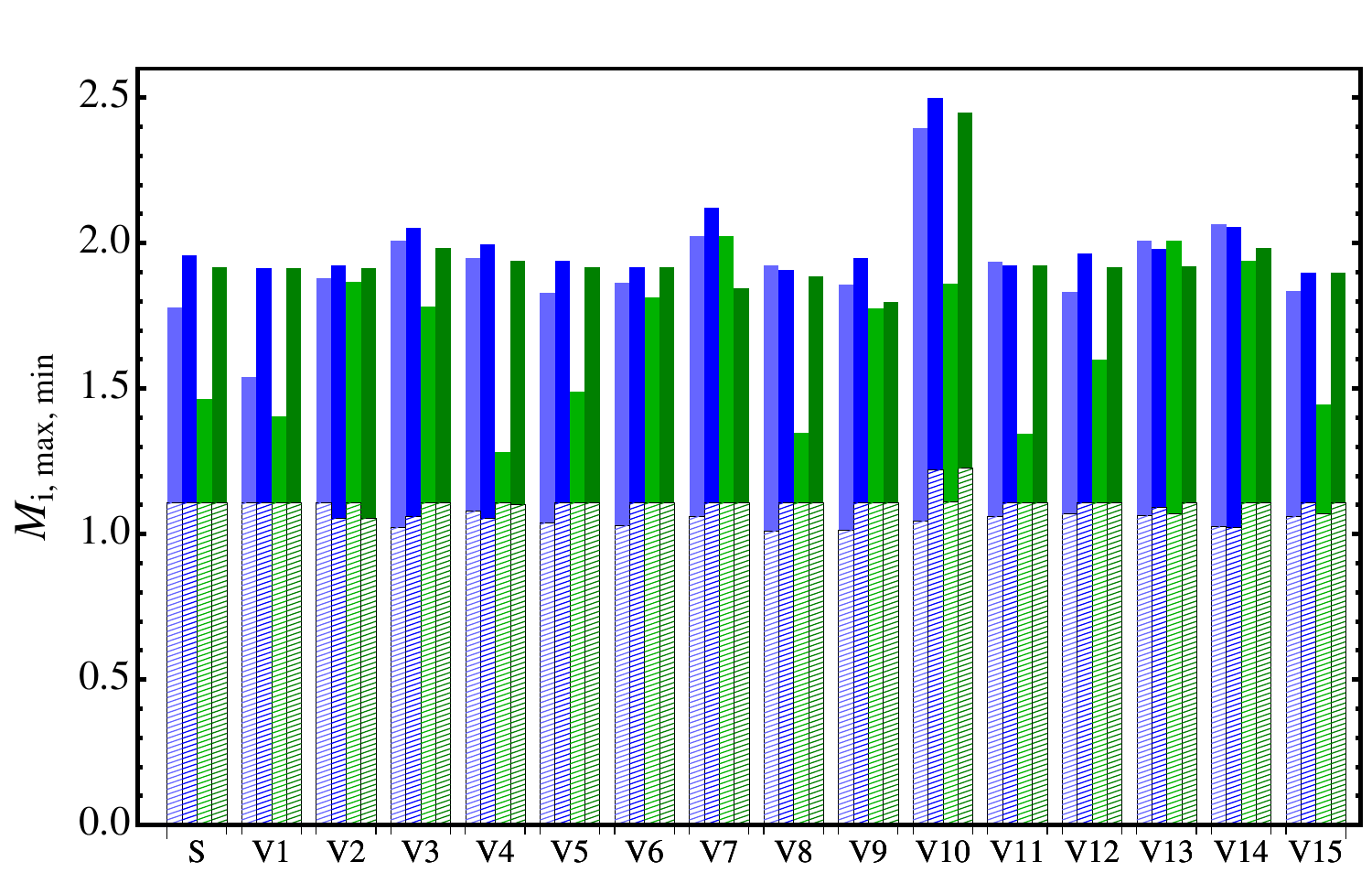}
  \includegraphics[width=0.5\textwidth]{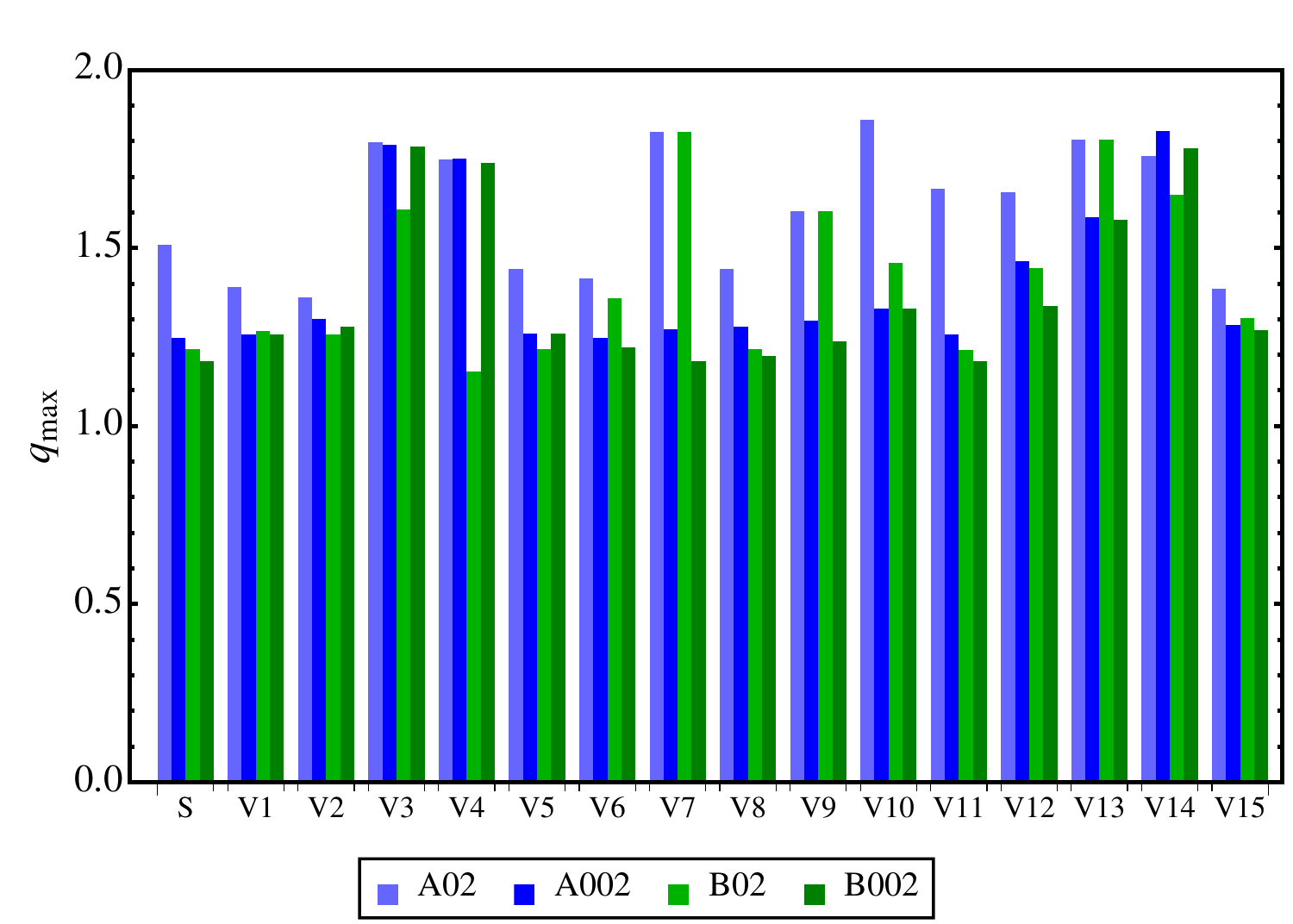}
  \caption{The total number of coalescing binary neutron star systems in the various
    Synthetic Universe population synthesis models, with the numbers
    with individual masses greater than $\geq1.5$, $1.75$, and
    $2M_\odot$ also marked (top), along with the maximum and minimum
    individual masses (middle) and maximum mass ratio (bottom) present
    in these models. Here ``S'' denotes
    the standard model and ``V$n$'' denotes the $n$th variation.  The
    four variants are denoted by different colored bars, where the
    letter (A or B) gives the submodel and ``02'' and ``002''
    correspond to solar metallicity and $0.1$ solar metallicity,
    respectively, the notation used on the Synthetic Universe
    webpage. In the top plot, the number of coalescing systems with
    individual masses $\geq1.5$, $1.75$, and $2M_\odot$ are marked
    with dense hatching, less dense hatching, and dotted hatching,
    respectively.}
  \label{fig:n_Mi_and_q}
\end{figure}

\begin{figure}[t]
    \includegraphics[width=0.5\textwidth]{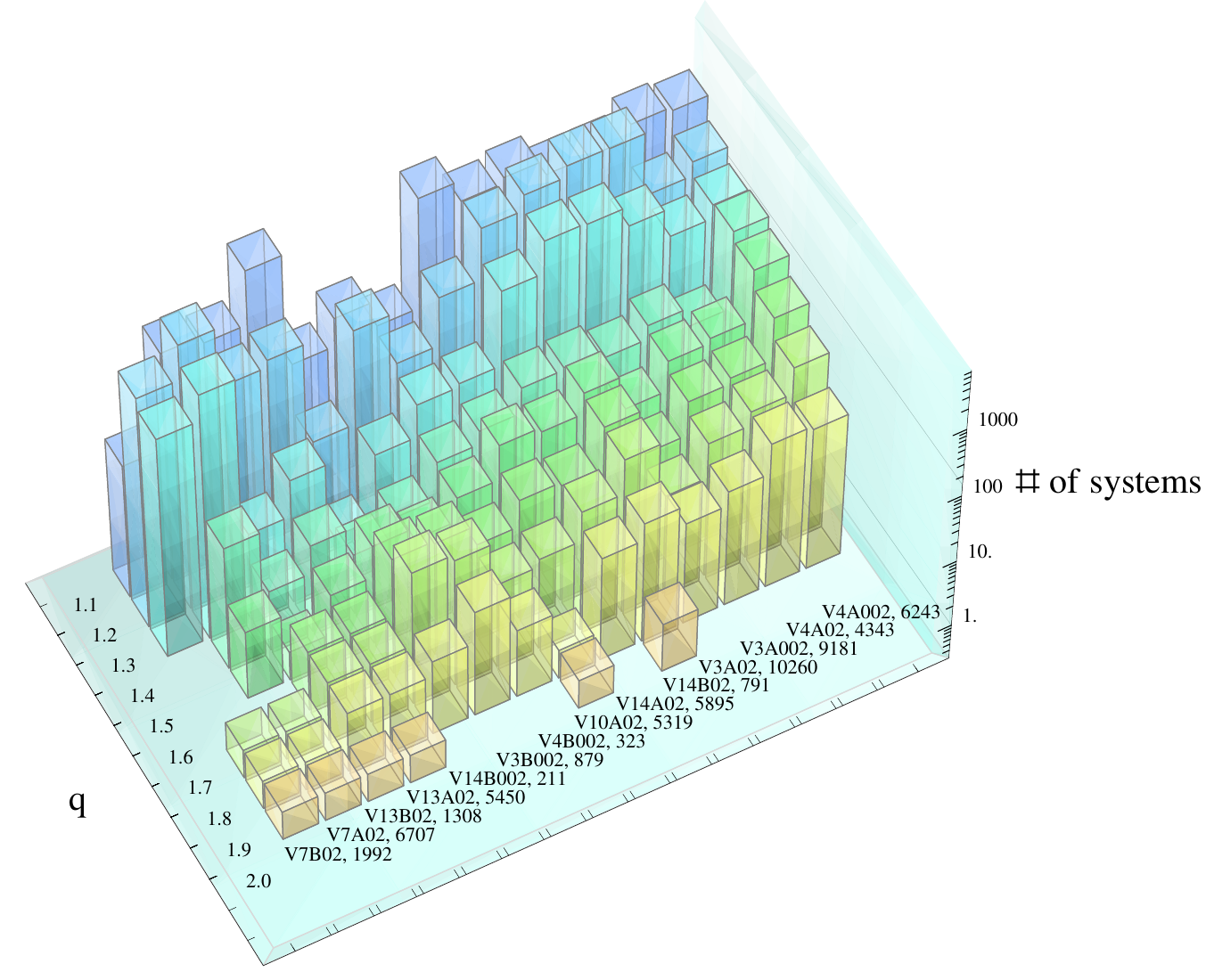}
    \caption{A histogram of mass ratio of coalescing systems for the
      various Synthetic Universe models that have a maximum (coalescing)
      mass ratio of $\geq 1.7$.  The number of coalescing systems in
      each model is given after the model's name (with the same notation
      as in Fig.~\ref{fig:n_Mi_and_q}). Each bin has size $0.1$ and is
      labeled by the maximum mass ratio present in the bin. The models
      are ordered to help clarify the plot.
      }
      \label{fig:mass_ratio}
\end{figure}

We show the total number of coalescing binaries in each of these
models in Fig.~\ref{fig:n_Mi_and_q}, marking the numbers of systems
with individual masses $\geq 1.5$, $1.75$, and $2M_\odot$. We also
show the maximum and minimum individual masses and mass ratio present
in the coalescing systems in that figure.  (Here we select the systems
that coalesce within $10$~Gyr from the formation of the binary, the
criterion for the population of potentially observable coalescing
binaries used in~\cite{DomBelFry12}.) We see that many of the models
predict at least some coalescing systems with individual masses close
to $2M_\odot$ (at least above $1.75M_\odot$). The minimum individual
masses present are mostly close to $1.1M_\odot$, though in a few
models they approach the minimum mass of $1M_\odot$, while in
Variation~$10$ (which assumes a delayed supernova engine), the minimum
mass in the low metallicity case is $1.2M_\odot$. The maximum total
mass is never more than $20\%$ smaller than twice the maximum
individual mass, and in almost half of the cases is within $5\%$ of it.

The minimum total mass is always very close to $2.2M_\odot$, except
for Variation~$1$, where it is $2.49M_\odot$ (for all four cases),
and Variation~$10$, where it is between $2.3$ and $2.5M_\odot$,
depending on the case. Indeed, mergers with a stable neutron star as
the final remnant may even be rather common, if the neutron star
maximum mass is $\gtrsim 2.5M_\odot$ as is assumed in all but
Variation 6: One finds hundreds to thousands of coalescing systems
with total masses below this limit in all but a handful of these
Synthetic Universe models. Moreover, these numbers are a significant
fraction of the total number of systems in each of the population
models (in a number of cases well above $90\%$, including the
standard case with solar metallicity and submodel B, where the
fraction is $99.88\%$). There are even tens to hundreds of coalescing
systems with total masses less than $2.25M_\odot$ in many models.  In
the models from~\cite{MinBel15}, which vary various initial
conditions using a base model quite similar to the standard model
from~\cite{DomBelFry12}, one does not find many of the more extreme
systems we consider here, since these models do not include the
choices for the binary evolution parameters that generate such systems. 

Since there is a wide range of individual masses in these models, one might expect
there to also be a wide range of mass ratios present, which we indeed find to be the case, as
illustrated in the bottom panel of Fig.~\ref{fig:n_Mi_and_q}.
We also show a histogram of the number of systems with different mass ratios for the models that have maximum
mass ratios $\geq1.7$ in Fig.~\ref{fig:mass_ratio}. The maximum mass ratio in the full (not just
the coalescing) population is $1.94$, found in Variation~$10$ with solar metallicity and either
submodel. The maximum mass ratio in the coalescing population is $1.86$ and is also found in Variation~$10$
with solar metallicity, though here just for submodel A. Note that this variation assumes a delayed supernova engine,
and thus may be unphysical, as discussed in~\cite{DomBelFry12}. We therefore note that Variation~$7$
(which assumes low supernova kicks) with solar metallicity gives a maximum mass ratio of $1.83$ in both the full and
coalescing populations for both submodels, as does Variation~$14$ (which assumes a weakly bound common envelope) for $0.1$
solar metallicity and submodel A.

While the very largest mass ratios are indeed uncommon even in these more extreme population models (see Fig.~\ref{fig:mass_ratio}),
two models predict $>50$ coalescing systems with mass ratios $\geq 1.7$ (Variation~$4$, which assumes a very weakly bound common envelope, with 
submodel A and either metallicity). These models are on the extreme side, but even the 
standard model predicts a single mass ratio $1.5$ coalescing system (with submodel A and solar metallicity). 

\subsection{Larger spins}
\label{sec:BNS_spins}
  
Compared to the spins of millisecond pulsars (where the largest spin
known is $716$~Hz~\cite{HesRanSta06}), the members of known binary
neutron stars are not spinning nearly so rapidly.\footnote{While PSR
  J1807$-$2500B (NGC 6445B) has a spin of $239$~Hz, it is at present
  unclear whether its companion is a neutron star or a white
  dwarf~\cite{LynFreRan12}. (The uncertainty about the nature of the
  companion is somewhat larger for this system than it is for most
  binary pulsars. This is also the most massive companion of a fully
  recycled neutron star.) Additionally, this system will not merge
  within a Hubble time, so it would not contribute directly to
  binary neutron star merger rate calculations.} The shortest spin
period observed so far is $22.7$~ms (corresponding to a frequency of
$44$~Hz) for the more massive star in the Double Pulsar, PSR
J0737$-$3039A, which is likely not to spin down too much by the time
the system merges, as discussed in Appendix~\ref{sec:spin_prediction}.
PSR J0737$-$3039A has a dimensionless spin of $j\in[0.02,0.03]$,
where the uncertainty comes from the uncertainty in the EOS. 
  
However, for a direct estimate of the typical spin in binary neutron
star systems, we are restricted by the fact that we currently only
know a small sample of the binary neutron star systems in our galaxy
(around twelve). Furthermore, the existence of the sizeable
population of rapidly spinning neutron stars, with $207$ known
pulsars with spins over $200$~Hz~\cite{ManHobTeo05}\footnote{This is
  out of $2525$ known pulsars. Also note that $20$ out of the $23$
  known pulsars in the rich globular cluster $47$~Tucanae have spins
  over $200$~Hz~\cite{Fre13}.}  suggests that there should be a
population of binary neutron stars where at least one star has a
significant spin at merger.  (Note that $200$~Hz corresponds to a
dimensionless spin of $\sim0.1$ for the EOSs considered in this
work.)  This would most likely be the heavier star, which could have
its spin increased by accretion from its companion when its
companion is still a post-main sequence star, a process known as
recycling (see, e.g.,~\cite{Tau11}). (The recycling process reduces
the star's external magnetic field, which will likely allow the
rapid spin obtained from accretion to persist until merger.)

However, one needs to accrete a fair amount of matter in order to
spin a neutron star up to high frequencies ($\sim 0.1M_\odot$ for
frequencies of $\sim 500$~Hz is quoted in Sec.~7 in~\cite{Tau11}),
and one does not expect to accrete lots of matter when the companion
is a neutron star progenitor. In particular, Tauris, Langer, and
Podsiadlowski~\cite{TauLanPod15} consider the ultra-stripped
supernova binary neutron star formation channel, which they claim is
the primary channel for forming binary neutron stars that will merge
within a Hubble time. In their calculations, they obtain a maximum
spin of $\sim 40$~Hz, assuming Eddington-limited accretion onto the
neutron star, and $\sim 90$~Hz, assuming three times Eddington
accretion (which they mention is easily obtained, and for which
there may be evidence in the inferred formation channels of other
systems); see their Sec.~6.1.1.

Additionally, MacLeod and Ramirez-Ruiz~\cite{MacRam15} find that
even though the neutron star only accretes $< 0.1M_\odot$ during the common
envelope phase in their simulations, the neutron star could still be spun up to
$\sim250$~Hz in the cases where one accretes close to $0.1M_\odot$.
Of course, if the neutron star experiences a significant
enough spindown after the accretion episode, then these high spins
will reduce substantially before merger. However, one expects the
magnetic field to be reduced by accretion, as mentioned above, and
for reasonably small periods and magnetic fields after the common
envelope phase (e.g., $\sim 4$~hours and $\sim 10^9$~G), the spin at
merger can still be $\sim 200$~Hz. See the next subsection for
further discussion of the issue of determining the spin at merger.

One can also imagine forming a double neutron star with a member
with a high spin through dynamical formation via binary-single (or
even binary-binary) interactions in dense stellar regions, such as
globular clusters (see, e.g.,~\cite{BenDow14,VerFre14}), where one
swaps out the companion that recycled the highly spinning neutron
star. This process is discussed as a likely formation channel for
the binary containing the $239$~Hz pulsar
J1807$-$2500B~\cite{LynFreRan12}, which resides in a globular
cluster, and whose companion may be another neutron star.  There is
even the exotic possibility of forming a binary where both neutron
stars have millisecond periods through this channel, or through the
double recycling scenario proposed by Sigurdsson and
Hernquist~\cite{SigHer92}, where the neutron stars' main sequence
companions are disrupted and this material recycles both neutron
stars.

\subsection{Neutron star spin predictions at merger}
\label{sec:spin_prediction}

In addition to considering the purely theoretical prospects for
relatively high spin in merging binary neutron stars, one can also
consider the spins at merger of known pulsars whose companion is a
neutron star. The simplest way to estimate the spin at merger is to
assume that the pulsar's observed spindown is due solely to magnetic
dipole radiation. This was already done for the fastest-spinning
neutron star known in a neutron star binary (J0737$-$3039A, the more
massive star in the Double Pulsar) by one of us in~\cite{Tic11}. There
it was found that this pulsar's spin will only decrease slightly, from
$44$ to $37$~Hz, in the $85$~Myr from now until the system merges. 
with this assumption.  Although the observed spindown is likely not
due entirely to magnetic dipole radiation, we will show that the
estimate for this particular pulsar is valid without this assumption
(with reasonable assumptions about the size of the pulsar's braking
index).

For magnetic dipole radiation the pulsar's braking index ($n :=
\nu\ddot{\nu}/\dot{\nu}^2$, where $\nu$ is the pulsar's spin
frequency) is $3$. However, all reliably measured values of the
braking index (only $8$) are less than $3$ (and can be as low as $\sim
1$); see, e.g.,~Table~I in~\cite{HamStoUrb15}.  Note that
all of the pulsars with reliably measured braking indices are much
younger than J0737$-$3039A, with ages of at most $\sim 10^4$~years, as
opposed to J0737$-$3039A's age of $\sim 10^8$~years.  If one looks at
these braking indices versus the pulsars' characteristic ages from the
ATNF catalogue~\cite{ManHobTeo05}, one sees that the significantly
older pulsars all have lower---but more uncertain---braking indices
than the younger ones. (See Fig.~7.5 in~\cite{Vig13} for an
illustration, albeit without error bars. They also plot less
well-determined braking indices for much older pulsars, finding quite
large values up to $\sim 10^5$, though they are also likely quite
uncertain.)
All these pulsars are also more slowly rotating than J0737$-$3039A,
with a largest frequency of $\sim30$~Hz for the Crab pulsar.
Nevertheless, it turns out that the prediction of J0737$-$3039A's spin
at merger is quite insensitive to the braking index
assumed, since the spindown timescale is considerably longer than the
time to merger.

Specifically, the period evolution of a pulsar with a generic
(constant) braking index $n$ is given by $\dot{P} = KP^{2-n}$, where
$K$ is a constant. We can solve this in terms of the pulsar's period
at $t = 0$, $P_0$, and express $K$ in terms of the pulsar's
characteristic age at $t = 0$, $\tau_0 := P_0/(2\dot{P}_0)$, giving
\begin{equation}
  \begin{split}
    P(t) &= P_0\left[1 + \frac{n - 1}{2}\frac{t}{\tau_0}\right]^{1/(n -
      1)}\\ &= P_0\left\{1 + \frac{1}{2}\frac{t}{\tau_0} + \frac{2 -
      n}{8}\left(\frac{t}{\tau_0}\right)^2 +
    O\left(\left[\frac{t}{\tau_0}\right]^3\right)\right\},
  \end{split}
\end{equation}
Here we have expanded to second order in $t/\tau_0$ to illustrate that
the correction term due to the braking index is quite small when
$t/\tau_0$ is small. In case of J0737$-$3039A, where $\tau_0 =
210$~Myr, so $t_\text{merger}/\tau_0 \simeq 0.4$, the corrections due
to possible deviations of the braking index from $3$ are $\lesssim
5\%$ (considering $n\in[0,5]$, where $n = 5$ corresponds to pure
gravitational wave damping). Additionally, for the very large
magnitude (and likely quite uncertain) braking indices found for
pulsars of about the age of J0737$-$3039A discussed above, the change
in the pulsar's spin until merger is negligible.

Of course, this calculation still assumes a constant braking index and
$K$. If, for instance, J0737$-$3039A has a large buried magnetic field
from the accretion episode that is thought to have spun it up (its
external field is only $6.3\times 10^9$~G, while its companion's is
$\sim 10^{12}$~G)~\cite{LorFreCar07}, and enough of this field becomes
unburied before the binary coalesces, then this could spin the pulsar
down far more than is predicted by the calculation above: Since the
spin-down rate goes as the square of the magnetic field, an increase
in the external field by an order of magnitude would decrease the
characteristic age $\tau$ by a factor of $100$, so one could easily
obtain a final spin that was smaller by a factor of $2$ or more if
this larger external field is present for $\sim 10$~Myr or longer. And
if a $\sim 10^{12}$~G field emerges completely, $\tau$ will decrease
to $\sim 10^4$~years, so such an external field could spin the star
down by an order of magnitude in only $\sim 1$~Myr.

However, this scenario relies on the pulsar's having accreted a small
enough amount of matter to allow a significant amount of field to
become unburied in the remaining time before merger. As illustrated
in, e.g., Fig.~1 of~\cite{YouCha95}, the unburial timescale is very
sensitive to the amount of accreted material (with factors of 2
leading to order of magnitude changes). Since there is no way of which
we are aware of estimating the amount of material accreted onto
J0737$-$3039A to within a factor of $2$, and the entire picture is
further complicated by the possibility of magnetic field decay (e.g.,
Ohmic decay), not just burial (see, e.g., Sec.~10.3.1 in~\cite{For12}
for a brief review), we do not pursue this possibility further here.

\section{Alternative derivation of a first integral to the Euler equation}
\label{sec:appendix_firstintegral}

We have already outlined a possible derivation of a first
integral to Euler's equation for rotating neutron stars, see
Sec.~\ref{sec:general_framework} and~\cite{Tic11,Tic12}.  In this
appendix we   obtain the first integral using a much shorter
derivation based on the Cartan identity, which   relates the Lie derivative operator $\mathcal{L}$ to the exterior derivative operator $\text{\bf{d}}$. In particular, for a  differential form $\vv \omega$ and a vector $\vv u$ one has
\begin{equation}
  \mathcal{L}_{\vv{u}}\vv\omega = \vv{u} \cdot  \text{\bf{d}}\vv\omega +
  \text{\bf{d}} (\vv{u} \cdot \vv\omega).
  \label{eq:Cartan_id}
\end{equation}
where    a dot denotes contraction between adjacent indices. We will now use the canonical momentum
\begin{equation}
\vv p = h \vv u = \text{\bf{d}}\phi+\vv w
  \label{eq:can_mom}
\end{equation}
as introduced in Eqs.~\eqref{eq:pdef} and
\eqref{eq:p_of_phi_w}, together with the additional
definition~\eqref{eq:kbardef}, to write
\begin{equation}
  \vv p = p^0 \vv{ \bar k} + \vv w = p^0 (\vv k + \vv V) ,
  \label{eq:pi_decomp}
\end{equation}
where in the last step we have used Eqs.~\eqref{eq:matterasm0} and
\eqref{eq:pdef}.

Using  $\vv u \cdot \vv u = -1$, the Euler
equation~(\ref{eq:Euler_h}) can be written  in the Carter-Lichnerowicz~\cite{Car79a,Lic67} form: 
\begin{equation}
  \vv u \cdot \text{\bf{d}} \vv p = 0 ,
  \label{eq:Euler_CLform}
    \end{equation}
Using
Eq.~(\ref{eq:matterasm0}), Eq.~(\ref{eq:Euler_CLform}) can be
rewritten as
\begin{equation}
  \vv k \cdot \text{\bf{d}} \vv p + \vv V \cdot \text{\bf{d}} \vv p = 0 .
\end{equation}
We now rewrite $\vv k \cdot \text{\bf{d}} \vv p$ using the Cartan
identity~\eqref{eq:Cartan_id} to obtain
\begin{equation}
  \mathcal{L}_{\vv k} \vv p - \text{\bf{d}}(\vv k \cdot \vv p) + \vv V
  \cdot\text{\bf{d}} \vv p= 0, \label{eq:Euler_CLform_mod1}
\end{equation}
where $\mathcal{L}_{\vv k} \vv p$ vanishes by assumption if the symmetry vector ${\vv k}$ Lie-derives the flow.  We could then obtain a first integral $\vv k \cdot \vv p = \text{const}$
for irrotational ($\text{\bf{d}} \vv p=0$) or corotational ($\vv V=0$) flow, since
then the last term is zero as well.  In the general case of
spinning neutron stars, however, that is not true.  Nevertheless,
we can again make use of our
assumptions~\eqref{eq:matterasm2},~\eqref{eq:matterasm3},
and~\eqref{eq:matrel3} (this time without projecting onto the
three dimensional slice), i.e.,
\begin{equation}
  \mathcal{L}_{\vv k}\text{\bf{d}}\phi \approx
  \mathcal{L}_{\vv {\bar k}} \vv w \approx \mathcal{L}_{\frac{\vv w}{
          p^0}} \vv w \approx 0,
  \label{eq:matterasms4d}
\end{equation}
where the latter two can be combined to get
\begin{equation}
  \mathcal{L}_{\frac{\vv p}{ p^0}} \vv w = \mathcal{L}_{\vv{ \bar k}}
 \vv w + \mathcal{L}_{\frac{\vv w}{ p^0}} \vv w \approx 0.
\end{equation}
Now it is possible to rewrite the first and the last term of
equation~\eqref{eq:Euler_CLform_mod1} with the goal of finding a
first integral.  Therefore, we can write $\mathcal{L}_{\vv k} \vv p =
\mathcal{L}_{\vv k}(\text{\bf{d}}\phi + \vv w) \approx
\mathcal{L}_{\vv k} \vv w$ to simplify the first term.  For the last
term, we will apply Cartan's identity once again and the
definitions from~\eqref{eq:can_mom} to write
\begin{equation}
  \vv V \cdot \text{\bf{d}}  \vv p = \vv V \cdot \text{\bf{d}} \vv w = \mathcal{L}_{\vv
    V} \vv w - \text{\bf{d}}(\vv V  \cdot \vv w).
\end{equation}
Substituting  these two terms into~\eqref{eq:Euler_CLform_mod1}
and  exploiting  linearity of the Lie derivative yields\begin{equation}
  \text{\bf{d}}(\vv k \cdot \vv p + \vv V \cdot \vv w) \approx \mathcal{L}_{\vv
    k+\vv V} \vv w = \mathcal{L}_{\vv{ \bar k + \frac{\vv w}{ p^0}}} \vv w
  \approx 0.
\end{equation}
which gives rise to an approximate first integral
\begin{equation}
  \vv k \cdot \vv p + \vv V \cdot\vv w \approx \text{const}
\end{equation}
In a final step, using the normalization
condition $\vv u \cdot \vv u = -1$, one can  straightforwardly   show that  $\vv k \cdot\vv p + \vv
 V \cdot \vv w = -\frac{h}{u^0}-\vv V \cdot \text{\bf{d}}\phi$, which
corresponds to equation~\eqref{eq:frist_integral}.

Notice, however, that the assumptions~(\ref{eq:matterasms4d}) used
in the derivation here are slightly stronger than the original
assumptions~\eqref{eq:matterasm2},~\eqref{eq:matterasm3}
and~\eqref{eq:matrel3}, which make assumptions only about
projections onto the three dimensional slice.

\section{Single CRV-stars}
\label{sec:single-CRV}

\subsection{Comparison with rigidly rotating stars}

As explained in~\cite{Tic12}, the particular choice of the angular
velocity $w^i$ given by~Eq.~\eqref{w_choice} leads to a negligible
shear, thus any substantial differential rotation can be neglected.
We provide corroborating evidence for this statement here.  For this
purpose we construct single rotating neutron stars in the CRV
approach and compare them with rigidly rotating stars. We compute
the rigidly rotating stars with the project \texttt{Nrotstar} of the
publicly available LORENE library~\cite{LORENE}.  \texttt{Nrotstar}
solves the Einstein equations with a self-consistent field method
and multi-domain spectral methods.  To compute CRV stars, we use
\SGRID and the formalism described in Sec.~\ref{sec:Method}.  In
particular, we take the approximate symmetry vector $\vv{k}$ to be
the timelike Killing vector $\vv{\partial_t}$. We also use
$n_A=n_B=26$, $n_\varphi=8$, and $n_\text{Cart}=22$.

\begin{figure}[t]
  \includegraphics[width=0.48\textwidth]{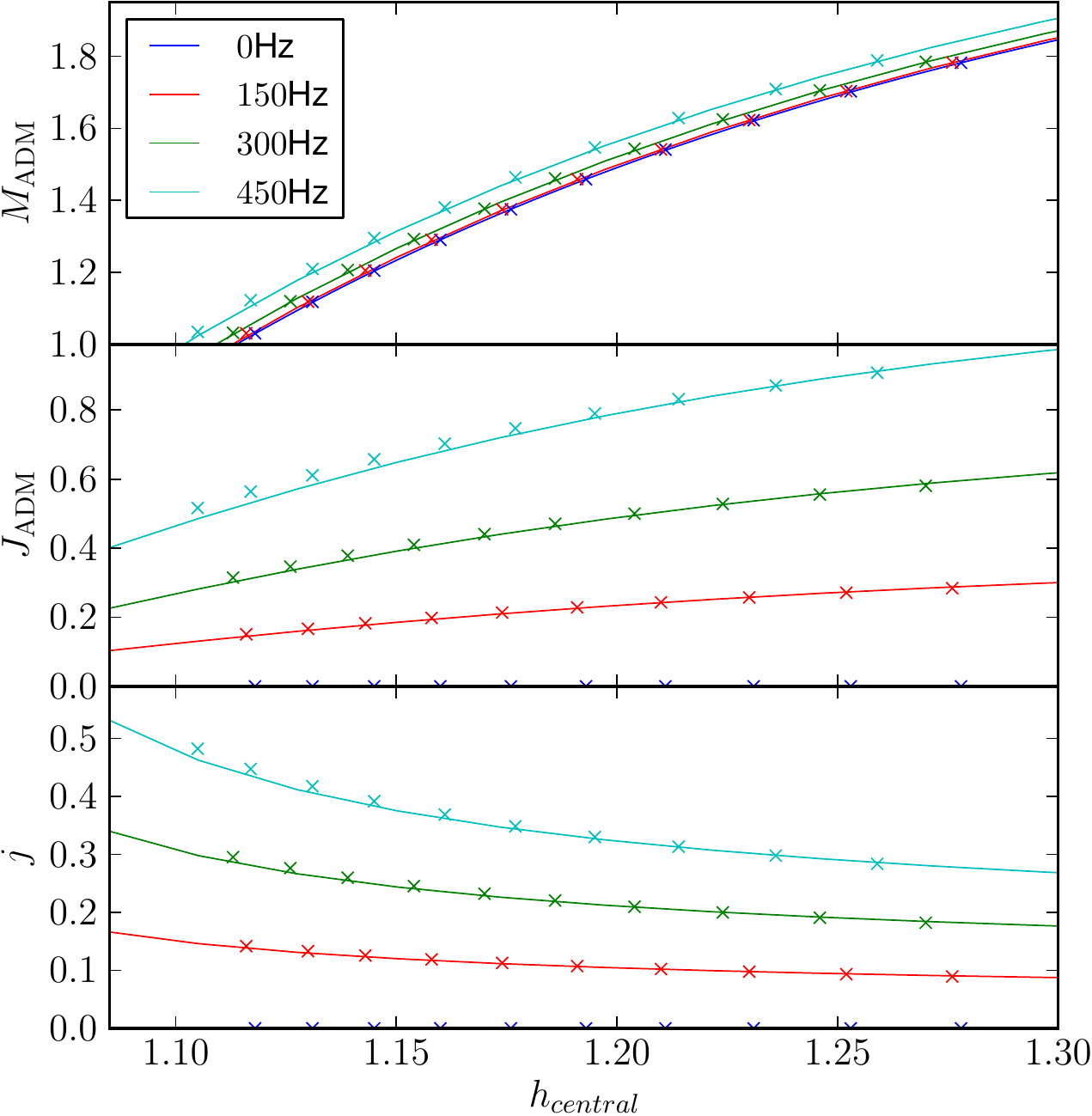}
  \caption{Comparison of single neutron stars with rigid rotation and
    CRV rotation using a simple polytropic EOS ($\Gamma2.72$).  The
    values for $\omega^z$ in the CRV-approach are 0.0, 0.005, 0.01,
    and 0.015. (These data are shown as crosses.)  The solid lines are
    computed for rigid rotating stars with the LORENE library.}
    \label{fig:sNS}
\end{figure}

In Fig.~\ref{fig:sNS} we present sequences of CRV and rigidly
rotating stars with the simple polytropic $\Gamma2.72$ EOS
for different central enthalpies (i.e. different baryonic masses).
There is no evidence that the result would be different for
piecewise polytropes.

Due to the fact that $\omega^z$ in the CRV approach does not agree
with the frequency measured by an observer at infinity (as already
outlined in~\cite{BerDieTic13}), we compute CRV configurations for a
constant $\omega^z$ and find a corresponding sequence of rigidly
rotating stars by choosing the frequency in \texttt{Nrotstar} such
that both methods give similar results. We obtain the \SGRID data by
varying the baryonic masses $M_{\text{b}} \in [1.1, 2.0]$ with a
spacing of $\Delta M_\text{b}=0.1$.  We clearly see that for
frequencies below $300$Hz, rigidly rotating data and CRV data agree
well.  However for larger angular momentum larger discrepancies
occur.  This can be caused by (i) the reduced accuracy for faster
rotating stars (see Sec.~\ref{sec:fast_rotating_pulsars}) and (ii)
the fact that the members of a sequence with different masses and
the same $\omega^z=\text{const}$ may not all correspond to the same
observable frequency.

\subsection{Empirical $\omega$-$j$-Relation}

\begin{figure}[t]
  \includegraphics[width=0.48\textwidth]{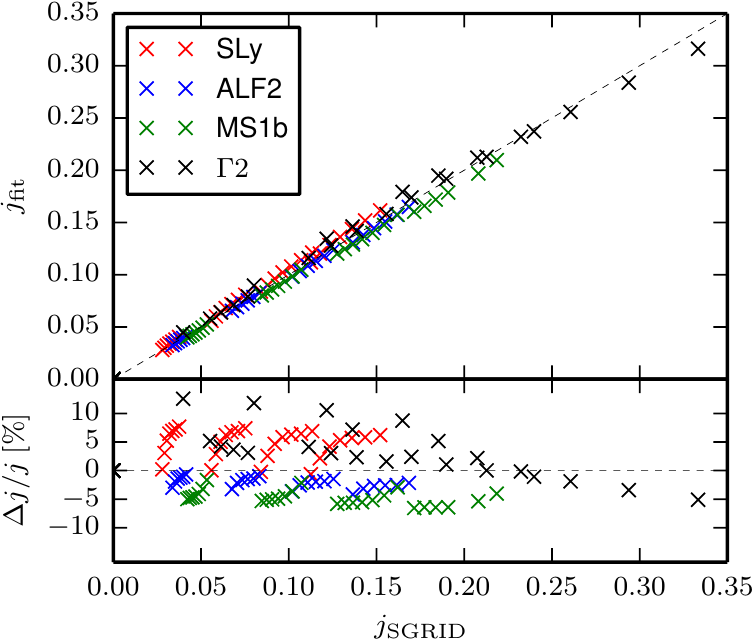}
  \caption{Comparing the spin computation according to
    Eq.~\eqref{eq:spin_estimate} ($j_\text{fit}$) with the spin output
    by \SGRID ($j_\text{\SGRID}$) for SLy (red), ALF2 (blue), MS1b (green), 
    and $\Gamma2$ (black).  Absolute values (upper panel) and fractional
    residuals (bottom panel). }
  \label{fig:sNS-EOS}
\end{figure}

In addition to the comparison with rigid rotating stars, we want to
answer the question of how the dimensionless spin of a single
CRV-star can be obtained from input parameters of the \SGRID code,
namely the EOS, the baryonic mass $M_\text{b}$, and the angular
velocity vector $\omega^i$.  A phenomenological model to obtain a
rough estimate of the star's spin value would reduce the
computational costs to find accurate initial data for particular
configurations.  For this reason we choose four EOS: SLy, ALF2,
MS1b, and $\Gamma2$ with baryonic masses in the range $[1.1,1.7]$,
spanning a range in the compactness of $\mathcal{C} \in
(0.09,0.20)$.

The dimensionless angular momentum $j$ of a neutron star is given by
\begin{equation}
  j = \frac{J_\text{ADM}}{M_\text{ADM}^2}= \frac{I \omega_{\rm
      obs}}{M_\text{ADM}^2}, \label{eq:dimspin_1}
\end{equation}
where $J_\text{ADM}$ and $M_\text{ADM}$ are the spacetime's ADM
angular momentum and mass, respectively, and $I$ the moment of
inertia of the star. 
Now $\omega_{\rm
    obs}$, the rotational period an observer at infinity would
measure, is not known \emph{a priori}, but probably depends linearly
on the angular velocity $\omega$, for slowly rotating neutron stars.
We thus recast Eq.~\eqref{eq:dimspin_1} in the form
\begin{equation}
  j= f(\mathcal{C},M_\text{b}) \omega \label{eq:dimspin:2}.
\end{equation}
Also, the gravitational mass of the single star (i.e.,
$M_\text{ADM}$) for this spacetime is not known in advance and is
thus absorbed in the function $f$.  We find with numerical
experiments the following expression:
\begin{equation}
j_\text{fit} = a_1 (1+ m_1 M_\text{b}) (1 + c_1 \mathcal{C} +c_2\mathcal{C}^2
  + c_3 \mathcal{C}^3 + c_4 \mathcal{C}^4)
  \omega, \label{eq:spin_estimate}
\end{equation}
where the parameters $a_1 = 88.8131$, $m_1=1.39522$, $c_1= - 19.003$,
$c_2 = 152.99$, $c_3 =-570.678$, $c_4= 806.896$ are obtained from a fit
of the data presented in Fig.~\ref{fig:sNS-EOS} and where we use the
compactness $\mathcal{C}$ of an irrotational star with the same
baryonic mass for simplicity.  Specifically, we computed five different values
 $\omega=(0.000,0.002,0.004,0.006,0.008)$ for each of the baryonic masses
$(1.1,1.2,1.3,1.4,1.5,1.6,1.7)$ and for the EOSs
SLy, ALF2, MS1b, and $\Gamma2$.  We want to stress that the resulting relation is
just empirical and probably does not represent any underlying physical properties.

\subsection{Rapidly rotating neutron stars}
\label{sec:fast_rotating_pulsars}

\begin{figure}[t]
   \includegraphics[width=0.48\textwidth]{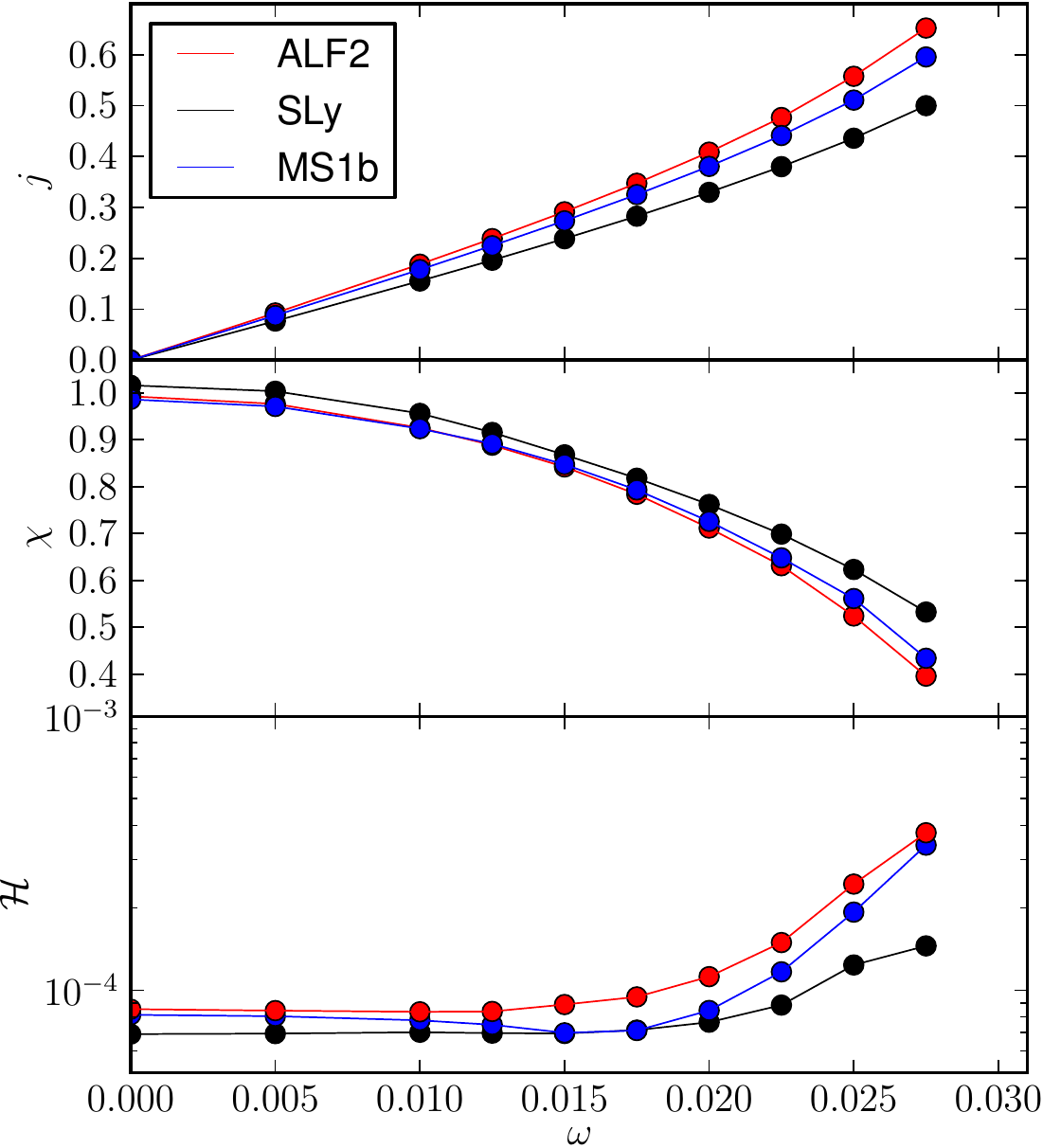}
   \caption{The dimensionless spin (top panel), mass shedding
     parameter (middle panel), and norm of the Hamiltonian constraint
     (bottom panel) versus the angular velocity $\omega$ for three EOS
     (SLy, ALF2, and MS1b).}
   \label{fig:sNS-highspin}
\end{figure}

The fastest spinning neutron star observed so far is PSR
J1748$-$2446ad, with a spin period of $1.4$~ms, corresponding to a
frequency of $716$~Hz~\cite{HesRanSta06}.  This corresponds to a
dimensionless spin of $j\in[0.3,0.6]$ (for the EOSs given in
Tab.~\ref{tab:listEOS}), where the uncertainty comes from our
ignorance of the EOS and the mass of the star.

Considering such systems, it is interesting to estimate the maximum
spin that can be achieved with \SGRID.  For this purpose we consider
once more the SLy, ALF2, and MS1b EOSs and single star
configurations.  The iteration process to achieve high spins for
realistic EOSs is as follows. We start with $\omega=0$ (zero spin,
i.e., a solution of the TOV-equation) and increase the angular
velocity in steps of $\Delta \omega= 0.005$ up to $\omega=0.01$ and
in smaller steps of $\Delta \omega = 0.0025$ up to $\omega=0.0275$.
Note that higher spins could be computed for lower resolutions, but
for higher resolutions the iteration procedure fails.
Depending on the EOS, maximum dimensionless spins of $0.5\leq
j_\text{max} \leq 0.7$ can be obtained (shown in the upper panel of
Fig.~\ref{fig:sNS-highspin}).  The middle panel shows the mass
shedding parameter~\cite{ShiTan11}
\begin{equation}
  \chi = \frac{\partial_r h |_\text{eq}}{\partial_r
    h|_\text{pole}} \label{eq:mass_shedding}
\end{equation}
which measures the deformation of the neutron star (caused by its
rotation) according to the derivative of the enthalpy at the star's
surface parallel or perpendicular to the symmetry
axis. Note that in binaries $\partial_r h |_\text{eq}$ is
evaluated along the line connecting the two stars' centers. 
The lower panel shows the $L^2$-norm Hamiltonian constraint in the
domain covering the neutron star up to $A_\text{max}=0.35$, i.e.,
including the star's surface, which is the most problematic region.
The Hamiltonian constraint grows for angular velocities $\omega \geq
0.02$.  We suggest that this is related to the deformed shape of the
neutron star, indicated by the decreasing $\chi$.  
  
\bibliography{Refs/refs}

\end{document}